\documentclass[prb,showpacs,preprintnumbers,amsmath,amssymb]{revtex4}

\usepackage{epsfig}
\usepackage{graphicx}
\usepackage{dcolumn}
\usepackage{bm}

\newcommand{\EQ}{\begin{equation}}
\newcommand{\EN}{\end{equation}}
\newcommand{\ea}{\end{eqnarray}}
\newcommand{\ba}{\begin{eqnarray}}

\newcommand{\bear}{\begin{eqnarray}}
\newcommand{\ear}{\end{eqnarray}}

\begin{document}

\title{Symmetry and spontaneous symmetry breaking of the Hubbard model on a square lattice ground state}
\author{J. M. P. Carmelo} 
\affiliation{GCEP-Centre of Physics, University of Minho, Campus Gualtar, P-4710-057 Braga, Portugal}

\date{25 March 2010}


\begin{abstract}
In this paper we consider the simplified form that a recently introduced general operator description 
of the Hubbard model on the square lattice with $N_a^2\gg 1$ sites, effective transfer integral $t$, and onsite repulsion $U$ 
has in a suitable one- and two-electron subspace. Such an operator description is
that consistent with the model exact global symmetry recently extended to $SO(3)\times SO(3)\times U(1)$.
Our study profits from the rotated-electron occupancy configurations in the original lattice 
that generate the energy eigenstates being described by occupancy configurations of
spin-$1/2$ spinons in a spin effective lattice, $\eta$-spin-$1/2$ $\eta$-spinons in a $\eta$-spin effective lattice, and 
$c$ fermions in a $c$ effective lattice. The latter three types of occupancy configurations
refer to state representations of the spin $SU(2)$ symmetry, $\eta$-spin $SU(2)$ symmetry,
and hidden $U(1)$ symmetry recently found in Ref. \cite{bipartite}, respectively. 
In the limit of very large number of lattice sites $N_a^2\gg 1$ that such a description 
refers to the emergence of the above three independent effective lattices simplifies 
the study of the effects of hole doping on the spin subsystem. 
In the one- and two-electron subspace the model refers to a square-lattice quantum liquid
that plays the same role for the Hubbard model on a square lattice as the
Fermi liquid for isotropic three-dimensional correlated perturbative models.
There is a large consensus that in the thermodynamic limit $N_a^2\rightarrow\infty$
long-range antiferromagnetic order occurs in the spin-density $m=0$ ground state of the
half-filled Hubbard model on the square lattice. Here we find that
the corresponding spontaneous symmetry breaking lowers the
symmetry of such a state from $SO(3)\times SO(3)\times U(1)$ for 
$N_a^2\gg 1$ large but finite to $[U(2)\times U(1)]/Z_2^2 \equiv [SO(3)\times U(1)\times U(1)]/Z_2$ for
$N_a^2\rightarrow\infty$. Moreover, we argue that the spin effective lattice being identical to the original lattice is a necessary 
condition for the occurrence of ground-state long-range antiferromagnetic order in
the limit $N_a^2\rightarrow\infty$. Consistently, strong evidence is provided that
for very small hole concentration $0<x\ll1$ the ground state has a short-range incommensurate-spiral spin order. (The related 
investigations of Ref. \cite{companion} provide evidence that a spin short-range order
exists for $0<x<x_*$, whereas for $x>x_*$ the ground state is a spin
disordered state. Here $x_* >0.23$ for approximately $U/4t>1.3$.)
Our results are of interest both for condensed-matter systems 
and ultra-cold fermionic atoms on an optical square lattice.
Elsewhere evidence is provided that upon addition of a weak three-dimensional
uniaxial anisotropy perturbation to the square-lattice quantum liquid, its short-range 
spin order coexists for $N_a^2\rightarrow\infty$, low temperatures, and a well-defined 
range of finite hole concentrations with a long-range superconducting order.
\end{abstract}
\pacs{71.10.Fd, 71.10.+w, 71.27.+a, 71.10.Hf, 71.30.+h}

\maketitle

\section{Introduction}

The Hubbard model on a square lattice 
is the simplest toy model for describing the effects of electronics 
correlations in the high-$T_c$ superconductors \cite{ARPES-review,2D-MIT} 
and their Mott-Hubbard insulators parent compounds \cite{LCO-neutr-scatt}.
In the case of such superconductors, addition of a weak three-dimensional (3D)
uniaxial anisotropy perturbation associated with a small 
effective transverse transfer integral $t_{\perp}$ is required to capture some of the
basic properties of their physics. 

The studies of this paper focus on the model on a square lattice. It has no exact solution, 
so that many open questions about its properties remain 
unsolved. A recent exact result, which 
applies to the model on any bipartite lattice \cite{bipartite}, 
is that for on-site repulsion $U>0$ its 
global symmetry is $SO(3)\times SO(3)\times U(1) =[SO(4)\times U(1)]/Z_2$. That
is an extension of the model well-known $SO(4)$ symmetry 
\cite{Lieb-89,Zhang}, which becomes explicit
provided that one describes the problem in
terms of the rotated electrons obtained from
the electrons by any of the unitary transformations
of the type considered in Refs. \cite{bipartite,Stein}. 
Another exact result useful for the studies of this paper
is that the ground state of the repulsive half-filled Hubbard model 
on a bipartite lattice is for finite number of lattice sites and spin 
density $m=0$ a spin singlet \cite{Lieb-89}.

The model can be experimentally realized with unprecedented 
precision in systems of ultra-cold fermionic atoms on an optical lattice 
with any geometry \cite{Zoller,cubic}. In the experimental context of cold-fermion optical lattices, 
Ref. \cite{optical-order} discusses the possibilities to approach the 
pseudogap or ordered phases by manipulating the scattering length or 
the strength of the laser-induced lattice potential. In turn, the large square lattices 
considered in Ref. \cite{half-filling} for the half-filled model
provide improved resolution of the GreenÕs function in 
momentum space, allowing a more quantitative 
comparison with time-of-flight optical lattice experiments. 
Such large scale determinant quantum Monte Carlo (DQMC) calculations provide 
as well useful information on the effective bandwidth, momentum distribution, 
and magnetic correlations of the half-Þlling model. 

\subsection{The model}

The studies of this paper profit from the simplified form that the general description of the Hubbard model on the 
square lattice with $N_a^2\equiv [N_a]^2\gg 1$ sites, effective nearest-neighbor transfer integral $t$, and onsite repulsion $U$ 
introduced in Ref. \cite{general} has in a suitable one- and two-electron subspace considered below. 
The hole concentration and spin density read $x= [N_a^2-N]/N_a^2$ and $m= [N_{\uparrow}-N_{\downarrow}]/N_a^2$,
respectively. The model on such a lattice with spacing $a$, $N_a\gg 1$ even, and lattice edge length $L=N_a\,a$ is given by,
\begin{equation}
\hat{H} = t\,\hat{T} + U\,[N_a^2-\hat{Q}]/2 
\, ; \hspace{0.25cm}
\hat{T} = -\sum_{\langle\vec{r}_j\vec{r}_{j'}\rangle}\sum_{\sigma}[c_{\vec{r}_j,\sigma}^{\dag}\,c_{\vec{r}_{j'},\sigma}+h.c.] \, ;
\hspace{0.25cm}
{\hat{Q}} = \sum_{j=1}^{N_a^2
}\sum_{\sigma =\uparrow
,\downarrow}\,n_{\vec{r}_j,\sigma}\,(1- n_{\vec{r}_j, -\sigma}) \, .
\label{H}
\end{equation}
Here $\hat{T}$ is the kinetic-energy operator in units of $t$ and ${\hat{Q}}$ is the operator that counts the 
number of electron singly occupied sites so that the operator ${\hat{D}}=[{\hat{N}}-{\hat{Q}}]/2$
counts the number of electron doubly
occupied sites, $n_{{\vec{r}}_j,\sigma} = c_{\vec{r}_j,\sigma}^{\dag} c_{\vec{r}_j,\sigma}$
where $-\sigma=\uparrow$ (and $-\sigma=\downarrow$)
for $\sigma =\downarrow$ (and $\sigma =\uparrow$), ${\hat{N}} = \sum_{\sigma}
{\hat{N}}_{\sigma}$, and ${\hat{N}}_{\sigma}=\sum_{j=1}^{N_a^2}
n_{{\vec{r}}_j,\sigma}$. 

The kinetic-energy operator $\hat{T}$ given in Eq. (\ref{H}) can be expressed in terms of the operators, 
\begin{eqnarray}
\hat{T}_0 & = & -\sum_{\langle\vec{r}_j\vec{r}_{j'}\rangle}\sum_{\sigma}[n_{\vec{r}_{j},-\sigma}\,c_{\vec{r}_j,\sigma}^{\dag}\,
c_{\vec{r}_{j'},\sigma}\,n_{\vec{r}_{j'},-\sigma} +
(1-n_{\vec{r}_{j},-\sigma})\,c_{\vec{r}_j,\sigma}^{\dag}\,
c_{\vec{r}_{j'},\sigma}\,(1-n_{\vec{r}_{j'},-\sigma})] \, ,
\nonumber \\
\hat{T}_{+1} & = & -\sum_{\langle\vec{r}_j\vec{r}_{j'}\rangle}\sum_{\sigma}
n_{\vec{r}_{j},-\sigma}\,c_{\vec{r}_j,\sigma}^{\dag}\,c_{\vec{r}_{j'},\sigma}\,(1-n_{\vec{r}_{j'},-\sigma}) \, ,
\nonumber \\
\hat{T}_{-1} & = & -\sum_{\langle\vec{r}_j\vec{r}_{j'}\rangle}\sum_{\sigma}
(1-n_{\vec{r}_{j},-\sigma})\,c_{\vec{r}_j,\sigma}^{\dag}\,
c_{\vec{r}_{j'},\sigma}\,n_{\vec{r}_{j'},-\sigma} \, ,
\label{T-op}
\end{eqnarray}
as $\hat{T}= \hat{T}_0 + \hat{T}_{+1} + \hat{T}_{-1}$. As discussed in Refs. \cite{bipartite,general,HO-04},
these three kinetic operators play an 
important role in the physics. The operator $\hat{T}_0$ does not change electron double 
occupancy whereas the operators $\hat{T}_{+1}$ and $\hat{T}_{-1}$ do it by $+1$ 
and $-1$, respectively.

\subsection{The general rotated-electron operator description}

The general operator representation introduced in Ref. \cite{general} applies for
$N_a^2\gg 1$ and refers to suitable quantum objects whose 
occupancy configurations generate the state representations of the group 
$SO(3)\times SO(3)\times U(1)$. Addition of 
chemical-potential and magnetic-field operator 
terms to the Hamiltonian lowers the model symmetry. 
Such terms commute with it so that its $4^{N_a^2}$ momentum and energy eigenstates
correspond to representations of that group for all values of $x$ and $m$.
It is shown in such a reference that all the physics of the model in the
whole Hilbert space can be obtained from that of the model
in the subspace spanned by the lowest-weight states (LWSs)
of both the $\eta$-spin $SU(2)$ and spin $SU(2)$ algebras.
In this paper we profit from the simpler properties that the
general operator description introduced in Ref. \cite{general} has in
a suitable {\it one- and two-electron subspace}. That subspace is spanned by 
a well-defined set of energy eigenstates, which for $U/4t>0$
are generated by simple momentum occupancy configurations of spin-less charge 
objects and spin-singlet objects, respectively. 

The general description introduced in Ref. \cite{general} refers
for $U/4t>0$ to a particular choice of the complete set of $4^{N_a^2}$ energy, momentum, $\eta$-spin,
$\eta$-spin projection, spin, and spin-projection eigenstates 
$\{\vert \Psi_{U/4t}\rangle\}$. In the limit $U/4t\rightarrow\infty$ such states correspond
to one of the many choices of sets $\{\vert\Psi_{\infty}\rangle\}$ of 
$4^{N_a^2}$ $U/4t$-infinite energy eigenstates. 
For the choice corresponding to the description of Ref. \cite{general}
there exists exactly one unitary operator ${\hat{V}}={\hat{V}}(U/4t)$
such that $\vert \Psi_{U/4t}\rangle ={\hat{V}}^{\dag}\vert\Psi_{\infty}\rangle$.
The point is that for most choices of the set of energy eigenstates $\{\vert\Psi_{\infty}\rangle\}$
the corresponding states $\vert \Psi_{U/4t}\rangle ={\hat{V}}^{\dag}\vert\Psi_{\infty}\rangle$
are not energy eigenstates for $U/4t>0$. For the description introduced in
that reference such states are energy eigenstates. The corresponding unitary operator ${\hat{V}}={\hat{V}}(U/4t)$ and
the rotated-electron operators,
\begin{equation}
{\tilde{c}}_{\vec{r}_j,\sigma}^{\dag} =
{\hat{V}}^{\dag}\,c_{\vec{r}_j,\sigma}^{\dag}\,{\hat{V}}
\, ; \hspace{0.35cm}
{\tilde{c}}_{\vec{r}_j,\sigma} =
{\hat{V}}^{\dag}\,c_{\vec{r}_j,\sigma}\,{\hat{V}}
\, ; \hspace{0.35cm}
{\tilde{n}}_{\vec{r}_j,\sigma} = 
{\tilde{c}}_{\vec{r}_j,\sigma}^{\dag}\,{\tilde{c}}_{\vec{r}_j,\sigma} \, ,
\label{rotated-operators}
\end{equation}
play a central role in that description.
The states $\vert \Psi_{U/4t}\rangle ={\hat{V}}^{\dag}\vert\Psi_{\infty}\rangle$
(one for each value of $U/4t>0$) that are 
generated from the same initial state $\vert\Psi_{\infty}\rangle$ 
belong to the same {\it $V$ tower}. 

Any operator ${\hat{O}}$ can 
be written in terms of rotated-electron creation and annihilation
operators as,
\begin{eqnarray}
{\hat{O}} & = & {\hat{V}}\,{\tilde{O}}\,{\hat{V}}^{\dag}
= {\tilde{O}}+ [{\tilde{O}},\,{\hat{S}}\,] + {1\over
2}\,[[{\tilde{O}},\,{\hat{S}}\,],\,{\hat{S}}\,] + ... 
=  {\tilde{O}}+ [{\tilde{O}},\,{\tilde{S}}\,] + {1\over
2}\,[[{\tilde{O}},\,{\tilde{S}}\,],\,{\tilde{S}}\,] + ... \, ,
\nonumber \\
{\hat{S}} & = & -{t\over U}\,\left[\hat{T}_{+1} -\hat{T}_{-1}\right] 
+ {\cal{O}} (t^2/U^2) \, ; \hspace{0.25cm}
{\tilde{S}} = -{t\over U}\,\left[\tilde{T}_{+1} -\tilde{T}_{-1}\right] 
+ {\cal{O}} (t^2/U^2) \, .
\label{OOr}
\end{eqnarray}
Here the operator ${\tilde{O}}={\hat{V}}^{\dag}\,{\hat{O}}\,{\hat{V}}$
has the same expression in terms of rotated-electron creation and 
annihilation operators as ${\hat{O}}$ in terms of electron creation and 
annihilation operators, respectively. The operator $\hat{S}$ appearing in Eq. (\ref{OOr})
is related to the unitary operator as ${\hat{V}}^{\dag} = e^{{\hat{S}}}$ and
${\hat{V}} = e^{-{\hat{S}}}$. Since for finite $U/4t$ values 
the Hamiltonian $\hat{H}$ of Eq. (\ref{H}) does not commute with 
the unitary operator ${\hat{V}} = e^{-{\hat{S}}}$, when expressed in terms 
of rotated-electron creation and annihilation operators it has an infinite 
number of terms. According to Eq. (\ref{OOr}) it then reads, 
\begin{equation}
{\hat{H}} = {\hat{V}}\,{\tilde{H}}\,{\hat{V}}^{\dag}
= {\tilde{H}} + [{\tilde{H}},\,{\tilde{S}}\,] + {1\over
2}\,[[{\tilde{H}},\,{\tilde{S}}\,],\,{\tilde{S}}\,] + ... \, .
\label{HHr}
\end{equation}
The commutator $[{\tilde{H}},\,{\tilde{S}}\,]$ does not vanish
except for $U/4t\rightarrow\infty$ so that ${\hat{H}} \neq {\tilde{H}}$ for finite values of $U/4t$. 

For $U/4t$ very large the Hamiltonian of Eq. (\ref{HHr}) 
corresponds in terms of rotated-electron creation and annihilation
operators to a simple rotated-electron $t-J$ model. In turn, the higher-order 
$t/U$ terms become increasingly important upon decreasing $U/4t$. They
generate effective rotated-electron hopping between second, 
third, and more distant neighboring sites. 
Indeed, the products of the kinetic operators
$\tilde{T}_0$, $\tilde{T}_{+1}$, and $\tilde{T}_{-1}$
contained in the higher-order terms of ${\tilde{S}} = -(t/U)\,[\tilde{T}_{+1} -\tilde{T}_{-1}] + {\cal{O}} (t^2/U^2)$
also appear in the Hamiltonian expression (\ref{HHr}) in terms of rotated-electron 
creation and annihilation operators of Eq. (\ref{rotated-operators}). In spite of the operators 
$\tilde{T}_0$, $\tilde{T}_{+1}$, and $\tilde{T}_{-1}$ generating 
only rotated-electron hopping between nearest-neighboring sites,
their products generate effective hopping between for instance second and
third neighboring sites. For instance, the Hamiltonian terms generated
up to fourth order in $t/U$ are within a unitary transformation the 
equivalent to the $t-J$ model with ring exchange and various correlated
hoppings \cite{HO-04}. The real-space distance in units of
the lattice spacing $a$ associated with the effective hopping between second and
third neighboring sites is for the model on the square lattice $\sqrt{2}\,a$ and $2\,a$, 
usually associated with transfer integrals $t'$ and $t''$, respectively. 
For instance, for the half filled Hubbard model on the square lattice in the subspace 
with both vanishing rotated-electron double occupancy and vanishing rotated-hole double occupancy the
interactions can be expressed completely by spin operators. The corresponding
Hamiltonian expression is given below in Section III-B up to fifth order in $t/U$. Some of the interactions 
in spin space in its $t^4/U^3$ terms refer indeed to pair of spins at second and third 
neighboring sites.

It follows that when expressed in terms of the
rotated-electron operators emerging from the specific
unitary transformation considered above the simple Hubbard model
(\ref{H}) as given in Eq. (\ref{HHr}) contains Hamiltonian terms associated with
higher-order contributions. Those can be effectively described by transfer 
integrals $t'$, $t''$, and of higher order. For hole concentration equal to or larger than zero 
both the model ground state and the excited states that 
span the one- and two-electron subspace considered below have vanishing rotated-electron double occupancy. 
Fortunately, for intermediate and large values of $U/4t$ obeying approximately 
the inequality $U/4t\geq u_0\approx 1.3$ and thus $t/U<0.2$, 
besides the original nearest-neighboring hopping processes  
only those involving second and third neighboring sites are
relevant for the square-lattice quantum liquid. That quantum liquid refers to 
the Hamiltonian of Eqs. (\ref{H}) and (\ref{HHr}) in the
one- and two-electron subspace. 
The value $U/4t=u_0\approx 1.302$ is that at which an important energy scale
$\Delta_0$ whose $U/4t$ dependence is studied below in Section III-C
reaches its maximum magnitude. 

Hence for approximately $U/4t\geq u_0$, out of the infinite terms on the right-hand-side
of Eq. (\ref{HHr}) only the first few Hamiltonian terms play
an active role in the physics of the Hubbard model on the square lattice
in the one- and two-electron subspace. Therefore, for intermediate and large values of $U/4t$ 
such a square-lattice quantum liquid can be mapped onto an effective $t-J$ model on
a square lattice with $t$, $t'=t'(U/4t)$, and $t''=t''(U/4t)$ transfer integrals. 
The role of the processes associated with 
$t'=t'(U/4t)$ and $t''=t''(U/4t)$ becomes increasingly important
upon decreasing the $U/4t$ value. 

\subsection{Three basic objects emerging from the rotated-electron description}

In addition to the spin $S_s$ and $\eta$-spin $S_{\eta}$ associated with
the spin $SU(2)$ and $\eta$-spin $SU(2)$ symmetries, respectively, of the model
global $SO(3)\times SO(3)\times U(1)=[SU(2)\times SU(2)\times U(1)]/Z_2^2$
symmetry, the eigenvalue $S_c$ of the generator of the hidden $U(1)$
symmetry found in Ref. \cite{bipartite} plays an important role in the
general  rotated-electron description of Ref. \cite{general}. It is such that
$2S_c$ is the number of rotated-electron singly occupied sites,
which is a good quantum number for $U/4t> 0$. 
We denote the $\eta$-spin and spin projection by $S^{x_3}_{\eta}= -[N_a^2-N]/2$ 
and $S^{x_3}_s= -[N_{\uparrow}-N_{\downarrow}]/2$, respectively. 
That the global symmetry of the Hubbard model on a bipartite lattice is 
$[SU(2)\times SU(2)\times U(1)]/Z_2^2$ rather than $SU(2)\times SU(2)\times U(1)$ is due to only state
representations such that $2S_{\eta}$, $2S_s$, and $2S_c$ are simultaneously
even or odd integer numbers being allowed \cite{bipartite}. The group
$SU(2)\times SU(2)\times U(1)$ has the same seven generators but four times 
more representations than the group $SO(3)\times SO(3)\times U(1)$.

Studies of the Hubbard model on the square lattice that rely on its transformation 
properties under symmetry operations \cite{Kampfer-05} should be extended to account for
the new hidden $U(1)$ symmetry recently found in Ref. \cite{bipartite}. Indeed
that the global symmetry of the Hubbard model on a square lattice and all 
other bipartite lattices is larger than 
$SO(4)$ and given by $SO(3)\times SO(3)\times U(1)$ is expected to 
have important physical consequences. 
In the case of the one-dimensional (1D) bipartite lattice the model
has an exact solution \cite{Lieb,Takahashi,Martins}. 
The main reason why its solution by the algebraic Bethe-ansatz
inverse scattering method \cite{Martins} was achieved only thirty years
after that of the coordinate Bethe ansatz \cite{Lieb,Takahashi} is
that it was expected that the charge and spin monodromy 
matrices had the same traditional ABCD form, found previously for the related 
1D isotropic spin $1/2$ Heinsenberg model \cite{ISM}. 
Such an expectation was that consistent with the occurrence
of a spin $SU(2)$ symmetry and a charge (and
$\eta$-spin) $SU(2)$ symmetry known long ago \cite{Zhang},
associated with a global $SO(4)=[SU(2)\times SU(2)]/Z_2$ 
symmetry. If that was the whole global symmetry of the 1D Hubbard model,
the charge and spin sectors would be associated with the
$\eta$-spin $SU(2)$ symmetry and spin $SU(2)$ symmetry,
respectively. A global $SO(4)=[SU(2)\times SU(2)]/Z_2$ symmetry
would then imply that the charge and spin monodromy 
matrices had indeed the same Faddeev-Zamolodchikov ABCD form.
Fortunately, Martins and Ramos
used an appropriate representation of the charge and spin monodromy 
matrices, which allows for possible {\it hidden symmetries} \cite{Martins}.
For the particular case of the bipartite 1D lattice the results of Ref. \cite{bipartite}
reveal that the hidden symmetry beyond $SO(4)$ is the charge global 
$U(1)$ symmetry found in that reference. For $U/4t>0$ the model charge and spin degrees of 
freedom are then associated with $U(2)=SU(2)\times U(1)$ and
$SU(2)$ symmetries, rather than with two $SU(2)$ symmetries,
respectively. The occurrence of such charge $U(2)=SU(2)\times U(1)$
symmetry and spin $SU(2)$ symmetry is behind
the different ABCDF and ABCD forms of the charge and spin monodromy 
matrices of Ref. \cite{Martins}, respectively.

Addition of chemical-potential and magnetic-field operator 
terms to the Hamiltonian (\ref{H}) lowers its symmetry. 
As mentioned above, such terms commute with it so that the global symmetry being
$[SU(2)\times SU(2)\times U(1)]/Z_2^2=SO(3)\times SO(3)\times U(1)$ 
implies that a set of $4^{N_a^2}$ independent rotated-electron occupancy configurations
generate the corresponding state representations of 
that global symmetry for all values of the electronic density $n=(1-x)$ and spin density 
$m$. Consistently, in Ref. \cite{bipartite} it is confirmed that the total number of 
such independent representations equals indeed the Hilbert-space dimension $4^{N_a^2}$. 
The rotated electron occupancy configurations  that generate such state representations are simpler 
to describe in terms of those of suitable related quantum objects. The eigenvalue
$S_c$ of the new hidden $U(1)$ symmetry controls the numbers of such objects,
which are as well good quantum numbers. Indeed the investigations of Ref. \cite{general} 
reveal the emergence within the rotated-electron description of three basic 
quantum objects: $M_s=2S_c$ spin-$1/2$ spinons, $M_{\eta}=[N_a^2-2S_c]$ 
$\eta$-spin-$1/2$ $\eta$-spinons, and $N_c=2S_c$ spin-less and $\eta$-spin-less 
charge $c$ fermions. The latter live on a lattice with $N_a^2 =[N_c+N_c^h]$ sites identical
to the original lattice. Here $N_c^h= [N_a^2-2S_c]$ gives the number of
$c$ fermion holes. 

The relation of such objects to the rotated electrons is as follows.
The $M_s=2S_c$ spin-$1/2$ spinons describe the spin degrees of freedom 
of the $2S_c$ rotated electrons that singly occupied sites. The $M_{\eta}=[N_a^2-2S_c]$
$\eta$-spin-$1/2$ $\eta$-spinons describe the $\eta$-spin degrees
of freedom of the $[N_a^2-2S_c]$ sites doubly occupied and unoccupied by rotated electrons. 
Specifically, the $\eta$-spinons of $\eta$-spin projection $-1/2$ and $+1/2$ refer to the sites doubly
occupied and unoccupied, respectively, by rotated electrons. The counting of the number of 
spinon and $\eta$-spinon independent occupancy configurations is an exercise fulfilled
in Ref. \cite{bipartite}. However, the internal structure of the specific spinon and $\eta$-spinon 
occupancy configurations of the $4^{N_a^2}$ $U/4t> 0$ energy eigenstates 
associated with the Ref. \cite{general} description is a complex problem,
which partially simplifies in the one- and two-electron subspace considered below.
In turn, in each rotated-electron configuration of a $U/4t> 0$ energy eigenstate, the
$N_c=2S_c$ $c$ fermions exactly occupy the same sites as the $2S_c$ rotated electrons
that singly occupy sites. Furthermore, the $N_c^h= [N_a^2-2S_c]$ sites unoccupied
by $c$ fermions are those doubly occupied and unoccupied by the rotated
electrons. The electronic charges of the $2S_c$ rotated electrons that
singly occupy sites are carried by the $N_c=2S_c$ $c$ fermions whereas their spins are
carried by the $M_s=2S_c$ spin-$1/2$ spinons.

For the general description of Ref. \cite{general} the rotated-electron occupancy configurations 
referring to (i) the spin degrees of freedom of the singly occupied and (ii) the $\eta$-spin
degrees of freedom of the unoccupied and doubly-occupied sites are independent. They 
correspond to the state representations of the spin $SU(2)$ symmetry $M_s=2S_c$ spin-$1/2$ spinons 
and $\eta$-spin $SU(2)$ symmetry $M_{\eta}=2S_c^h$ $\eta$-spin-$1/2$ $\eta$-spinons, 
respectively. In turn, the $U(1)$ symmetry state representations refer to the relative 
occupancy configurations of the $2S_c$ rotated-electron singly-occupied sites and $2S_c^h$ 
rotated-electron unoccupied and doubly-occupied sites. Those correspond to the $c$ fermion
occupancy configurations. 

Within the LWS representation of the general operator description introduced
in Ref. \cite{general}, the $c$ fermion creation operator has the following
expression in terms of the rotated-electron operators of Eq. (\ref{rotated-operators}),
\begin{equation}
f_{\vec{r}_j,c}^{\dag} =
{\tilde{c}}_{\vec{r}_j,\uparrow}^{\dag}\,
(1-{\tilde{n}}_{\vec{r}_j,\downarrow})
+ e^{i\vec{\pi}\cdot\vec{r}_j}\,{\tilde{c}}_{\vec{r}_j,\uparrow}\,
{\tilde{n}}_{\vec{r}_j,\downarrow} 
\, ; \hspace{0.35cm}
f_{\vec{q}_j,c}^{\dag} =
{1\over {\sqrt{N_a^2}}}\sum_{j'=1}^{N_a^2}\,e^{+i\vec{q}_j\cdot\vec{r}_{j'}}\,
f_{\vec{r}_{j'},c}^{\dag} \, .
\label{fc+}
\end{equation}
Here we have introduced the corresponding $c$ fermion momentum-dependent
operators as well and $e^{i\vec{\pi}\cdot\vec{r}_j}$ is $\pm 1$ depending on which
sub-lattice site $\vec{r}_j$ is on. The use of Eq. (\ref{rotated-operators}) allows the expression 
of the $c$ fermion operators (\ref{fc+}) in terms of electron operators. This involves 
the electron - rotated-electron unitary operator ${\hat{V}}$.
The $c$ momentum band is studied in Ref. \cite{companion} and has the same
shape and momentum area as the electronic first-Brillouin zone.

In turn, the three spinon local operators
$s^l_{\vec{r}_j}$ and three $\eta$-spinon local operators $p^l_{\vec{r}_j}$
such that $l=\pm,z$ are given by,
\begin{equation}
s^l_{\vec{r}_j} = n_{\vec{r}_j,c}\,q^l_{\vec{r}_j} \, ; \hspace{0.25cm}
p^l_{\vec{r}_j} = (1-n_{\vec{r}_j,c})\,q^l_{\vec{r}_j} \, , 
\hspace{0.15cm} l =\pm,x_3 \, .
\label{sir-pir}
\end{equation}
Here $q^{x_3}_{\vec{r}_j}$ and $q^{\pm}_{\vec{r}_j}= q^{x_1}_{\vec{r}_j}\pm i\,q^{x_2}_{\vec{r}_j}$
are the rotated quasi-spin operators $q^{l}_{\vec{r}_j}=s^l_{\vec{r}_j}+p^l_{\vec{r}_j}$. (We denote the Cartesian coordinates 
by $x_1,x_2,x_3$.) In addition,
\begin{equation}
n_{\vec{r}_j,c} = f_{\vec{r}_j,c}^{\dag}\,f_{\vec{r}_j,c} \, ,
\label{n-r-c}
\end{equation}
is the $c$ fermion local density operator.
In terms of rotated-electron creation and annihilation
operators the rotated quasi-spin operators read,
\begin{equation}
q^+_{\vec{r}_j} = s^+_{\vec{r}_j} + p^+_{\vec{r}_j} = ({\tilde{c}}_{\vec{r}_j,\uparrow}^{\dag}
- e^{i\vec{\pi}\cdot\vec{r}_j}\,{\tilde{c}}_{\vec{r}_j,\uparrow})\,
{\tilde{c}}_{\vec{r}_j,\downarrow} \, ; \hspace{0.25cm}
q^-_{\vec{r}_j} = (q^+_{\vec{r}_j})^{\dag} \, ;
\hspace{0.25cm}
q^{x_3}_{\vec{r}_j} = s^{x_3}_{\vec{r}_j} + p^{x_3}_{\vec{r}_j} = {1\over 2} - {\tilde{n}}_{\vec{r}_j,\downarrow} \, .
\label{rotated-quasi-spin}
\end{equation}

\subsection{Two qualitatively different types of spinon (and $\eta$-spinon) configurations and our notation}

An important result of Ref. \cite{general} is that a well-defined number of 
spin-$1/2$ spinons and $\eta$-spin-$1/2$ $\eta$-spinons remain
invariant under the electron - rotated-electron unitary transformation.
We call those independent spinons and independent $\eta$-spinons,
respectively. In the one- and two-electron subspace considered in this
paper there are no $\eta$-spin-projection $-1/2$ $\eta$-spinons.
The independent spinons and the independent $\eta$-spin-projection 
$+1/2$ $\eta$-spinons have vanishing energy and momentum. 
There are no independent spinons in $m=0$ ground states.

The spinons that are not invariant 
under the electron - rotated-electron unitary transformation are
confined within spin-neutral $2\nu$-spinon composite $s\nu$ fermions. Here $\nu=1,2,...$
is the number of spinon pairs confined within a spin-neutral $2\nu$-spinon composite $s\nu$ fermion. 
The description of Ref. \cite{general} used in our studies refers to a number of
sites $N_a^2\gg 1$ very large but finite. Within such a description all $m=0$ ground-state 
spinons are confined within spin-neutral
two-spinon $s1$ fermions. The condition
that in an excited state spinons are deconfined is that they are invariant
under the electron - rotated-electron unitary transformation.
For instance, a spin-triplet excited state involves two independent
spinons that correspond to an isolated mode below a continuum of
two-$s1$-fermion-hole excitations. 

For a given state, the values of the numbers $L_{\eta,\,\pm 1/2}$ of independent 
$\pm 1/2$ $\eta$-spinons and $L_{s,\,\pm 1/2}$ of independent $\pm 1/2$ spinons 
are fully determined by those of the 
$\eta$-spin $S_{\eta}$ and $\eta$-spin projection $S_{\eta}^{x_3}=-x\,N_a^2$ 
and spin $S_{s}$ and spin projection $S_{s}^{x_3}=-m\,N_a^2$, respectively, as follows,
\begin{equation}
L_{\alpha} = [L_{\alpha,-1/2}+L_{\alpha,+1/2}]=2S_{\alpha} \, ;
\hspace{0.25cm}
L_{\alpha,\,\pm 1/2} = [S_{\alpha}\mp S_{\alpha}^{x_3}]
\, ; \hspace{0.25cm} \alpha = \eta \, , s \, .
\label{L-L}
\end{equation}
The invariance of such independent $\eta$-spinons (and spinons)
stems from the off diagonal generators of the $\eta$-spin
(and spin) algebra, which flip their $\eta$-spin (and spin), commuting with 
the unitary operator $\hat{V}$. Hence it follows from Eq. (\ref{OOr})
that such generators have for $U/4t>0$ the same expressions in terms of electron and
rotated-electron operators.

Alike the spin-$1/2$ spinons, the $\eta$-spin-$1/2$ $\eta$-spinons that are
not invariant under the electron - rotated-electron unitary transformation
are confined within $\eta$-spin-neutral $2\nu$-$\eta$-spinon composite $\eta\nu$ fermions.
Within the notation of Ref. \cite{general} also used in this paper, $\nu = 1,...,C_{\eta}$
and $\nu = 1,...,C_{s}$ are the numbers of $\eta$-spinon and spinon pairs
confined within a composite $\eta\nu$ fermion and a $s\nu$ fermion, respectively. Their maximum 
values $C_{\eta}$ and $C_{s}$, respectively, are 
given in Eq. (\ref{N-h-an}) of Appendix A.
In principle there are some occupancy configurations of 
such objects that generate the exact momentum and energy eigenstates  
for $U/4t>0$, yet the detailed internal structure of such configurations
is a complex problem \cite{general}. 

Accordingly to the studies of Ref. \cite{general} that refer to $N_a^2\gg1$ very
large but finite the $m=0$ and $x\geq 0$ ground states are spin singlets with $N_{s1}=M_s/2=S_c=N/2$
two-spinon $s1$ fermions, $N_c=2S_c=N=(1-x)N_a^2$ $c$ fermions, $L_{\eta,+1/2}=[N_a^2-2S_c]=x\,N_a^2$
$\eta$-spinon-projection $+1/2$ independent $\eta$-spinons, no independent spinons, no
$s\nu$ fermions with $\nu>1$ spinon pairs, and no $\eta\nu$ fermions. It is found below that for 
the Hubbard model on the square lattice in the one- and two-electron subspace considered in the 
studies of this paper only the $c$ fermions and the two-spinon $s1$ fermions play an active role.  
In the two quantum liquids referring to (i) $x=0$ and $m=0$ and (ii) $x>0$ and $m=0$, respectively, considered
in our studies all the ground-state $M_s=2S_c$ spinons are for $N_a^2\gg1$ very large but finite confined within 
$N_{s1}=M_s/2$ spin-neutral two-spinon $s1$ fermions. Spinon confinement and the corresponding
deconfined degrees of freedom in second-order phase transitions is a problem of
physical interest \cite{Senthil-04,Anders-07,Anders-10}. 

In summary, the degrees of freedom of the rotated-electron occupancy configurations of each 
of the sets of $N_{a_{\eta}}^2=[N_a^2-2S_c]$ and  $N_{a_{s}}^2=2S_c$ sites of the original 
lattice that generate the energy eigenstates separate into
two types of configurations. A first type of occupancy configurations are those
of the $c$ fermions associated with the 
operators $f_{\vec{r}_j,c}^{\dag}$ of Eq. (\ref{fc+}) where $j=1,...,N_a^2$.
Such occupancy configurations correspond to the state representations
of the global $U(1)$ symmetry found in Ref. \cite{bipartite}. The $c$
fermions live on a $c$ effective lattice, which is identical to the original lattice. 
The $N_c = N_{a_{s}}^2=2S_c$ $c$ fermions occupy the sites singly occupied by the
rotated electrons. The $N_c^h = N_{a_{\eta}}^2=[N_a^2-2S_c]$ rotated-electron doubly-occupied
and unoccupied sites are those unoccupied by the $c$ fermions. The
$c$ fermion occupancy configurations describe the relative positions
in the original lattice of the $N_{a_{\eta}}^2=[N_a^2-2S_c]$ sites of the
$\eta$-spin effective lattice and $N_{a_{s}}^2=2S_c$ sites
of the spin effective lattice.

Consistently, the remaining degrees of freedom of rotated-electron
occupancies of the sets of $N_{a_{\eta}}^2=[N_a^2-2S_c]$ and  
$N_{a_{s}}^2=2S_c$ original-lattice sites correspond to a second
type of occupancy configurations. Those are 
associated with the $\eta$-spin $SU(2)$ symmetry
and spin $SU(2)$ symmetry representations,
respectively. The occupancy configurations of the set of
$N_{a_{\eta}}^2=[N_a^2-2S_c]$ sites of the
$\eta$-spin effective lattice and set of $N_{a_{s}}^2=2S_c$ sites
of the spin effective lattice are independent. The former configurations 
refer to the operators $p^l_{\vec{r}_j}$ of Eq. (\ref{sir-pir}), which
act only onto the $N_{a_{\eta}}^2=[N_a^2-2S_c]$ sites of
the $\eta$-spin effective lattice. The latter configurations
correspond to the operators 
$s^l_{\vec{r}_j}$ given in the same equation, which
act onto the $N_{a_{s}}^2=2S_c$ sites of
the spin effective lattice, respectively.
This is assured by the operators $(1-n_{\vec{r}_j,c})$
and $n_{\vec{r}_j,c}$ in their expressions provided
in that equation, which play the role of projectors
onto the $\eta$-spin and spin effective lattice, respectively.

The spin-singlet two-spinon composite local $s1$ fermions of the
description of Ref. \cite{general} live on their own $s1$ effective lattice.
One occupied site of such a $s1$ effective lattice involves two sites of the spin effective
lattice. Consistently, for spin density $m=0$ and spin $S_s=0$ states of the 
one- and two-electron subspace considered below the number of sites of the $s1$ effective lattice reads
$N_{a_{s1}}^2= N_{a_{s}}^2/2=S_c$. As justified in Refs. \cite{general,companion0},
for $N_a^2\rightarrow\infty$ both the spin and $s1$ effective lattices 
are square lattices with spacing $a_s$ and $a_{s1}$, respectively, related 
as $a_{s1} \sqrt{2}\,a_s$. Some further basic information on the general description 
introduced in Ref. \cite{general} needed for the studies of this paper is 
provided in Appendix A. 

\subsection{The goals and organization of this paper}

Our general aim is to show that in the one- and two-electron subspace 
considered below the physics of the Hubbard model on the square lattice
simplifies and refers to the square-lattice quantum liquid further studied
in Ref. \cite{companion}. The set of energy eigenstates that span such a subspace 
are generated by momentum occupancy configurations of the $c$ and $s1$ fermions.
Evidence that such a subspace is associated with nearly the whole spectral weight generated
by applying one- and two-electron operators onto $m=0$ and $x\geq 0$ ground states
is provided. It is based on the relative amount of such a weight generated by application
onto these states of $c$ and $\alpha\nu$ fermion
operators. The ground-state configurations of such objects were studied in Ref. \cite{general}. 
Although part of our results are argued on phenomenological grounds, they emerge naturally
from the scenario provided by the interplay of symmetry and the suitable quantum-object operator 
description used in this paper. It is found that for the Hubbard model on the square lattice in the 
one- and two-electron subspace only the $c$ and $s1$ fermions play an active role. Hence one may neglect the
remaining $\alpha\nu$ fermion branches considered in Ref. \cite{general}. The states of such a subspace
may have none or one spin-neutral four-spinon $s2$ fermion but such an object has vanishing momentum and 
vanishing energy. In Ref. \cite{companion} important physical quantities of the square-lattice quantum liquid 
introduced here are expressed in terms of $c$ and $s1$ fermion energy dispersions and velocities. 
It is confirmed in Sections II-A and II-B that for simple one- and two-electron operators ${\hat{O}}$ 
the leading operator term ${\tilde{O}}$ on the right-hand side of Eq. (\ref{OOr}) 
generates nearly the whole spectral weight. For such operators the terms of the general expression 
(\ref{OOr}) containing commutators involving the related operator $\hat{S}={\tilde{S}}$
are found to generate very little spectral weight.
Hence one can reach a quite faithful representation of
such operators by expressing them in terms of the
$c$ and $s1$ fermion operators.

In this paper strong evidence is found that provided that in the thermodynamic limit $N_a^2\rightarrow\infty$
a long-range antiferromagnetic order occurs in the ground state of the related isotropic spin-$1/2$ Heisenberg
model on the square lattice, a similar long range order sets in in that limit in the ground state
of the half-filled Hubbard model on the square lattice for $U/4t>0$. However,
it is not among our goals providing a mathematical proof that in the thermodynamic limit $N_a^2\rightarrow\infty$,
$x=0$, $m=0$, and vanishing temperature $T=0$ such a long-range antiferromagnetic order sets in. Although 
there is no such a proof, there is a large consensus that it should be so in both the isotropic
spin-$1/2$ Heisenberg model on the square lattice \cite{Hirsch85,Peter-88,Sandvik}
and in the half-filled Hubbard on the square lattice \cite{half-filling,2D-A2,bag-mech,2D-NM,Hubbard-T*-x=0,Kopec}. 
One of our aims is though providing useful physical information on how the occurrence of a long-range antiferromagnetic order
and a short-range spin order at $x=0$ and $x>0$, respectively, in the thermodynamic limit $N_a^2\rightarrow\infty$,
$m=0$, and vanishing temperature $T=0$ is related to the properties of the spin effective lattice.
Indeed, one of our goals is to show that within the operator description used in this paper
the study of the effects of hole doping on the spin subsystem simplifies. That simplification follows from the independence that the 
state representations of the spin $SU(2)$ symmetry, $\eta$-spin $SU(2)$ symmetry, and hidden $U(1)$ 
symmetry recently found in Ref. \cite{bipartite}, respectively, have
within the present description. Such state representations correspond to independent occupancy configurations of
the spin-$1/2$ spinons in the spin effective lattice, $\eta$-spin-$1/2$ $\eta$-spinons in the $\eta$-spin effective lattice, and 
$c$ fermions in the $c$ effective lattice, respectively. Specifically, at $m=0$ spin density one of the main effects 
of hole doping is on the number $M_s=2S_c$ of sites of the spin effective lattice, which at $x=0$ and
$x>0$ hole concentrations equals and differs from that of the original lattice, respectively. (We recall
that for $x>0$ the spin effective lattice is well defined only for $N_a^2\gg 1$ very large or infinite \cite{general}.)

The microscopic processes corresponding to the effective transfer integrals $t'=t'(U/4t)$ and $t''=t''(U/4t)$ 
of the Hamiltonian (\ref{HHr}) expressed in terms of creation and annihilation rotated-electron operators
are needed to characterize the type of order of the square-lattice quantum liquid. 
This holds for instance concerning the short-range incomensurate-spiral spin order considered in 
Section III-C. The qualitative changes occurring in the spin effective lattice due
to hole doping are behind processes that destroy long-range antiferromagnetic order
not being active and being active at $x=0$ and for $x>0$, respectively. 
In this paper strong evidence is provided that
for $0<x\ll1$ the ground state has a short-range incomensurate-spiral spin order. 
The related investigations of Ref. \cite{companion} provide evidence that a spin short-range order
exists for $0<x<x_*$, whereas for $x>x_*$ the ground state is a spin
disordered state without short-range spin order. Here $x_*$ is a critical hole concentration whose
magnitude is for approximately $U/4t>1.3$ an increasing function of $U/4t$.
For the intermediate $U/4t$ values of interest for the square-lattice quantum
liquid studies of that reference it reads $x_*=0.23$ for $U/4t\approx 1.3$ and $x_*=0.28$ 
for $U/4t\approx 1.6$. 

Another goal of this paper is defining the symmetry of the $m=0$ and $x=0$ ground state
after its symmetry is spontaneously broken in the limit $N_a^2\rightarrow\infty$, upon the
emergence of the long-range antiferromagnetic order. It is found that at $x=0$, $m=0$, and 
vanishing temperature $T=0$ that state symmetry is broken from $SO(3)\times SO(3)\times U(1)$ for  
$N_a^2\gg 1$ large but finite to $[U(2)\times U(1)]/Z_2^2 \equiv [SO(3)\times U(1)\times U(1)]/Z_2$ 
for $N_a^2\rightarrow\infty$. Finally, the $U/4t$ dependence of the energy order parameters of the $x=0$
antiferromagnetic order and $0<x\ll1$ short-range incomensurate-spiral spin order are
issues also addressed in this paper.   
   
The paper is organized as follows. The introduction and basic information on the
general operator description introduced in Ref. \cite{general} is given in Section I. 
In Section II a suitable one- and two-electron subspace is considered. The form that
the spin and $s1$ effective lattices have in it as well as the corresponding elementary 
excitations are issues also addressed in that section. In Section III it is shown that 
for $N_a^2\rightarrow\infty$ our results are consistent with a Mott-Hubbard insulating 
ground state with long-range antiferromagnetic order at half filling and a ground state with 
short-range spin order for a well-defined range of finite hole concentrations. 
Strong evidence is given that for $0<x\ll1$ and intermediate and large values of $U/4t$ the 
short-range spin order has an incommensurate-spiral character. 
The $U/4t$ dependence of the energy order parameters is also addressed.
Finally, Section III contains the concluding remarks.
        
\section{The square-lattice quantum liquid: A two-component fluid
of charge $c$ fermions and spin-neutral two-spinon $s1$ fermions}

In this section we consider a suitable one- and two-electron subspace and
study the form that the spin and $s1$ effective lattices have in it. To achieve our goals, 
in the following we address the problem of the generation of the one- and 
two-electron spectral weight in terms of processes of $c$ fermions and spinons.
The picture that emerges is that of a two-component quantum liquid
of charge $c$ fermions and spin neutral two-spinon $s1$ fermions.
It refers to the square-lattice quantum liquid introduced in this paper. 

\subsection{The one- and two-electron subspace}

Let $\vert \Psi_{GS}\rangle$ be
the exact ground state for $x\geq 0$ and $m=0$ and
${\hat{O}}$ denote an one- or two-electron operator.
Then the state,
\begin{equation}
{\hat{O}}\vert \Psi_{GS}\rangle = \sum_j C_j \vert \Psi_{j}\rangle 
\, ; \hspace{0.5cm} C_j = \langle\Psi_{j}\vert{\hat{O}}\vert \Psi_{GS}\rangle \, ,
\label{1-2-subspace}
\end{equation}
generated by application of ${\hat{O}}$ onto that ground
state is contained in the general one- and two-electron subspace.
This is the subspace spanned by the set of energy eigenstates
$\vert \Psi_{j}\rangle$ such that the corresponding coefficients $C_j $ are not vanishing.
Such a subspace must contain all excitations ${\hat{O}}\vert \Psi_{GS}\rangle$
with the operator ${\hat{O}}$ being any of a well defined set of operators. It includes
the creation and annihilation electron operator and
the whole set of simple two-electron operators.
For $x\geq 0$ and $m=0$ the $c$ and $s1$ fermion occupancies of the ground
states $\vert \Psi_{GS}\rangle$ are found in Ref. \cite{general}.

Our goal here is finding what occupancy configurations of the objects of the description
of such a reference generate a set of excited energy eigenstates $\{\vert \Psi_{j}\rangle\}$
such that $\sum_j \vert C_j\vert^2\approx 1$ for ${\hat{O}}$ being a creation and
annihilation electron operator and all simple two-electron operators. That set of states must contain
nearly the whole spectral weight of the above one- and two-electron excitations. Their
general form is provided below in Section II-E. It refers to a particular case of the
momentum eigenstates considered in Ref. \cite{general}, which in general are not
energy eigenstates. Fortunately, the set of such states that span the one- and two-electron
subspace are found to be energy eigenstates. For an initial $x> 0$ (and $x=0$) and $m=0$ ground state 
$\vert \Psi_{GS}\rangle$ evidence is provided in the following that such states have 
excitation energy $\omega<2\mu$ (and $\omega<\mu^0$). 
The inequality $\omega<2\mu$ applies to some range of finite hole concentrations
$x>0$ and spin density $m=0$. In turn, for the $\mu=0$ and $m=0$ absolute ground state with the
chemical-potential zero level at the middle of the Mott-Hubbard gap
the smallest energy required for creation
of either one rotated-electron doubly occupied site or one rotated-hole doubly occupied site is
instead given by the energy scale $\mu^0$. This justifies why
for such initial ground state the above inequality
$\omega<2\mu$ is replaced by $\omega<\mu^0$.

From the use of the invariance under the electron - rotated-electron
unitary transformation of the independent $\pm 1/2$ spinons \cite{general},
one finds that the number $L_{s} = [L_{s,-1/2}+L_{s,+1/2}]=2S_{s}$ of Eq. (\ref{L-L})
for $\alpha =s$ of such objects 
generated by application of $\cal{N}$-electron operators 
onto a ground state is exactly restricted to the following range,
\begin{equation}
L_s = [L_{s,\,-1/2} + L_{s,\,+1/2}] = 2S_s = 0,1,2,...,{\cal{N}} 
\, . \label{srcsL}
\end{equation}
It follows that for the one- and two-electron
subspace only the values  
$L_s =[L_{s,\,-1/2} + L_{s,\,+1/2}] = 2S_s = 0,1,2$ are
allowed. Such a restriction is exact for both the model on the square and 1D lattices,
as well as for any other bipartite lattice.

For a finite number $\nu\geq 2$ of spinon pairs the $s\nu$ fermions created onto 
a $x\geq 0$ and $m=0$ ground state have vanishing energy and momentum \cite{general}.
A vanishing spin density $m=0$ refers to a vanishing magnetic field $H=0$.
Hence such objects obey the criterion $\epsilon_{s\nu} = 2\nu\mu_B\,H = 0$ of
Eq. (\ref{invariant-V}) of Appendix A, so that they are invariant under 
the electron - rotated-electron unitary transformation. Therefore,
for $U/4t>0$ they correspond to the same occupancy 
configurations in terms of both rotated electrons and electrons.
That reveals that such $s\nu$ fermions describe
the spin degrees of freedom of a number $2\nu=4,6,...,2C_s$ of electrons.
It follows that nearly the whole spectral weight 
generated by application onto a $x\geq 0$ and $m=0$ ground state of 
$\cal{N}$-electron operators refers to a subspace 
spanned by energy eigenstates with numbers in
the following range,
\begin{equation}
[L_{s,\,-1/2} + L_{s,\,+1/2} +2C_s - 2B_s] =0,1,2,...,{\cal{N}} 
\, ; \hspace{0.35cm} C_s=\sum_{\nu=1}^{C_s}\nu \,N_{s\nu}
\, ; \hspace{0.35cm} B_s=\sum_{\nu=1}^{C_s}N_{s\nu} \, . 
\label{srs0}
\end{equation}
Note that owing to the above invariance of the $s\nu$ fermions with $\nu\geq 2$ spinon
pairs, provided that ${\cal{N}}/N_a^2
\rightarrow 0$ and $B_s/N_a^2\rightarrow 0$ for $N_a^2
\rightarrow\infty$ the number $B_s=\sum_{\nu}N_{s\nu}$ is a good quantum number.
(This is a generalization of the subspace (A) defined in Ref. \cite{general}.) 
Consistently, the $x> 0$ (and $x=0$ and $\mu=0$) and $m=0$ ground state 
and the set of excited states of energy $\omega<2\mu$
(and $\omega<\mu^0$) that span the one- and two-electron subspace considered here 
have no $-1/2$ $\eta$-spinons, $\eta\nu$ fermions, and $s\nu'$ fermions with $\nu'\geq 3$ spinon pairs so that 
$N_{\eta\nu}=0$ and $N_{s\nu'}=0$ for $\nu'\geq 3$. Summation over 
the set of states that span such a subspace gives indeed
$\sum_j \vert C_j\vert^2\approx 1$ for the coefficients of the one- or two-electron excitation
${\hat{O}}\vert \Psi_{GS}\rangle$ of Eq. (\ref{1-2-subspace}). This holds for ${\hat{O}}$
being the electron creation or annihilation operator or any of the simple two-electron
operators. Consistently, there is both for the
model on the 1D and square lattices an extremely
small weight corresponding mostly to states with $N_{s3}=1$, which is neglected within
the use of the one- and two-electron subspace considered here.
(Note that while the number restrictions of Eq. (\ref{srcsL}) are exact,
those of Eq. (\ref{srs0}) are a very good approximation. This is why $\sum_j \vert C_j\vert^2\approx 1$
rather than $\sum_j \vert C_j\vert^2= 1$ for the $j$ summation running over the 
set of ${\cal{N}}=1,2$ states that span the one- and two-electron subspace as defined here.)

Thus, according to Eq. (\ref{srs0}) the numbers of independent $\pm 1/2$ spinons and 
that of $s2$ fermions of the excited states that span such a subspace are restricted to the 
following ranges,
\begin{eqnarray}
L_{s,\,\pm 1/2} & = & 0,1  \, ;
\hspace{0.35cm} N_{s2} = 0 \, ,
\hspace{0.25cm}{\rm for}\hspace{0.25cm} {\cal{N}} = 1 \, ,
\nonumber \\
2S_s + 2N_{s2} & = & [L_{s,\,-1/2} + L_{s,\,+1/2} + 2N_{s2}] =
0,1,2   \, ,
\hspace{0.25cm}{\rm for}\hspace{0.25cm} {\cal{N}} = 2 \, .
\label{srs-0-ss}
\end{eqnarray}
Here ${\cal{N}}=1,2$ refers to any of the ${\cal{N}}$-electron
operators ${\hat{O}}$ whose application onto the ground
state $\vert \Psi_{GS}\rangle$ generates the above excited states,
as given in Eq. (\ref{1-2-subspace}). Furthermore, the number of $c$ fermions
and the number of $s1$ fermions read $N_c =N=(1-x)\,N_a^2$ and
$N_{s1} = [N/2-2N_{s2}-S_s]=(1-x)\,N_a^2/2 -[2N_{s2}+S_s]$, respectively.
If in addition we restrict our
considerations to the LWS-subspace of the one- and
two-electron subspace \cite{general}, then $L_{s,\,-1/2}=0$ 
in Eq. (\ref{srs-0-ss}), whereas
the values $L_{s,\,+1/2} = 0,1$ for $N_{s2} = 0$ and
${\cal{N}} = 1$ remain valid and in $[2S_s + 2N_{s2}] =0,1,2$
one has that $2S_s=L_{s,\,+1/2}$ for ${\cal{N}} = 2$.
\begin{table}
\begin{tabular}{|c|c|c|c|c|c|c|c|c|c|c|c|c|c|c|} 
\hline
numbers & charge & +1$\uparrow$el. & -1$\downarrow$el. & +1$\downarrow$el. & -1$\uparrow$el. & singl.spin & 
tripl.spin & tripl.spin & tripl.spin & $\pm$2$\uparrow\downarrow$el. & +2$\uparrow$el. & -2$\downarrow$el. & +2$\downarrow$el. & -2$\uparrow$el. \\
\hline
$\delta N_c^h$ & $0$ & $-1$ & $1$ & $-1$ & $1$ & $0$ & $0$ & $0$ & $0$ & $\mp 2$ & $-2$ & $2$ & $-2$ & $2$ \\
\hline
$N_{s1}^h$ & $0$ & $1$ & $1$ & $1$ & $1$ & $2$ & $2$ & $2$ & $2$ & $0$ & $2$ & $2$ & $2$ & $2$ \\
\hline
$\delta N_{\uparrow}$ & $0$ & $1$ & $0$ & $0$ & $-1$ & $0$ & $1$ & $-1$ & $0$ & $\pm 1$ & $2$ & $0$ & $0$ & $-2$ \\
\hline
$\delta N_{\downarrow}$ & $0$ & $0$ & $-1$ & $1$ & $0$ & $0$ & $-1$ & $1$ & $0$ & $\pm 1$ & $0$ & $-2$ & $2$ & $0$ \\
\hline
$L_{s,\,+1/2}$ & $0$ & $1$ & $1$ & $0$ & $0$ & $0$ & $2$ & $0$ & $1$ & $0$ & $2$ & $2$ & $0$ & $0$ \\
\hline
$L_{s,\,-1/2}$ & $0$ & $0$ & $0$ & $1$ & $1$ & $0$ & $0$ & $2$ & $1$ & $0$ & $0$ & $0$ & $2$ & $2$ \\
\hline
$N_{s2}$ & $0$ & $0$ & $0$ & $0$ & $0$ & $1$ & $0$ & $0$ & $0$ & $0$ & $0$ & $0$ & $0$ & $0$ \\
\hline
$S_s$ & $0$ & $1/2$ & $1/2$ & $1/2$ & $1/2$ & $0$ & $1$ & $1$ & $1$ & $0$ & $1$ & $1$ & $1$ & $1$ \\
\hline
$\delta S_c$ & $0$ & $1/2$ & $-1/2$ & $1/2$ & $-1/2$ & $0$ & $0$ & $0$ & $0$ & $\pm 1$ & $1$ & $-1$ & $1$ & $-1$ \\
\hline
$\delta N_{s1}$ & $0$ & $0$ & $-1$ & $0$ & $-1$ & $-2$ & $-1$ & $-1$ & $-1$ & $\pm 1$ & $0$ & $-2$ & $0$ & $-2$ \\
\hline
$\delta N_{a_{s1}}$ & $0$ & $1$ & $0$ & $1$ & $0$ & $0$ & $1$ & $1$ & $1$ & $\pm 1$ & $2$ & $0$ & $2$ & $0$ \\
\hline
\end{tabular}
\caption{The deviations $\delta N_c^h=-2\delta S_c$ and numbers $N_{s1}^h=[2S_s+2N_{s2}]$ of Eq. (\ref{deltaNcs1}) for the 
fourteen classes of one- and two-electron excited states of the $x>0$ and $m=0$ ground state that span the one- and two-electron 
subspace considered in this paper, corresponding electron number deviations 
$\delta N_{\uparrow}$ and $\delta N_{\downarrow}$, and independent-spinon numbers $L_{s,\,+1/2}$ and 
$L_{s,\,-1/2}$ and $s2$ fermion numbers $N_{s2}$ restricted to the values provided in Eq. (\ref{srs-0-ss}).
The spin $S_s$ and deviations $\delta S_c$, $\delta N_{s1}=[\delta S_c-S_s-2N_{s2}]$, and $\delta N_{a_{s1}}=[\delta S_c +S_s]$ 
of each excitation are also provided.}
\label{table1}
\end{table} 

As shown in Ref. \cite{general}, the numbers of independent $\eta$-spinons ($\alpha=\eta$) and 
independent spinons ($\alpha=s$) $L_{\alpha,\,\pm 1/2}$, the total numbers of $\eta$-spinons 
($\alpha=\eta$) and spinons ($\alpha=s$) $M_{\alpha,\,\pm 1/2} = [L_{\alpha,\,\pm 1/2} + C_{\alpha}]$,
the number of sites of the spin effective lattice $N_{a_{s}}^2=2S_c$, the number of sites of the $\eta$-spin
effective lattice $N_{a_{\eta}}^2=[N_a^2-2S_c]$, the number of $c$ fermions $N_c=2S_c$,
and the number of $c$ fermion holes $N_c^h=[N_a^2-2S_c]$ are good quantum numbers of
the Hubbard model on the square lattice. The good news is that in the one- and
two-electron subspace considered here the numbers $N_{a_{s1}}^2$, $N_{s1}$, $N_{s1}^h$, 
and $N_{s2}$ are also good quantum numbers of such a model.
The number of sites, unoccupied sites, and occupied sites of the $c$ and $s1$ effective
lattices equal those of discrete momentum values, unfilled momentum values, and
filled momentum values of the $c$ and $s1$ bands, respectively.
From straightforward manipulations of Eqs. (\ref{N*}) and (\ref{N-h-an}) of
Appendix A for $\alpha\nu =s1$ we find
that the number $N_{a_{s1}}^2 = [N_{s1} + N^h_{s1}]$ of $s1$ effective lattice sites and thus of
$s1$ band discrete momentum values is in general given by,
\begin{equation}
N_{a_{s1}}^2 = [S_c + S_{s}+\sum_{\nu=3}^{C_{s}}(\nu-2)N_{s\nu}] \, .
\label{N-a-s1-D}
\end{equation}
Hence for the one- and two-electron subspace considered in this paper for which $N_{s\nu}=0$ 
for $\nu\geq 3$ one has that $N_{a_{s1}}^2 = [S_c + S_{s}]$ is a good quantum number.  
Consistently, for spin $S_s=0$ that subspace is a subspace (A) as defined in 
Ref. \cite{general}. It follows that $N_{s1}=[S_c-2N_{s2}]$, $N_{s1}^h=2N_{s2}$, and $N_{s2}$ and hence $N_{a_{s1}}^2=[N_{s1}+N_{s1}^h]=S_c$
are good quantum numbers. Furthermore, for the remaining spin values $S_s=1/2$ and $S_s=1$ such a subspace is
a subspace (B) as defined in that reference. Hence $N_{s1}=[S_c-S_s]$, $N^h_{s 1} =2S_s$, and $N_{a_{s1}}^2=
[S_c+S_s]$ are good quantum numbers. It then follows from the general properties of the operator description 
of Ref. \cite{general} that for the Hubbard model on the square lattice in the one- and two-electron
subspace the $s1$ fermion band microscopic momenta ${\vec{q}}$ are good quantum numbers as well.

Moreover, use of the general expressions of $s1$ band discrete momentum values and
number $N_{s1}^h=[N_{a_{s1}}^2 -N_{s1}]$ of $s1$ band unfilled momentum values with
$N_{a_{s1}}^2$ given in Eq. (\ref{N-a-s1-D}) 
together with the restrictions in the values of the numbers of Eqs. (\ref{srcsL}), (\ref{srs0}), and (\ref{srs-0-ss})
and the exact result proved in Ref. \cite{general} that one-electron (and two-electron) excitations have
no overlap with excited states with none and two (and one) $s1$ band holes 
(and hole) reveals that nearly the whole one- and two-electron spectral weight is contained in the subspace 
spanned by states whose deviation $\delta N_c^h$ in the number of $c$ band holes
and number $N_{s1}^h$ of $s1$ band holes are given by, 
\begin{eqnarray}
\delta N_c^h & = & -2\delta S_c = - \delta N = 0, \mp 1, \mp 2 \, ,
\nonumber \\
N_{s1}^h & = & S_c + S_{s} - N_{s1} = 2S_s+2N_{s2} 
\nonumber \\
& = & \pm (\delta N_{\uparrow} - \delta N_{\downarrow})
+ 2L_{s,\,\mp 1/2} + 2N_{s2} = L_{s,\,-1/2} + L_{s,\,+1/2} + 2N_{s2} = 0, 1, 2 \, .
\label{deltaNcs1}
\end{eqnarray}
Here $\delta N$ is the deviation relative to the initial ground state in the number of electrons, 
$\delta N_{\uparrow}$ and $\delta N_{\downarrow}$ are those in
the number of spin-projection $\uparrow$ and $\downarrow$ 
electrons, respectively, $N_{s2}$ is the number of the excited-state $s2$
fermions, and $L_{s,\,\pm1/2}$ is that of independent
spinons of spin projection $\pm1/2$. We emphasize that for an initial 
$m=0$ and $S_s=0$ ground state the numbers $N_{s2}=L_{s,\,\pm1/2}=N_{s1}^h =0$ vanish \cite{general}.

The deviations $\delta N_c^h$ and numbers $N_{s1}^h$ 
of Eq. (\ref{deltaNcs1}) for the fourteen classes of one- and two-electron excited states of the $x>0$ 
and $m=0$ ground state that span the one- and two-electron subspace, 
corresponding electron number deviations $\delta N_{\uparrow}$ 
and $\delta N_{\downarrow}$, and independent-spinon numbers $L_{s,\,+1/2}$ and 
$L_{s,\,-1/2}$ and $s2$ fermion numbers 
$N_{s2}$ restricted to the values provided in Eq. (\ref{srs-0-ss})
are given in Table \ref{table1}. For $N_{s2}=1$ spin-siglet excited states  
the $s2$ effective lattice has a single site and the corresponding $s2$ band a single vanishing 
discrete momentum value, $\vec{q} = 0$, occupied by the $s2$ fermion \cite{general}. 
We recall that such a $s2$ fermion is invariant under the electron - rotated-electron unitary transformation 
and thus has vanishing energy, consistently with the invariance condition of Eq. (\ref{invariant-V}) 
of Appendix A for $\alpha\nu=s2$ and vanishing magnetic field $H=0$.

As mentioned above, the initial $x>0$ and $m=0$ ground states of the one- and two-electron subspace 
have zero holes in the $s1$ band so that $\delta N_{s1}^h=N_{s1}^h$ for the excited states. 
This follows from all $M_s=2S_c=N$ spinons being confined within the two-spinon bonds of
the $N_{s1}=M_s/2$ $s1$ fermions. The one- and two-electron subspace  
is spanned by the states of Table \ref{table1} generated by creation or annihilation 
of $\vert\delta N_c^h\vert=0,1,2$ holes in the
$c$ momentum band and $N_{s1}^h=0,1,2$ holes in the $s1$ band 
plus small momentum and low energy particle-hole processes in the $c$ band. 
The charge excitations of $x>0$ and $m=0$ initial ground states
consist of a single particle-hole process in the $c$ band of arbitrary momentum and 
energy compatible with its momentum and energy bandwidths, 
plus small-momentum and low-energy $c$ fermion particle-hole processes.
Such charge excitations correspond to state
representations of the global $U(1)$ symmetry and
refer to the type of states denoted by ``charge" in the table.
The one-electron spin-doublet excitations correspond 
to the four types of states denoted by ``$\pm1\sigma$el." in
Table \ref{table1} where $+1$ and $-1$ denotes creation and
annihilation, respectively, and $\sigma =\uparrow ,\downarrow$.
The spin-singlet and spin-triplet excitations refer to the four types of states 
denoted by ``singl.spin" and ``tripl.spin" in the table.
The two-electron excitations whose electrons are in a spin-singlet
configuration and those whose two created or annihilated electrons are 
in a spin-triplet configuration correspond to the five types of 
states ``$\pm 2\uparrow\downarrow$el." and ``$\pm 2\sigma$el." of that table  
where $+2$ and $-2$ denotes creation and
annihilation, respectively, of two electrons.

For the Hubbard model on the square lattice such fourteen types of
states are energy eigenstates and nearly exhaust the whole one- and two-electron spectral
weight. Excited states of classes other than those of the table contain
nearly no one- and two-electron spectral weight. Such a weight analysis
applies to the 1D Hubbard model as well.
For the corresponding quantum liquid describing the 
Hamiltonian (\ref{H}) in the one- and two-electron 
subspace, the numbers $2S_c$,
$2S_{\eta}$, $2S_s$, and $-2S_s^{x_3}$ associated with the global
$SO(3)\times SO(3)\times U(1)$ symmetry are given by,
\begin{equation}
2S_c = (1-x)\,N_a^2
 \, ; \hspace{0.25cm}
2S_{\eta} = -2S_{\eta}^{x_3}= x\,N_a^2
 \, ; \hspace{0.25cm}
2S_s = (1-x)\,N_a^2
 - 2[N_{s1}+2N_{s2}] \, ; \hspace{0.25cm}
-2S_s^{x_3} = m\,N_a^2
 = 2S_s - 2L_{s,-1/2} \, .
\label{NN-SSS}
\end{equation}
For such a quantum liquid the number of sites of the spin effective
lattice $N_{a_s}^2=2S_c$ and corresponding spacing $a_s$
given in Eq. (\ref{a-alpha}) of Appendix A read,
\begin{equation}
N_{a_{s}}^2
 = (1-x)\,N_a^2
\, ; \hspace{0.35cm}
a_{s} = {a\over \sqrt{1-x}} \, ,
\hspace{0.25cm} (1-x)> 1/N_a^2 \, ,
\label{NNCC}
\end{equation}
respectively. Its sites refer to those singly occupied by rotated electrons in the
original lattice. In turn, the sites of the $\eta$-spin lattice also introduced in 
Ref. \cite{general} refer to the sites doubly occupied and unoccupied by rotated 
electrons in the original lattice.

For the Hubbard model in the one- and two-electron subspace
the concept of a $\eta$-spin lattice considered in Ref. \cite{general}
is useless. Indeed, for that subspace such a lattice either is empty ($x>0$) or does not exist ($x=0$).
This is because the $\eta$-spin degrees of freedom
of the states that span that subspace are the same as those of
the $C_{\eta}=\sum_{\nu=1}^{C_{\eta}}\nu\,N_{\eta\nu} =(N_a^2/2-S_c-S_{\eta})=0$ vacuum $\vert 0_{\eta};N_{a_{\eta}}^2\rangle$ of Eq. (\ref{vacuum})
of Appendix A. For states for which $S_c=N_a^2/2$ and thus $N_{a_{\eta}}^2=S_{\eta}=0$
the $\eta$-spin lattice does not exist and thus the spin effective lattice is identical to
the original lattice. This is argued below in Section III-C to be a necessary condition 
for a spontaneous symmetry breaking and emergence of a long-range antiferromagnetic
order to occur in the ground state as $N_a^2\rightarrow\infty$. In turn, for $S_c<N_a^2/2$ the $\eta$-spin degrees of freedom 
correspond to a single occupancy configuration of the $N_{a_{\eta}}^2$ 
independent $+1/2$ $\eta$-spinons. Such objects are invariant under the electron - rotated-electron
transformation and thus play the role of unoccupied sites of the $\eta$-spin lattice \cite{general}.
Indeed, for the one- and two-electron subspace the corresponding number of $\eta$-spin lattice
occupied sites $C_{\eta}=\sum_{\nu=1}^{C_{\eta}}\nu\,N_{\eta\nu}$ vanishes.
Only for states and subspaces for which $N_{a_{\eta}}^2/N_a^2=[1-2S_c/N_a^2]$ is finite and the inequality $0<C_{\eta}<N_{a_{\eta}}^2$ holds
is the concept of a $\eta$-spin effective lattice useful. For a given state, $C_{\eta}+C_{s}$ is the
number of sites of the original lattice whose electron occupancy configurations are not invariant under
the electron - rotated-electron unitary transformation. For the states that span the one- and two-electron 
subspace the numbers $C_{\eta}$ and $C_{s}$ are given by $C_{\eta} = 0$ and $C_s=[N_{s1}+2N_{s2}]$, 
respectively. For $x=0$ (and $x=1$) one finds $N_{a_{\eta}}^2=0$ and $N_{a_{s}}^2= N_a^2$ (and 
$N_{a_{\eta}}^2= N_a^2$ and $N_{a_{s}}^2= 0$), so that there is no
$\eta$-spin (and no spin) effective lattice and the
spin (and $\eta$-spin) effective lattice equals
the original lattice.

As further confirmed below, for the one- and two-electron subspace considered here only the $c$ and $s1$
fermions play an active role. Straightforward manipulations of the general
expressions given in Eq. (\ref{N-a-s1-D}) and Eqs. (\ref{N*})-(\ref{a-a-nu}) of Appendix A and related expressions
provided in Ref. \cite{general}
reveal that for that subspace the number $N_{a_{s1}}^2$
of $s1$ band discrete momentum values, 
$N_{s1}$ of $s1$ fermions, and $N^h_{s 1}$ of $s1$ fermion holes are given by,
\begin{equation}
N_{a_{s1}}^2 = [N_{s1} + N^h_{s1}] = 
[S_c + S_s] \, ;
\hspace{0.25cm}
N_{s1} = [S_c - S_s -2N_{s2}]  \, ;
\hspace{0.25cm}
N^h_{s 1} = [2S_s +2N_{s2}] =0,1,2 \, ,
\label{Nas1-Nhs1}
\end{equation}
respectively. In turn,
the corresponding $c$ effective lattice, $c$ momentum band,
and $c$ fermion numbers read,
\begin{equation}
N_{a_{c}}^2 = [N_{c} + N^h_{c}] = 
N_{a}^2  \, ;
\hspace{0.25cm}
N_{c} = 2S_c = (1-x)\,N_{a}^2  \, ;
\hspace{0.25cm}
N^h_{c} = x\,N_{a}^2 \, .
\label{Nac-Nhc}
\end{equation}
 
\subsection{Confirmation for 1D that most one-electron spectral weight is generated 
by processes obeying the ranges of Eqs. (\ref{srs0}), (\ref{srs-0-ss}), and (\ref{deltaNcs1})}

Above the transformation laws under the electron - rotated-electron transformation
of the $\alpha\nu$ fermions were used to show that nearly the whole spectral weight  
of the one- and two-electron excitations of the Hubbard model on the square
lattice is generated by processes obeying the ranges of Eqs. (\ref{srs0}), (\ref{srs-0-ss}), 
and (\ref{deltaNcs1}). Similar results apply to the 1D model. 

The terms of one- or two-electron operators ${\hat{O}}$ 
on the right-hand side of the first equation of (\ref{OOr}) 
that generate the excitations ${\hat{O}}\vert \Psi_{GS}\rangle$ 
of Eq. (\ref{1-2-subspace}) may be expressed
in terms of the of $c$ fermion operators, spinon operators, and $\eta$-spinon
operators given in Eqs. (\ref{fc+})-(\ref{rotated-quasi-spin}).
This is done by use of the operator relations provided in Eq. (\ref{c-up-c-down}) 
of Appendix A. Concerning the contributions to the ${\hat{O}}$ expression provided in Eq. (\ref{OOr}) containing
commutators involving the operator ${\tilde{S}} = -(t/U)\,[\tilde{T}_{+1} -\tilde{T}_{-1}] 
+ {\cal{O}} (t^2/U^2)$, to fulfill such a task one takes into account that independently of their form, the additional operator terms 
${\cal{O}} (t^2/U^2)$ of higher order are products of the kinetic operators 
$\tilde{T}_0$, $\tilde{T}_{+1}$, and $\tilde{T}_{-1}$ of Eq. (\ref{T-op}).

From such an analysis one finds that the elementary processes associated with
the one- and two-electron subspace number value ranges of Eqs. (\ref{srs0}), (\ref{srs-0-ss}), 
and (\ref{deltaNcs1}) are fully generated by 
the leading-order operator ${\tilde{O}}$. In turn, the processes generated by 
the operator terms containing commutators involving the operator ${\tilde{S}}$
refer to excitations whose number value ranges are different
from those of Eqs. (\ref{srs0}), (\ref{srs-0-ss}), and (\ref{deltaNcs1}). This confirms that
such processes generate very little one- and two-electron
spectral weight, consistently with the exact number restrictions of Eq. (\ref{srcsL}),
the approximate number restrictions of Eq. (\ref{srs0}), and the
results of Ref. \cite{general}. 

For the Hubbard model on the 1D lattice also often 
considered in the studies of that reference, the spectral-weight
distributions can be calculated explicitly by the pseudofermion
dynamical theory associated with the model exact solution, 
exact diagonalization of small chains, and other methods. 
From the use of the same arguments as for the model on the square
lattice one finds that at 1D the one- and two-electron subspace
considered in this paper corresponds to the same number
deviations and numbers as for the square lattice. In addition, the relative 
one-electron spectral weight generated by different types of 
microscopic processes can be studied by means of the above methods.
That program is fulfilled in Ref. \cite{1EL-1D}.
The results of that reference confirm the dominance of
the processes associated with the number value ranges provided in
Eqs. (\ref{srs-0-ss}) and (\ref{deltaNcs1}). They refer spacifically to operators ${\hat{O}}$ and ${\tilde{O}}$ 
that are electron and rotated-electron, respectively, creation or
annihilation operators. Such studies confirm that the operator ${\tilde{O}}$
generates {\it all} processes associated with the number value ranges
of Eqs. (\ref{srs-0-ss}) and (\ref{deltaNcs1}) and number values of
Table \ref{table1}. In addition, it also generates some of the non-dominant processes. That
is confirmed by the weights given in Table 1 of Ref.
\cite{1EL-1D}, which correspond to the dominant processes
associated with only these ranges. The small missing
weight refers to excitations whose number value ranges
are not those of Eqs. (\ref{srs-0-ss}) and (\ref{deltaNcs1}) but whose
weight is also generated by the operator ${\tilde{O}}$.
Indeed, that table refers to $U/4t\rightarrow\infty$
so that ${\hat{O}}={\tilde{O}}$ and the operator terms of 
the ${\hat{O}}$ expression provided in
Eq. (\ref{OOr}) containing commutators involving the 
operator ${\tilde{S}}$ vanish. 

On the other hand, for finite values of $U/4t$ all dominant processes associated 
with the number value ranges of Eqs. (\ref{srs-0-ss}) and (\ref{deltaNcs1}) 
and number values provided in Table \ref{table1} 
are generated by the operator ${\tilde{O}}$. In turn,
the small spectral weight associated with excitations whose
number value ranges are different from those are generated both
by that operator and the operator terms of the ${\hat{O}}$ expression of
Eq. (\ref{OOr}) containing commutators involving the 
operator ${\tilde{S}}$. For the model on the 1D lattice the small 
one-electron spectral weight generated by the non-dominant 
processes is largest at half filling and $U/4t\approx 1$. 

For $x\geq 0$ the one- and two-electron subspace is spanned
by states with vanishing rotated-electron double
occupancy. This holds both for the Hubbard model on
the 1D and square lattices. Generalization of the results to the range $x\leq 0$ reveals
that then such a subspace is spanned by states with vanishing rotated-hole double
occupancy. That property combined with the particle-hole
symmetry explicit at $x=0$ and $\mu=0$, implies that
the relative spectral-weight contributions from different types of one-electron addition
excitations  given in Fig. 2 of Ref. \cite{1EL-1D} for the 1D model at half filling  
leads to similar corresponding relative weights for   
half-filling one-electron removal. Analysis of that figure 
confirms that for the corresponding one-electron removal
spectrum the dominant processes associated 
with the number value ranges of Eqs. (\ref{srs-0-ss}) and (\ref{deltaNcs1}) and 
number values given in Table \ref{table1}
refer to the states called in figure 1 holon - 1 $s1$ hole states. Their
minimum relative weight of about $0.95$ is reached
at $U/4t\approx 1$. For other hole concentrations $x>0$ 
and values of $U/4t$ the relative weight of such states is always 
larger than $0.95$, as confirmed from analysis of 
Figs. 1 and 2 and Table 1 of Ref. \cite{1EL-1D}.

For the Hubbard model on the square lattice the
explicit derivation of one- and two-electron spectral
weights is a more involved problem. The number value ranges
of Eqs. (\ref{srs0}), (\ref{srs-0-ss}), and (\ref{deltaNcs1}) 
and number values provided in Table \ref{table1} 
for the one- and two-electron subspace also apply,
implying similar results for the relative spectral weights
of one- and two-electron excitations. 
Consistently and as mentioned above, there is an exact selection rule valid both for the
Hubbard model on a 1D and square lattices that
confirms that $N_{s1}^h=1$ and $N_{s1}^h =0,2$
for one-electron excitations and two-electron excitations,
respectively. It follows from the expression provided in Eq. (\ref{Ns1h-general}) of Appendix A 
for the quantum number $P^h_{s1}=e^{i\pi N^h_{s1}}=e^{i\pi N} = \pm 1$. It
reveals that for $\delta N=\pm 1$ (and $\delta N=\pm 0$ and $\delta N=\pm 2$) 
excited states the number $N^h_{s1}$ of holes in the $s1$ 
band must be always an odd (and even) integer. Here  $\delta N$
is the deviation in the electron number $N$ under a transition from
a $x>0$ and $m=0$ ground state to such excited states.
This implies that one-electron (and two-electron) excitations do not 
couple to excited states with two holes (and one hole) in the $s1$ band. 
Indeed, for such excitations one has that $e^{i\pi \delta N} = -1$
(and $e^{i\pi \delta N} = +1$), so that  $e^{i\pi \delta N^h_{s1}} = -1$
(and $e^{i\pi \delta N^h_{s1}} = +1$).

\subsection{The spin and $s1$ effective lattices
for the one- and two-electron subspace}

According to the restrictions and numbers values of
Eqs. (\ref{srs-0-ss}) and (\ref{deltaNcs1}) and Table \ref{table1}, the states
that span the one- and two-electron subspace may involve none 
or one $s2$ fermion. As confirmed in Ref. \cite{companion},
it is convenient to express the one- and two-electron excitation spectrum relative
to initial $x>0$ and $m=0$ ground states in terms of the deviations in the numbers
of $c$ effective lattice and $s1$ effective lattice unoccupied sites. Those are given 
explicitly in Eq. (\ref{deltaNcs1}) and Table \ref{table1}. Note that the number of $s1$ fermions 
provided in Eq. (\ref{Nas1-Nhs1}) can be written as 
$N_{s1}=[(1-x)N_a^2/2-S_s -2N_{s2}]$ 
where $S_s=0$ for $N_{s2}=1$ and $S_s=0,1/2,1$ for $N_{s2}=0$.

As discussed above, for $N_{s2}=1$ spin-singlet excited energy eigenstates the single $s2$ fermion has vanishing energy 
and momentum and consistently with Eq. (\ref{invariant-V}) of Appendix A,
for vanishing magnetic field $H=0$ it is invariant under the electron - rotated-electron
unitary transformation. Therefore, the only effect of its creation
and annihilation is in the numbers of occupied and unoccupied sites of the $s1$ effective lattice.
Its creation can then be accounted for merely by small changes in the occupancies of the 
discrete momentum values of the $s1$ band, as discussed below. Hence the only 
composite object whose internal occupancy configurations in the spin effective lattice are 
important for the physics of the Hamiltonian (\ref{H}) in the one- and two-electron subspace 
is the spin-neutral two-spinon $s1$ fermion and related spin-singlet two-spinon $s1$ bond particle
\cite{general,companion0}.

It is confirmed below that for the Hubbard model in the one- and two-electron 
subspace and alike for the $s2$ fermion,
the presence of independent spinons is felt through the numbers of occupied and unoccupied sites of
the $s1$ effective lattice. In turn, the number of independent
$+1/2$ $\eta$-spinons equals that of the unoccupied 
sites of the $c$ effective lattice. Therefore, when acting onto that
subspace, the Hubbard model refers to a two-component quantum liquid that
can be described only in terms of $c$ fermions and $s1$ fermions. 

For $x>0$ and states belonging to the one- and two-electron 
subspace the spin effective lattice has a number of sites $N_{a_{s}}^2 =(1-x)\,N_a^2$. Its
value is smaller than that of the original lattice. The lattice constant $a_s$ provided in Eq. (\ref{a-alpha}) 
of Appendix A for $\alpha = s$ reads $a_{s} \approx a/\sqrt{1-x}$ for such states, 
as given in Eq. (\ref{NNCC}). It is such that the area $L^2$ of the system is preserved. 
Any real-space point within the spin effective lattice corresponds to 
the same real-space point in the system original lattice. 
Except for a suitable phase factor, a local $s1$ fermion has the same
internal structure as the corresponding $s1$ bond-particle \cite{companion,general}.
The $s1$ fermion spinon occupancy configurations considered in Ref. \cite{companion0}
are expected to be a good approximation 
provided that the ratio $N_{a_s}^2/N_a^2$ and thus the electronic density $n=(1-x)$
are finite in the limit $N_a^2\rightarrow\infty$. This is met for the hole concentration
range $x\in (0,x_*)$ where $x_*<1$ for which according to the studies of Ref. \cite{companion}
the maximum magnitude of the $s1$ fermion spinon pairing energy is finite.

Within the $N_a^2\gg 1$ limit that the description used in the studies of this paper
refers to, there is for the states that span the one- and two-electron subspace 
commensurability between the real-space distributions of the $N_{a_{s1}}^2\approx N_{s1}$
sites of the $s1$ effective lattice and the $N_{a_{s}}^2\approx 2N_{s1}$ 
sites of the spin effective lattice. For $(1-x)\geq 1/N_a^2$ and $N_a^2
\gg 1$ the spin effective lattice has $N_{a_s}^2=(1-x)\,N_a^2$ sites and from the use of 
the expression given in Eq. (\ref{Nas1-Nhs1}) for the number of $s1$ effective
lattice sites $N_{a_{s1}}^2$ and Eq. (\ref{a-a-nu}) of Appendix A for the
corresponding spacing $a_{s1}$ we find,
\begin{equation}
a_{s1} = a_s\,\sqrt{{2\over1+{2S_s\over (1-x)N_a^2}}}\approx \sqrt{2}\,a_s\, \left(1-{2S_s\over 2(1-x)}{1\over N_a^2}\right)
\approx \sqrt{2}\,a_s \, , \hspace{0.25cm} S_s =0, {1\over 2}, 1 \, .
\label{a-a-s1-sube}
\end{equation}

For $N^h_{s1}=0$ states such as the $x\geq 0$ and $m=0$ ground states
the square spin effective lattice has two
well-defined sub-lattices, which we call sub-lattice A and B, respectively. 
As discussed in Ref. \cite{companion0}, for $N^h_{s1}=1,2$ 
states the spin effective lattice has two bipartite lattices as well,
with one or two extra sites corresponding to suitable boundary conditions. 
The two spin effective sub-lattices have spacing $a_{s1} =\sqrt{2}\,a_s$. 
The fundamental translation vectors of the sub-lattices A and B read,
\begin{equation}
{\vec{a}}_{s1} = {a_{s1}\over\sqrt{2}}({\vec{e}}_{x_1}+{\vec{e}}_{x_2})
\, , \hspace{0.25cm} 
{\vec{b}}_{s1} = -{a_{s1}\over\sqrt{2}}({\vec{e}}_{x_1}-{\vec{e}}_{x_2}) \, ,
\label{a-b-s1}
\end{equation}
respectively. Here ${\vec{e}}_{x_1}$ and ${\vec{e}}_{x_2}$ are the unit vectors
and $x_1$ and $x_2$ Cartesian coordinates. As confirmed in Ref. \cite{companion0}, 
the vectors given in Eq. (\ref{a-b-s1}) are the fundamental translation vectors 
of the $s1$ effective lattice.

In the case of $x\geq 0$, $m=0$, and $N^h_{s1}=0$ ground states 
whose $s1$ momentum band is full and all $N_{a_{s1}}^2$ sites of the $s1$ effective lattice 
are occupied we consider that the square root $N_{a_s}$ of the number $N_{a_s}^2$ 
of sites of the spin effective lattice is an integer. It then follows that the spin effective lattice is a
square lattice with $N_{a_s}\times N_{a_s}$ sites. Thus the square root $N_{a_{s1}}$ of
the number $N_{a_{s1}}^2$ of sites of the $s1$ effective lattice cannot in general
be an integer number yet $N_{a_{s1}}^2$ is. However, within the $N_a^2\gg 1$ limit considered here
we use the notation $N_{a_{s1}}^2$ for the number of sites of the $s1$ effective lattice. 

\subsection{The quantum liquid of $c$ fermions and $s1$ fermions: Why only 
such objects play an active role?}

For the one- and two-electron subspace considered in this paper the 
number $N_{a_{s1}}^2$ of sites of the $s1$ effective lattice and $s1$ band discrete momentum values, 
$N_{s1}$ of $s1$ fermions, and $N^h_{s 1}$ of $s1$ fermion holes have expressions
given in Eq. (\ref{Nas1-Nhs1}). The corresponding numbers of the $c$ effective lattice and $c$ band
are provided in Eq. (\ref{Nac-Nhc}). For that subspace the $s1$ band is either full
and thus is filled by $N_{s1}=N_{a_{s1}}^2 =2S_c =N_{a_s}^2$ $s1$
fermions or has one or two holes.  
Furthermore, one-electron and two-electron excitations have
no overlap with excited states with two holes and one hole 
in the $s1$ band, respectively. Specifically, 
excited states with a single hole
in the $s1$ band correspond to one-electron
excitations and those with $N^h_{s 1} =0,2$ 
holes in that band refer to
two-electron excitations, as given in Table \ref{table1}. Excited states with 
$N^h_{s1} =3$ (and $N^h_{s 1} =4$) holes
in the $s1$ momentum band correspond to very little 
one-electron (and two-electron) spectral weight
and are ignored within the use of the one- and
two-electron subspace considered here.

We now justify why the square-lattice quantum liquid corresponding to the
Hubbard model on the square lattice in the one- and two-electron subspace
may be described only by $c$ and $s1$ fermions on their $c$ and $s1$
effective lattices, respectively.
According to the number value ranges of Eqs. (\ref{srs-0-ss}) and (\ref{deltaNcs1}) and 
number values provided in Table \ref{table1}, the
one- and two-electron subspace is spanned by excited
states having either none $N_{s2}=0$ or one $N_{s2}=1$
spin-neutral four-spinon $s2$ fermion. $N_{s2}=1$ spin-singlet excited
states have no independent spinons. One then finds 
from the use of Eq. (\ref{N-h-an}) of Appendix A for $\alpha\nu =s2$ 
that $N^h_{s2}=0$ so that such states have no holes in the $s1$ momentum band 
and thus $N_{a_{s2}}^2=1$.
This means that for such states the $s2$ fermion occupies a $s2$ 
band with a single vanishing momentum value. 
Since the $s2$ fermion under consideration has both vanishing
momentum and energy and is invariant
under the electron - rotated-electron unitary transformation, the only explicit effect of its creation  
is onto the numbers of occupied and unoccupied sites of 
the $s1$ effective lattice and corresponding numbers of 
$s1$ band $s1$ fermions and $s1$ fermion holes. Specifically, according to the
expressions provided in Eq. (\ref{Nas1-Nhs1}) and number values of Table \ref{table1}, 
the deviations $\delta S_c =\delta S_s =0$ and $\delta N_{s2}=1$ generated by a state
transition involving creation of one $s2$ fermion lead to deviations in the number
of $s1$ fermions and $s1$ fermion holes given by 
$\delta N_{s1} = -2\delta N_{s2}=-2$ and
$\delta N^h_{s1} = 2\delta N_{s2}=2$, respectively.
 
Moreover, the ranges of Eqs. (\ref{srs-0-ss}) and (\ref{deltaNcs1}) and 
number values of Table \ref{table1} confirm that such $N_{s2}=1$ excited states 
have zero spin, $S_s=0$. According to Eq. (\ref{Nas1-Nhs1}),
the number of holes in the $s1$ band is $N^h_{s 1} = 2N_{s2} =2$
for such states, in contrast to $N^h_{s 1}=0$ for the initial ground state.
In turn, the number $N_{a_{s 1}}^2$ of sites of the $s1$ effective lattice
remains unaltered. Following the annihilation of two $s1$ fermions and
creation of one $s2$ fermion, two unoccupied sites appear in 
the $s1$ effective lattice. As a result
two holes emerge in the $s1$ band as well. The emergence of these 
unoccupied sites and holes involves two virtual processes where 
(i) two $s1$ fermions are annihilated and
four independent spinons are created and (ii)
the latter independent spinons are annihilated
and the $s2$ fermion is created. 

Hence the only explicit net effect of the creation of a single vanishing-energy and zero-momentum
$s2$ fermion is the annihilation of two $s1$ fermions and corresponding emergence
of two holes in the $s1$ band and two unoccupied sites in the $s1$ effective lattice. Therefore, 
in the case of the one- and two-electron subspace one can
ignore that object in the theory provided that the corresponding changes in the
$s1$ band and $s1$ effective lattice occupancies are accounted for.
Within neutral $s1$ fermion particle-hole processes of transitions between
two excited states with a single $s2$ fermion, two of the four spinons of such 
an object are used in the motion of $s1$ fermions around in the $s1$ effective 
lattice. Indeed, such two spinons play the role of unoccupied sites of that lattice 
\cite{general,companion0}, consistently with the expression
$N^h_{s 1} = 2N_{s2}$ given in Eq. (\ref{Nas1-Nhs1}) for $2S_s=0$.

Also the $L_s =2S_s$ independent spinons play the role of unoccupied sites of
the $s1$ effective lattice. Again, this is consistent with the expression
$N^h_{s 1} = L_s =2S_s=1,2$ provided in Eq. (\ref{Nas1-Nhs1}) for
the number of unoccupied sites of the $s1$ effective lattice and of $s1$ fermion 
holes of the corresponding $N_{s2}=0$ excited states.
As given in Eqs. (\ref{srs-0-ss}) and (\ref{deltaNcs1}) and 
Table \ref{table1}, the one- and two-electron subspace $L_s$ allowed
values are $L_s =[L_{s\,,-1/2}+L_{s\,,+1/2}]=2S_s=0,1,2$. For 
$2S_s=1,2$ one has that $N_{s2} =0$. 
Now in contrast to creation of a single $s2$ fermion, a deviation $\delta 2S_s=1,2$ 
generated by a transition from the ground state to such $2S_s=1,2$ 
excited states may lead to deviations in the numbers
of occupied and unoccupied sites of the $s1$
effective lattice and corresponding $s1$ fermion and $s1$
fermion holes that do not obey the
usual equality $\delta N_{s1}= -\delta N^h_{s1}$.
Indeed, in the present case $2\delta S_c=\pm 1$ for 
$\delta N^h_{s 1} = 2\delta S_s=1$ and $2\delta S_c=0,\pm 2$
for $\delta N^h_{s 1} = 2\delta S_s=2$. Hence 
according to the expressions provided in Eq. (\ref{Nas1-Nhs1}),
such deviations lead to deviations in the numbers
of occupied and unoccupied sites of the $s1$
effective lattice and corresponding numbers of $s1$ fermions and
$s1$ fermion holes. Those read
$\delta N_{s1} = [\delta S_c-\delta S_s]$ and
$\delta N^h_{s1} = \delta 2S_s$, respectively.
It follows that the total number of sites and thus of
discrete momentum values of the $s1$ band may  
change under such transitions. This leads to an additional deviation
$\delta N_{a_{s1}}^2 = [\delta S_c +\delta S_s]$.
As given in Table \ref{table1}, for one-electron excited states one has that
$\delta N^h_{s 1} = 2\delta S_s=1$
and $2\delta S_c=\pm 1$. As a result,
$\delta N_{s1} = \pm 1/2-1/2=-1,0$
and $\delta N_{a_{s1}}^2 = \pm 1/2 +1/2=0,-1$.
In turn, for $N_{s2}=0$ two-electron excited states one has 
$\delta N^h_{s 1} = 2\delta S_s=2$
and $2\delta S_c=0,\pm 2$ . Thus
$\delta N_{s1} = -1, (\pm 1-1)=-2,-1,0$
and $\delta N_{a_{s1}}^2 = 1,(\pm 1+1)=0,1,2$. 

For the $s1$ fermion operators $f^{\dag}_{{\vec{q}},s1}$ and $f_{{\vec{q}},s1}$,
excitations that involve changes $\delta N_{a_{s1}}^2 = [\delta S_c +\delta S_s]$
in the number of sites and discrete momentum values of the $s1$ effective
lattice and $s1$ band, respectively, correspond to transitions between different quantum problems. Indeed,
such operators act onto subspaces spanned by neutral states, which conserve 
$S_c$, $S_s$, and $N_{a_{s1}}^2$. In turn, the generator of a non-neutral excitation
is the product of two operators. The first operator makes small changes 
in the $s1$ effective lattice or corresponding $s1$ momentum band.
Such changes follows the above deviations $\delta N_{a_{s1}}^2 = [\delta S_c +\delta S_s]$.
The second operator is a $s1$ fermion operator or a product of such operators appropriate to
the excited-state subspace.

Also the vanishing momentum and energy $L_{\eta,+1/2}=x\,N_a^2$ independent $+1/2$
$\eta$-spinons are invariant under the electron - rotated-electron unitary 
transformation. Their creation or annihilation may be accounted for 
by small suitable changes in occupancies of the $c$ effective lattice and $c$ 
momentum band. For $x>0$ and the one- and two-electron subspace 
considered here such independent $+1/2$
$\eta$-spinons correspond to a single
occupancy configuration associated with
the $\eta$-spin vacuum $\vert 0_{\eta};N_{a_{\eta}}^2\rangle$
of Eq. (\ref{vacuum}) of Appendix A. In turn, the degrees of freedom
of the rotated-electron occupancies of such $x\,N_a^2$ sites of the original 
lattice associated with the $U(1)$ symmetry
refer to the unoccupied
sites of the $c$ effective lattice of Eq. (\ref{Nac-Nhc}) and
corresponding $c$ band holes. Hence the number $L_{\eta,+1/2}=x\,N_a^2$ 
of independent $+1/2$ $\eta$-spinons equals that $N^h_c=x\,N_a^2$
of unoccupied sites of the $c$ effective lattice and
corresponding $c$ band holes.
This confirms that the deviations $\delta L_{\eta,+1/2}=(\delta x)\,N_a^2$
originated by creation and annihilation of 
independent $+1/2$ $\eta$-spinons within the one- and two-electron subspace
has no effects on the physics other than the corresponding 
deviation $\delta N^h_c=(\delta x)\,N_a^2$
in the number of unoccupied sites of the $c$ effective lattice and $c$ band holes. 

Spin-singlet excitations generated by application onto a $m=0$ and $x\geq 0$
initial ground state of the operator $f^{\dag}_{0,s2}\,f_{{\vec{q}},s1}\,f_{{\vec{q}}\,',s1}$ where ${\vec{q}}$ and
${\vec{q}}\,'$ are the momenta of the two emerging $s1$ fermion holes are
neutral states which conserve $S_c$, $S_s$, and $N_{a_{s1}}^2$. (See Table \ref{table1}.) 
The implicit role of the $s2$ fermion creation operator
$f^{\dag}_{0,s2}$ is exactly canceling the contributions of the
annihilation of the two $s1$ fermions of momenta ${\vec{q}}$ and ${\vec{q}}\,'$   
to the commutator $[\hat{q}_{s1\,x_1},\hat{q}_{s1\,x_2}]$ of the $s1$ translation generators
in the presence of the fictitious magnetic field ${\vec{B}}_{s1}$ of Eq. (\ref{A-j-s1-3D}) of
Appendix A. This ensures that the overall excitation is neutral. Since the $s2$ fermion has vanishing
energy and momentum and the $s1$ band and its number $N_{a_{s1}}^2$ of
discrete momentum values remain unaltered, one can effectively consider that the generator
of such an excitation is $f_{{\vec{q}},s1}\,f_{{\vec{q}}\,',s1}$ and omit the $s2$
fermion creation operator. Its only role is ensuring that the overall excitation is neutral
and the two components of the $s1$ fermion microscopic momenta can be specified. It follows
that for the one- and two-electron subspace the operators $f_{{\vec{q}},s1}\,f_{{\vec{q}}\,',s1}$,
$f^{\dag}_{{\vec{q}}\,',s1}\,f^{\dag}_{{\vec{q}},s1}$, $f^{\dag}_{{\vec{q}},s1}\,f_{{\vec{q}}\,',s1}$,
and $f_{{\vec{q}},s1}\,f^{\dag}_{{\vec{q}}\,',s1}$ generate neutral excitations.

In summary, when acting onto the one- and two-electron
subspace considered in Section I-A, the Hubbard 
model on a square lattice refers to a two-component quantum liquid described
in terms of two types of objects on the corresponding effective lattices
and momentum bands: The charge $c$ fermions and spin-neutral two-spinon $s1$ fermions.
The one- and two-electron subspace can be divided into smaller subspaces
that conserve $S_c$ and $S_s$. Those are spanned by
states of general form given below in Section II-E.
When expressed in terms of $c$ and $s1$ fermion operators, the
Hubbard model on a square lattice in the one- and two-electron
subspace is the square-lattice quantum liquid further studied in Ref. \cite{companion}.
The presence of independent $+1/2$ spinons or of a composite $s2$ 
fermion is accounted for by the values of the occupied and unoccupied sites numbers of
the $s1$ effective lattice and corresponding $s1$ fermion and $s1$ fermion
holes. In turn, the number of independent
$+1/2$ $\eta$-spinons equals that of the unoccupied 
sites of the $c$ effective lattice and $c$ band holes. Otherwise, the
presence of vanishing momentum and energy independent spinons or of 
a single spin-neutral four-spinon $s2$ fermion as well as that of independent $+1/2$ $\eta$-spinons 
has no explicit direct effects on the physics. This property follows from all such objects being
invariant under the electron - rotated-electron unitary transformation \cite{general}.
      
The quantum-liquid $c$ fermions are $\eta$-spinless and spinless
objects without internal degrees of freedom and 
structure whose effective lattice 
is identical to the original lattice. For the complete set of $U/4t>0$ 
energy eigenstates that span the Hilbert space the occupied sites (and unoccupied
sites) of the $c$ effective lattice correspond to those
singly occupied (and doubly occupied and unoccupied)
by the rotated electrons. The corresponding $c$ band
has the same shape and momentum area as the
first Brillouin zone.

In contrast, the quantum-liquid composite spin-neutral two-spinon $s1$ fermions 
have internal structure and the definition of the $s1$ effective lattice in terms of
both the original lattice and the spin effective lattice as well
as the spinon occupancy configurations that describe
such objects is for the one- and two-electron subspace 
a more complex problem \cite{companion0}. It is simplified by the
property of the states that span such a subspace
that the corresponding $s1$ effective lattice has none, one, or two 
unoccupied sites. 

\subsection{The $c$ and $s1$ fermion momentum values and the energy eigenstates}
      
Here we provide the specific form that the momentum energy eigenstates considered  
in Ref. \cite{general} have in the one- and two-electron subspace. Such states refer
to a complete set of states in the full Hilbert space. In general they are not energy
eigenstates. Fortunately, in the one- and two-electron subspace such momentum energy eigenstates
are as well energy eigenstates. This confirms the usefulness of the square-lattice
quantum liquid that refers to that subspace.
      
The $s1$ band discrete momentum values ${\vec{q}}_j$ where $j=1,...,N_{a_{s1}}^2$ 
are the conjugate of the real-space coordinates ${\vec{r}}_j$ of the $s1$ effective lattice
for which also $j=1,...,N_{a_{s1}}^2$. The same applies to the $c$ band discrete momentum values 
${\vec{q}}_j$ and the $c$ effective lattice real-space coordinates ${\vec{r}}_j$ 
where in both cases $j=1,...,N_{a}^2$. (The latter lattice is identical to the original lattice.)
The $c$ translation generators ${\hat{{\vec{q}}}}_c$
commute with both the Hamiltonian and momentum operator for the
whole Hilbert space \cite{general}. This is why the $c$ band discrete momentum values 
are good quantum numbers. In turn, the $s1$ translation generators ${\hat{{\vec{q}}}}_{s1}$ in the 
presence of the fictitious magnetic field ${\vec{B}}_{s1}$ of Eq. (\ref{A-j-s1-3D}) of Appendix A
do not commute in general
with the Hamiltonian of the Hubbard model on the square lattice. Combining the 
results of Ref. \cite{general} with the specific properties of that model in the one- and two-electron subspace 
reveals that in the neutral subspaces of such a subspace 
the $s1$ translation generators ${\hat{{\vec{q}}}}_{s1}$ do commute with both the Hamiltonian and momentum operator. 
This is why for the present square-lattice quantum liquid the $s1$ fermion discrete momentum values 
$\vec{q}=[{q}_{x1},{q}_{x2}]$ are good quantum numbers and thus are conserved. 
The $c$ and $s1$ translation generators read \cite{general},
\begin{equation}
{\hat{{\vec{q}}}}_c  = \sum_{{\vec{q}}}{\vec{q}}\, \hat{N}_c ({\vec{q}})
\, ; \hspace{0.35cm}
{\hat{{\vec{q}}}}_{s1} = \sum_{{\vec{q}}}{\vec{q}}\, \hat{N}_{s1} ({\vec{q}}) \, .
\label{m-generators}
\end{equation}
Here $\hat{N}_{c}({\vec{q}})$ and $\hat{N}_{s1}({\vec{q}})$ are the momentum 
distribution-function operators,
\begin{equation}
\hat{N}_{c}({\vec{q}}) = f^{\dag}_{{\vec{q}},c}\,f_{{\vec{q}},c} \, ;
\hspace{0.35cm}
\hat{N}_{s1}({\vec{q}}) = f^{\dag}_{{\vec{q}},s1}\,f_{{\vec{q}},s1} \, ,
\label{Nc-s1op}
\end{equation}
respectively. For the Hubbard model on the square lattice in the one- and two-electron
subspace the expression of the momentum operator simplifies. It reads,
\begin{equation}
\hat{{\vec{P}}} = {\hat{{\vec{q}}}}_c + {\hat{{\vec{q}}}}_{s1} \, .
\label{P-c-alphanu}
\end{equation}
Indeed, we recall that in it the $c2$ fermion, independent $\pm 1/2$ spinons,
and independent $+1/2$ $\eta$-spinons have vanishing momentum.

Since in contrast to the $c$ fermions, the $s1$ fermions have internal structure,
how is the $s1$ fermion momentum ${\vec{q}}$ related to the two underlying spinons?
Independent spinons carry no momentum and are invariant under the electron - rotated-electron unitary 
transformation \cite{general}. On the other hand, within the LWS
representation of the spin $SU(2)$ algebra \cite{general}, the spin-down
spinon of the spin-singlet two-spinon the $s1$ fermion of
momentum ${\vec{q}}$ carries momentum ${\vec{q}}$ and its 
spin-up spinon carries momentum $-{\vec{q}}$.
In turn, within the highest-weight state (HWS) representation of that algebra, 
its spin-down spinon carries momentum $-{\vec{q}}$ and its spin-up spinon carries momentum ${\vec{q}}$.
Alike in Ref. \cite{general}, here we use the LWS representation, so that the spin-singlet two-spinon
$s1$ fermions of momenta (i) ${\vec{q}}$ and (ii) $-{\vec{q}}$ involve (i) a spin-down
spinon of momentum ${\vec{q}}$ and a spin-up spinon of momentum $-{\vec{q}}$
and (ii) a spin-down spinon of momentum $-{\vec{q}}$ and a spin-up spinon of momentum ${\vec{q}}$,
respectively. 

Within the LWS representation, a one-electron removal excitation breaks the 
spin-singlet spinon pair of a $s1$ fermion before the annihilation of a spin-down electron.
The spin-down spinon of momentum ${\vec{q}}$ is then removed within the electron.
The uncompensated spin-up spinon momentum $-{\vec{q}}$ is associated with that of a hole emerging  
in the $s1$ band at momentum ${\vec{q}}$. Indeed, the
latter spinon decays into that momentum $-{\vec{q}}$ $s1$ band hole and
a vanishing-momentum spin-up independent spinon. 
Hence one-electron excitations break spinon bond pairs whose spinons had momenta ${\vec{q}}$
and $-{\vec{q}}$ corresponding to their relative motion in the pair. 

In turn, spinon pair breaking under spin excitations or excitations involving
removal of two electrons with the same spin projection 
may introduce an extra momentum contribution
that corresponds to the motion of the center of mass of the broken spin-singlet spinon pair.  
Under such excitations there emerge two holes in the $s1$ band at momenta ${\vec{q}}\,'$ and ${\vec{q}}\,''$
of the general form,
\begin{equation}
{\vec{q}}\,' = {\vec{q}} + {1\over 2}\delta {\vec{q}}
\, ; \hspace{0.35cm}
{\vec{q}}\,'' = -{\vec{q}} + {1\over 2}\delta {\vec{q}} \, .
\label{qq-dqdq}
\end{equation}
Here,
\begin{equation}
{\vec{q}} = {1\over 2}[{\vec{q}}\,' -{\vec{q}}\,'']
\, ; \hspace{0.35cm}
\delta {\vec{q}} = {\vec{q}}\,' +{\vec{q}}\,'' \, .
\label{q-dq}
\end{equation}
where ${\vec{q}}$ corresponds to the spinon relative motion in the pair
and $\delta {\vec{q}}$ refers to the motion of the center of mass of the 
spinon pair. 

That for the square-lattice quantum liquid both the $c$ and $s1$ fermion discrete momentum values 
are good quantum numbers confirms the suitability of the present description in terms
of occupancy configurations of the $c$ and $s1$ effective lattices. Indeed, the
$c$ and $s1$ band discrete momentum values are the conjugate of the real-space 
coordinates of the $c$ and $s1$ effective lattice, respectively. 
Are the approximations used in the construction of the $s1$ effective lattice inconsistent
with the $s1$ band discrete momentum values being good quantum numbers?
The answer is no. Indeed, such approximations concern the relative positions of 
the $j=1,...,N_{a_{s1}}^2$ sites of the $s1$ effective lattice \cite{general,companion,companion0}. 
They are only directly related to the shape of the $s1$ band boundary. They do not
affect the $s1$ band discrete momentum values being good quantum numbers.
At $x=0$ the spin effective lattice is identical to the original square lattice and
the $s1$ effective lattice is one of its two sub-lattices. Consistently, at $x=0$ and $m=0$
the boundary of the $s1$ momentum band is accurately known. Indeed, then
the $s1$ band coincides with an antiferromagnetic reduced Brillouin 
zone of momentum area $2\pi^2$ such that $\vert q_{x_1}\vert+\vert q_{x_2}\vert\leq\pi$ \cite{companion}.    
In turn, it is known that for $x>0$ and $m=0$ the the $s1$ band boundary encloses a 
smaller momentum area $(1-x)2\pi^2$ yet its precise shape remains an open issue.
The related problems of the $c$ and $s1$ momentum bands and corresponding energy dispersions and velocities
are studied in Ref. \cite{companion}.     

For a number of sites $N_a^2\gg 1$ very large but finite that the description used
in the studies of this paper refers to a $m=0$ ground state is both for 
$x=0$ and $x>0$ a spin-singlet state \cite{general}. For $m=0$ and $x=0$ this agrees with
a theorem introduced and proved in Ref. \cite{Lieb-89}. The corresponding
one- and two-electron subspace considered in this paper may be divided
into a well-defined set of smaller subspaces spanned by neutral states. Such
states conserve the eigenvalue $S_c=[1/2](1-x)N_a^2$ of the generator
of the hidden $U(1)$ symmetry and the spin $S_s=0,{1\over 2},1$ and thus
conserve as well the number of sites of the $s1$ effective lattice
$N_{a_{s1}}^2=[S_c+S_s]$. The set of energy eigenstates 
that span such subspaces are particular cases of the general momentum
eigenstates $\vert \Phi_{U/4t}\rangle$ studied in Ref. \cite{general}. The use
of the general expression of such states leads to the following
corresponding general form for the energy eigenstates $\vert \Psi_{U/4t}\rangle =\vert \Phi_{U/4t}\rangle$  
that span the one- and two-electron subspace considered here,
\begin{equation}
\vert \Psi_{U/4t}\rangle = \frac{({\hat{S}}^{\dag}_{s})^{L_{s,\,-1/2}}}{
\sqrt{{\cal{C}}_{s}}}\vert \Phi_{LWS;U/4t}\rangle 
\, ; \hspace{0.5cm} {\cal{C}}_{s} = \delta_{L_{s,\,-1/2},\,0} +
\prod_{l=1}^{L_{s,\,-1/2}}l\,[\,L_{s}+1-l\,] = 1, 2, 4 \, .
\label{non-LWS}
\end{equation}
The LWS appearing in this equation reads,
\begin{equation}
\vert \Psi_{LWS;U/4t}\rangle = [\vert 0_{\eta};N_{a_{\eta}}^2\rangle]
[\prod_{{\vec{q}}\,'}f^{\dag}_{{\vec{q}}\,',s1}\vert 0_{s1};N_{a_{s}^2}\rangle]
[\prod_{{\vec{q}}}f^{\dag}_{{\vec{q}},c}\vert GS_c;0\rangle] \, ; \hspace{0.25cm} f^{\dag}_{{\vec{q}}\,',s1} =
{\hat{V}}^{\dag}\,{\mathcal{F}}^{\dag}_{{\vec{q}}\,',s1}\,{\hat{V}} \, ; \hspace{0.25cm} f^{\dag}_{{\vec{q}},c} =
{\hat{V}}^{\dag}\,{\mathcal{F}}^{\dag}_{{\vec{q}},c}\,{\hat{V}} \, .
\label{LWS-full-el}
\end{equation}
Here ${\mathcal{F}}^{\dag}_{{\vec{q}}\,',s1}$ and ${\mathcal{F}}^{\dag}_{{\vec{q}},c}$ are the
creation operators of a $U/4t\rightarrow\infty$ $s1$ fermion of momentum ${\vec{q}}\,'$ and $c$ fermion of
momentum ${\vec{q}}$, respectively \cite{general}. Moreover, $\vert 0_{\eta};N_{a_{\eta}}^2\rangle$ is
the $\eta$-spin $SU(2)$ vacuum associated with $N_{a_{\eta}}^2$ independent $+1/2$ $\eta$-spinons,
$\vert 0_{s};N_{a_{s}}^2\rangle$ is the spin $SU(2)$ vacuum associated with $N_{a_{s}}^2$ 
independent $+1/2$ spinons, and $\vert GS_c;0\rangle$ is the $c$ $U(1)$
vacuum. Such three vacua are invariant under the electron - rotated-electron unitary
transformation, refer to the model global
$[SU(2)\times SU(2)\times U(1)]/Z_2^2 =SO(3)\times SO(3)\times U(1)$ symmetry
\cite{bipartite}, and appear in the theory vacuum of Eq. (\ref{vacuum}) of Appendix A. 
(In that equation, $\vert GS_c;2S_c\rangle=\prod_{{\vec{q}}}f^{\dag}_{{\vec{q}},c}\vert GS_c;0\rangle$.)

The more general states $\vert \Phi_{U/4t}\rangle$ considered in Ref. \cite{general} involve 
occupancy configurations of the remaining $\eta\nu$ fermion and $\nu>1$ $s\nu$ fermion
branches and $-1/2$ $\eta$-spinon occupancies absent in the expressions
given in Eqs. (\ref{non-LWS}) and (\ref{LWS-full-el}). Importantly, 
the results of that reference concerning the subspaces A and B defined
in it confirm that the states of form (\ref{non-LWS}) and (\ref{LWS-full-el})
are indeed energy eigenstates. Since they span all subspaces of the one- and two-electron
subspace that conserve $S_c$ and $S_s$, they span the
latter subspace as well. In contrast, the momentum eigenstates generated by
simple occupancy configurations of $c$ and $\alpha\nu$ fermions and independent
$\eta$-spinons and spinons of the larger set of  
states $\{\vert \Phi_{U/4t}\rangle\}$ considered in Ref. \cite{general}
are not in general energy eigenstates. As
justified in that reference, the energy eigenstates are superpositions
$\vert \Psi_{U/4t}\rangle = \sum_l C_l\,\vert \Phi_{U/4t;l}\rangle$
of a set of such states $\{\vert \Phi_{U/4t;l}\rangle\}$ with the same momentum eigenvalue. 
We recall that states with a single $s2$ fermion have
also the general form provided Eqs. (\ref{non-LWS}) and (\ref{LWS-full-el}). 
As discussed above, the presence of that vanishing-energy,
vanishing-momentum, and spin-neutral four-spinon object is accounted for
the values of the numbers $[N^h_{s 1}-2S_s]=[S_c - S_s - N_{s1}]=2N_{s2}=0,2$
of Eq. (\ref{Nas1-Nhs1}).

The energy eigenstates $\vert \Psi_{U/4t}\rangle$ of general form (\ref{non-LWS})
that span the one- and two-electron subspace have
numbers $N_{s2}=N_{a_{s2}}^2=0,1$ and $N_{s1}\approx N_{a_{s1}}^2$ such that
$N_{s1}^h=[N_{a_{s1}}^2-N_{s1}]=0,1,2$. Hence according to the results of Ref. \cite{general}
the lattice spacing $a_{s1} \approx \sqrt{2}\,a_s=\sqrt{2/(1-x)}\,a$ of Eq. (\ref{a-a-s1-sube})
is directly related to the fictitious magnetic-field length $l_{s1}$ associated with the 
field of Eq. (\ref{A-j-s1-3D}) of Appendix A. Indeed, in that subspace one has that
$\langle n_{\vec{r}_j,s1}\rangle\approx 1$ and such a fictitious magnetic field reads
${\vec{B}}_{s1} ({\vec{r}}_j) \approx \Phi_0\sum_{j'\neq j}\delta ({\vec{r}}_{j'}-{\vec{r}}_{j})\,{\vec{e}}_{x_3}$.
It acting on one $s1$ fermion differs from zero only at the positions
of other $s1$ fermions. In the mean field approximation one replaces it
by the average field created by all $s1$ fermions at position $\vec{r}_j$. This gives,
${\vec{B}}_{s1} ({\vec{r}}_j) \approx \Phi_0\,n_{s1} (\vec{r}_j)\,{\vec{e}}_{x_3}
\approx \Phi_0\,[N_{a_{s1}}^2/L^2]\,{\vec{e}}_{x_3}=[\Phi_0/a_{s1}^2]\,{\vec{e}}_{x_3}$. 
One then finds that the number $N_{a_{s1}}^2$
of the $s1$ band discrete momentum values equals $[B_{s1}\,L^2]/\Phi_0$. In addition, the
$s1$ effective lattice spacing $a_{s1}$ is expressed in terms to the fictitious 
magnetic-field length $l_{s1}\approx a/\sqrt{\pi(1-x)}$ as $a_{s1}^2=2\pi\,l_{s1}^2$.
This is consistent with each $s1$ fermion having a flux
tube of one flux quantum on average attached to it. 

As further discussed in Ref. \cite{companion}, for the present one- and two-electron subspace
the $s1$ fermion problem is then related to the Chern-Simons theory  
\cite{Giu-Vigna}. Indeed the number of flux quanta being one is consistent with the
$s1$ fermion and $s1$ bond-particle wave functions obeying Fermi and Bose statistics, respectively. 
Hence the composite $s1$ fermion consists of two spinons in a spin-singlet configuration plus an infinitely thin flux tube attached 
to it. Thus, each $s1$ fermion appears to carry a fictitious magnetic solenoid
with it as it moves around in the $s1$ effective lattice.
     
\section{Long-range antiferromagnetic order and short-range spiral-incommensurate spin order 
for $x=0$ and $0<x\ll 1$, respectively}

Here we profit from the rotated-electron description used in the studies of this paper 
to address issues related to the occurrence for $m=0$, zero temperature $T=0$, 
and $N_a^2\rightarrow\infty$ of a long-range antiferromagnetic 
order and a short-range spiral-incommensurate spin order at $x=0$ and for $0<x\ll 1$, 
respectively. The emergence as $N_a^2\rightarrow\infty$
of a long-range antiferromagnetic order in the $x=0$ and $m=0$ ground state 
is associated with a spontaneous symmetry breaking. We argue that a condition
for emergence of such a long-range antiferromagnetic order is that the spin
effective lattice is identical to the original lattice. Such a condition is not
met for small hole concentrations $0<x\ll1$.

\subsection{Extension of the Mermin and Wagner Theorem to the
half-filled Hubbard model for $U/4t>0$}

It is well known that for $U/4t\gg 1$ the spin degrees of freedom of the half-filled Hubbard model 
on a square lattice may be described by an isotropic spin-$1/2$ Heinsenberg
model on a square lattice. It follows that the Mermin and 
Wagner Theorem \cite{MWT} is valid for the former
model at half filling and $U/4t\gg 1$. The theorem states 
that then there is no long-range antiferromagnetic order 
for finite temperatures $T>0$ and $N_a^2\rightarrow\infty$. 

Let us provide evidence that the Mermin and Wagner Theorem 
applies to the half-filled Hubbard model on a square lattice
for all values $U/4t>0$. The possibility of such an extension 
to $U/4t>0$ is strongly suggested by evidence involving
the transformation laws of the spin configurations under the 
electron - rotated-electron unitary transformation.
We recall that in terms of the rotated electrons as defined in
Section I and Ref. \cite{general}, the occupancy
configurations that generate the energy eigenstates $\vert \Psi_{U/4t}\rangle ={\hat{V}}^{\dag}\vert\Psi_{\infty}\rangle$
are the same for all finite values $U/4t>0$. Moreover,
such rotated-electron occupancy configurations equal
those that generate the corresponding energy eigenstates $\vert\Psi_{\infty}\rangle$
in terms of electrons  in the $U/4t\rightarrow\infty$ limit. 

The rotated-electron configurations that generate the energy eigenstates $\vert \Psi_{U/4t}\rangle$
that span the one- and two-electron subspace defined in this paper are much more complex
than those associated with the simple form (\ref{non-LWS}) and (\ref{LWS-full-el}) in terms of
$c$ and $s1$ fermion operators. Indeed, the expression of their generators is in terms
of rotated-electron creation operators is an involved problem. This follows from the 
expression of the spin-neutral two-spinon $s1$ fermion operators not being simple in terms
of the rotated-electron operators \cite{companion,companion0}. This reveals that
the electronic occupancy configurations that in the $U/4t\rightarrow\infty$ limit generate such energy eigenstates
correspond to an involved problem as well. For $U/4t\rightarrow\infty$
the $c$ fermion holes, spinons, and $\eta$-spinons are the ``quasicharges'', spins, and pseudospins,
respectively, of Ref. \cite{Ostlund-06}. For the Hubbard model on the square lattice in the one- and two-electron subspace
the energy bandwidth of the $s1$ fermion dispersion vanishes for $U/4t\rightarrow\infty$
and the $c$ fermion dispersion has in that limit the simple form
$\epsilon_c ({\vec{q}}) = -2t \sum_{i=1}^2\,[\cos (q_{x_i})-\cos (q_{Fcx_i})]$ in terms of the $c$ band momentum
components \cite{companion}. Here $q_{Fcx_i}$ where $i=1,2$ are the components of the $c$ Fermi
momentum ${\vec{q}}_{Fc}$ defined in Ref. \cite{companion}.
 
The use below of the following two properties provides strong evidence that 
the Mermin and Wagner Theorem holds for the half-filled Hubbard model on the
square lattice for $U/4t>0$:
I) The $x=0$ and $m=0$ absolute ground state is in the limit $N_a^2\rightarrow\infty$ invariant under the electron - rotated-electron
unitary transformation \cite{general,companion}. Hence the occurrence for $N_a^2\rightarrow\infty$
of long-range antiferromagnetic order as $U/4t\rightarrow\infty$, associated with that of the isotropic spin-$1/2$ Heisenberg
model, implies the occurrence for $N_a^2\rightarrow\infty$ of that long-range order for $U/4t>0$ as well;
II) Since in terms of rotated electrons single and double occupancy are good quantum
numbers for $U/4t>0$, the rotated-electron occupancy configurations that generate
the energy eigenstates are more ordered than the corresponding electron occupancy
configurations. It follows that the lack of long-range antiferromagnetic
order in terms of the spins of the rotated electrons implies as well 
a lack of such an order in terms of the spins of the electrons whose
occupancy configurations generate the same states. 

The rotated-electron operator description of Ref. \cite{general} has been constructed to inherently 
the electron occupancy configurations that for $U/4t\rightarrow\infty$ generate an energy eigenstate
$\vert\Psi_{\infty}\rangle$ being identical to the rotated-electron occupancy configurations that for $U/4t>0$ generate
the energy eigenstates $\vert \Psi_{U/4t}\rangle ={\hat{V}}^{\dag}\vert\Psi_{\infty}\rangle$
belonging to the corresponding $V$ tower. Hence concerning the original-lattice 
rotated-electron occupancies, the Mermin and Wagner Theorem applies 
to all finite values of $U/4t>0$: That for the occupancy configurations
of the rotated-electron spins there is no long-range antiferromagnetic
order for temperatures $T>0$ and $N_a^2\rightarrow\infty$ is an exact result. The above property II
then implies that the lack of long-range antiferromagnetic order of the rotated-electron spins 
for $U/4t>0$ and $T>0$ implies a similar
lack of such an order for the spins of the original electrons. Indeed, for finite values of $U/4t$ the
spin occupancy configurations are more ordered 
for the rotated electrons than for the electrons. 

This is also consistent with the emergence of a long-range antiferromagnetic
order in the Hubbard model on a square lattice in the thermodynamic limit $N_a^2\rightarrow\infty$
at hole concentration $x=0$, temperature $T=0$, and $U/4t>0$. Indeed, there is a large 
consensus that in the thermodynamic limit $N_a^2\rightarrow\infty$
long-range antiferromagnetic order occurs in the ground state of the related isotropic spin-$1/2$ Heisenberg
model on the square lattice \cite{Hirsch85,Peter-88,Sandvik}. This implies that in the thermodynamic limit 
$N_a^2\rightarrow\infty$ a similar long-range order sets in in the ground state of the half-filled 
Hubbard model on the square lattice at large $U/4t\gg 1$ values. Moreover, for $U/4t>0$ a similar ground-state order occurs in
that limit in the ground state of the latter model, in terms of the spins of the rotated electrons. 
That according to the above property I the $x=0$ and $m=0$ absolute ground state is in the limit 
$N_a^2\rightarrow\infty$ invariant under the electron - rotated-electron
unitary transformation then implies that in the thermodynamic limit $N_a^2\rightarrow\infty$
and for $U/4t>0$ a long-range antiferromagnetic order sets in in the ground state of the 
half-filled Hubbard model on the square lattice in terms of the spins of the electrons as well.
This agrees with many previous studies of that model, as 
for instance those of Refs. \cite{half-filling,2D-A2,bag-mech,2D-NM,Hubbard-T*-x=0,Kopec}. As 
discussed below, there is strong evidence that both for $x=0$, $T>0$, and $U/4t>0$ 
and for $0<x\ll 1$, $T\geq 0$, and $U/4t>0$ such a ground-state long-range order is replaced by a 
ground-state short-range spiral-incommensurate spin order. 

\subsection{$x=0$ and $m=0$ ground-state symmetry for $N_a^2\rightarrow\infty$
and a necessary condition for its spontaneous symmetry breaking}

Our above arguments provide strong evidence that provided that in the thermodynamic limit $N_a^2\rightarrow\infty$
a long-range antiferromagnetic order occurs in the ground state of the related isotropic spin-$1/2$ Heisenberg
model on the square lattice, a similar long range order sets in in that limit in the ground state
of the half-filled Hubbard model on the square lattice for $U/4t>0$.
However, our goal is not providing a mathematical proof that in the thermodynamic limit $N_a^2\rightarrow\infty$
long-range antiferromagnetic order occurs in the half-filled Hubbard model on 
the square lattice. Although there is no such a proof, there is a large 
consensus that it should be so \cite{half-filling,2D-A2,bag-mech,2D-NM,Hubbard-T*-x=0,Kopec}. 
An appropriate measure of such a long-range order is the square of the staggered magnetization, 
\begin{equation}
\langle\Psi_{GS}\vert{\vec{\hat{M}}}_s^2\vert\Psi_{GS}\rangle =
\langle\Psi_{GS}\vert\left({1\over N_a^2}\sum_{j=0}^{N_a^2-1} \varepsilon_j {\vec{\hat{s}}}_{\vec{r}_j}\right)^2\vert\Psi_{GS}\rangle 
= {1\over N_a^2}\sum_{j=0}^{N_a^2-1} \varepsilon_{0}\,\varepsilon_{j}\,C_j \, .
\label{stag-magn-vec}
\end{equation}
Here the spin operator ${\vec{\hat{s}}}_{\vec{r}_j}$ refers to the spin of an electron at
the site of real-space coordinate $\vec{r}_j$ in the original lattice,
$C_j$ is the spin correlation function in the spin-singlet $S_s=0$ and $x=0$ ground state 
$\vert\Psi_{GS}\rangle$ of the model for $N_a^2\gg 1$ very large but finite, 
\begin{equation}
C_j = \langle\Psi_{GS}\vert{\vec{\hat{s}}}_{\vec{r}_{0}}\cdot{\vec{\hat{s}}}_{\vec{r}_{j}}\vert\Psi_{GS}\rangle \, ,
\label{C-j}
\end{equation}
and $\varepsilon_j  = +1$ if $j$ refers to the sub-lattice A. Otherwise 
$\varepsilon_j  = -1$. This just compensates the sign of the antiferromagnetic 
correlation function (\ref{C-j}). For $N_a^2\gg 1$ finite and even the $x=0$ and $m=0$ 
ground state $\vert\Psi_{GS}\rangle$ appearing in Eqs. (\ref{stag-magn-vec}) and (\ref{C-j}) 
has zero momentum.

The spin operator ${\vec{\hat{s}}}_{\vec{r}_j}$ appearing in Eqs. (\ref{stag-magn-vec}) and (\ref{C-j}) 
has operator Cartesian components ${\hat{s}}^{x_1}_{\vec{r}_j}$, ${\hat{s}}^{x_2}_{\vec{r}_j}$,
and ${\hat{s}}^{x_3}_{\vec{r}_j}$ and refers to the spin of an electron at
the site of real-space coordinate $\vec{r}_j$. In turn, we denote the corresponding
rotated spin or spinon operator by ${\vec{s}}_{\vec{r}_j}$ and its operator Cartesian components
by $s^{x_1}_{\vec{r}_j}$, $s^{x_2}_{\vec{r}_j}$, and $s^{x_3}_{\vec{r}_j}$. It refers to the spin 
of a rotated electron at the site of real-space coordinate $\vec{r}_j$ whose components
appear in Eqs. (\ref{sir-pir}) and (\ref{rotated-quasi-spin}). Our spinons are
such rotated spins.

Under the emergence of long-range antiferromagnetic order in the limit $N_a^2\rightarrow\infty$,
the square of the staggered magnetization $\langle\Psi_{GS}\vert{\vec{\hat{M}}}_s^2\vert\Psi_{GS}\rangle$ extrapolates 
to a finite asymptotic absolute value $\vert C_{\infty}\vert$ of the correlation function (\ref{C-j}),
\begin{equation}
\lim_{N_a^2\rightarrow\infty} \langle\Psi_{GS}\vert{\vec{\hat{M}}}_s^2\vert\Psi_{GS}\rangle = \vert C_{\infty}\vert \, .
\label{stag-magn-t-limit}
\end{equation}
The related magnetic structure factor $S (\vec{k})$ is the Fourier transform of that correlation function,
\begin{equation}
S (\vec{k}) = \sum_{j=0}^{N_a^2-1} e^{i\vec{k}\cdot\vec{r}_j} C_j  \, .
\label{S-k}
\end{equation}
It can be measured directly in neutron scattering experiments 
\cite{LCO-neutr-scatt,companion}. As a result of its form (\ref{S-k}),
$S (\vec{\pi})= N_a^2 \langle{\vec{\hat{M}}}_s^2\rangle$ will grow linearly with 
the number of sites $N_a^2$ if there is long-range antiferromagnetic order for
$N_a^2\rightarrow\infty$,
\begin{equation}
{S (\vec{\pi}) \over N_a^2} =   \langle\Psi_{GS}\vert{\vec{\hat{M}}}_s^2\vert\Psi_{GS}\rangle = {m_{AF}^2\over 3} + {\cal{O}}(1/N_a)  \, .
\label{S-pi-pi}
\end{equation}
Here the sub-lattice magnetization $m_{AF}=\lim_{N_a^2\rightarrow\infty}3\langle{\vec{\hat{M}}}_s^2\rangle=3\vert C_{\infty}\vert$ 
plays the role of antiferromagnetic order parameter.

The large scale DQMC calculations of Ref. \cite{half-filling} provide 
useful information on the effective bandwidth, momentum distribution, and magnetic correlations 
of the half-filled Hubbard model on the square lattice. They employ the DQMC method, which provides an 
approximation-free solution of the such a model on square lattices large 
enough to use finite-size scaling to, for example, reliably 
extract the sub-lattice magnetization $m_{AF}=3\vert C_{\infty}\vert$
as a function of $U/4t$. Such Monte Carlo calculations 
as well the random-phase approximation results of Ref. \cite{bag-mech}
reveal that $m_{AF}$ vanishes for $U/4t\rightarrow 0$ and is an increasing
function of $U/4t$ that for approximately $U/4t\approx 2$ saturates to the value 
$m_{AF}^{HM}\approx 0.614$ of the isotropic spin-$1/2$ Heisenberg model on the square
lattice determined in Ref. \cite{Sandvik}. The quantum Monte Carlo DQMC results of Ref. \cite{half-filling} are an
improvement of the corresponding results of Monte Carlo simulations
of Ref. \cite{2D-A2}, which predicted a lower value for the saturated $m_{AF}$.
Below we relate the sub-lattice magnetization 
$m_{AF}$ to an energy scale that can as well be used as antiferromagnetic order
parameter. Combining that relation with the behavior $m_{AF}= m_{AF}^{HM}\approx 0.614$ 
for $U/4t\gg 1$, we find the following approximate limiting behaviors,
\begin{eqnarray}
m_{AF} & \approx & {8^2 t\over U}\,e^{-\pi\sqrt{4t\over U}} \, ,
\hspace{0.25cm} U/4t\ll 1 \, ,
\nonumber \\
& \approx & m_{AF}^{HM} - {1\over 4}\left({2 - U/4t\over 2\ln 2}\right)^2  \, ,
\hspace{0.25cm} u_0 \leq U/4t\leq 2 \, ,
\nonumber \\
& \approx  &  m_{AF}^{HM} \, ,
\hspace{0.25cm}  U/4t> 2 \, .
\label{m-AF}
\end{eqnarray}
where
\begin{equation}
m_{AF}^{HM} \approx {2e^1\over \pi^2}  +  {1\over 4}\left({2 - u_0\over 2\ln 2}\right)^2 \approx 0.6142 \, .
\label{M-AF-HM}
\end{equation}
The expression given in Eq. (\ref{m-AF}) for $u_0 \leq U/4t\leq 2$ is also
valid for $0\leq (u_0-U/4t)\ll 1$ where $u_0$ is a $U/4t$ value found below to read $u_0\approx 1.302$. Alike 
in Ref. \cite{half-filling},  Eqs. (\ref{m-AF}) and  (\ref{M-AF-HM}) refer to units where the classical 
N\'eel state has $m_{AF}=1$. In Ref. \cite{Sandvik} units are used where that state has
$m_{AF}=1/2$. In the latter units the parameter (\ref{M-AF-HM}) reads
instead $\approx 0.3071$, consistently with the results of that reference.

When expressed in terms of rotated-electron creation and annihilation operators, the 
Hamiltonian of the Hubbard model on the square lattice (\ref{H}) has an infinite number of terms,
as given in Eq. (\ref{HHr}). For the half-filled model in the one- and two-electron subspace
with both no rotated-electron double occupancy and no rotated-hole double occupancy the 
$c$ fermion band is full for the ground state. Moreover, the excitations involving the emergence of
$c$ fermion holes are gapped. In the one- and two-electron subspace the model
Hamiltonian (\ref{HHr}) may then be expressed only in terms of spinon operators $\vec{s}_{\vec{r}_{j}}$ 
whose operator components $s^l_{\vec{r}_j}$ are given in Eq. (\ref{sir-pir}). Consistently with
the related results of Ref. \cite{Stein}, one finds that up to fifth order in $t/U$ the Hamiltonian (\ref{HHr})
has in terms of such spin operators the following form,
\begin{eqnarray}
H  & = & {t^2\over 2U}\sum_{\langle j_1 j_2\rangle} (\vec{s}_{\vec{r}_{j_1}}\cdot\vec{s}_{\vec{r}_{j_2}}-1)
- {2t^4\over U^3}\sum_{\langle j_1 j_2\rangle} (\vec{s}_{\vec{r}_{j_1}}\cdot\vec{s}_{\vec{r}_{j_2}}-1) 
\nonumber \\
& + & {t^4\over 2U^3}\sum_{j_1,j_2,j_3} D_{j_1,j_2} D_{j_2,j_3} (\vec{s}_{\vec{r}_{j_1}}\cdot\vec{s}_{\vec{r}_{j_3}}-1)
\nonumber \\
& + & {t^4\over 8U^3}\sum_{j_1,j_2,j_3,j_4} D_{j_1,j_2} D_{j_2,j_3} D_{j_3,j_4}D_{j_4,j_1}
(1-\vec{s}_{\vec{r}_{j_1}}\cdot\vec{s}_{\vec{r}_{j_2}}
\nonumber \\
& - & \vec{s}_{\vec{r}_{j_1}}\cdot\vec{s}_{\vec{r}_{j_3}}
-\vec{s}_{\vec{r}_{j_1}}\cdot\vec{s}_{\vec{r}_{j_4}}-\vec{s}_{\vec{r}_{j_2}}\cdot\vec{s}_{\vec{r}_{j_3}}
-\vec{s}_{\vec{r}_{j_2}}\cdot\vec{s}_{\vec{r}_{j_4}}-\vec{s}_{\vec{r}_{j_3}}\cdot\vec{s}_{\vec{r}_{j_4}})
\nonumber \\
& + & {5t^4\over 8U^3}\sum_{j_1,j_2,j_3,j_4} D_{j_1,j_2} D_{j_2,j_3} D_{j_3,j_4}D_{j_4,j_1}
[(\vec{s}_{\vec{r}_{j_1}}\cdot\vec{s}_{\vec{r}_{j_2}})(\vec{s}_{\vec{r}_{j_3}}\cdot\vec{s}_{\vec{r}_{j_4}})
\nonumber \\
& + & (\vec{s}_{\vec{r}_{j_1}}\cdot\vec{s}_{\vec{r}_{j_4}})(\vec{s}_{\vec{r}_{j_2}}\cdot\vec{s}_{\vec{r}_{j_3}})
- (\vec{s}_{\vec{r}_{j_1}}\cdot\vec{s}_{\vec{r}_{j_3}})(\vec{s}_{\vec{r}_{j_2}}\cdot\vec{s}_{\vec{r}_{j_4}})] \, .
\label{Heff}
\end{eqnarray}
Here $\langle j_1 j_2\rangle$ refers to a summation running over nearest-neighboring sites
and $D_{j,j'}=1$ for the real-space coordinates ${\vec{r}}_j$ and ${\vec{r}}_{j'}$
corresponding to nearest-neigboring sites and $D_{j,j'}=0$ otherwise.
Analysis of the interactions in spin space of the Hamiltonian (\ref{Heff})  
reveals that some of its terms do not introduce frustration whereas other do. 
However, at half filling the spin interactions of the Hamiltonian (\ref{Heff}) including those
of all higher order contributions do not destroy the sub-lattice magnetization $m_{AF}$.
They merely destabilize the classical N\'eel state, lessening the sub-lattice magnetization 
from its classical magnitude $m_{AF}=1$.

That as obtained by different authors and methods \cite{half-filling,2D-A2,bag-mech,Hirsch85}
the sub-lattice magnetization $m_{AF}$ of Eqs. (\ref{S-pi-pi}) and (\ref{m-AF})
is indeed finite for the Hubbard model on the square lattice at $U/4t>0$ provides strong evidence that for 
$N_a^2\rightarrow\infty$ the spin correlation function (\ref{C-j}) has long-range antiferromagnetic 
order. In contrast to 1D, the quantum fluctuations associated with the interactions in spin space 
of the Hamiltonian (\ref{Heff}) and its higher-order terms are not strong enough to destroy it.
In turn, the exact Mermin-Wagner Theorem \cite{MWT} implies that at finite temperatures thermal 
fluctuations destroy such an order of the square-lattice model. 

A stronger confirmation is obtained by the scaling of the spectrum
itself. It is an illustration of the mechanism of spontaneous symmetry
breaking. Anderson was the first to point out that the spontaneous
symmetry breaking mechanism that occurs in the thermodynamic
limit $N_a^2\rightarrow\infty$ involves a whole tower of low-lying energy eigenstates of
the finite system \cite{Ander-collap}. They collapse in that limit onto the ground state. 
One may investigate which energy eigenstates couple to the 
exact finite $N_a^2\gg 1$ and $x=0$ and $m=0$ ground state $\vert\Psi_{GS}\rangle$
via the operator of the staggered magnetization ${\hat{M}}_s^{l}$. Here $l=\pm, x_3$. 
We insert a complete set of energy eigenstates as follows,
\begin{equation}
\langle \Psi_{GS}\vert({\hat{M}}_s^{l})^2\vert\Psi_{GS}\rangle =
\sum_i \langle\Psi_{GS}\vert{\hat{M}}_s^{l}\vert \Psi_{i}\rangle\langle\Psi_{i}\vert{\hat{M}}_s^{l}\vert \Psi_{GS}\rangle =
\sum_i \vert\langle\Psi_{GS}\vert{\hat{M}}_s^{l}\vert \Psi_{i}\rangle\vert^2 
\, ; \hspace{0.25cm} l = \pm , x_3 \, .
\label{stag-magn}
\end{equation}
Only excited energy eigenstates $\vert \Psi_{i}\rangle$ with momentum 
$\vec{k}=\vec{\pi}$ and quantum numbers $S_{\eta}=0$, $S_c=N_a^2/2$,
$S_s= 1$, and $S_s^{x_3}=0,\pm 1$ corresponding to $l=x_3,\pm$ contribute 
to the sum of Eq. (\ref{stag-magn}). We emphasize that the quantum numbers $S_{\eta}=0$
and $S_c=N_a^2/2$ remain unchanged and thus are the same as for
the ground state $\vert\Psi_{GS}\rangle$. We denote by $\vert \Psi_{1T}\rangle$
the lowest $S_s=1$, $S_{\eta}=0$, $S_c=N_a^2/2$, and $\vec{k}=[\pi,\pi]$ spin-triplet state
whose excitation energy behaves as $1/N_a^2$ for finite
$N_a^2\gg 1$. For the range $U/4t>u_0\approx 1.3$ of interest for the studies of Ref. \cite{companion}
the contribution from such a lowest spin triplet state 
is by far the largest: For instance for approximately $U/4t>2$ the matrix-element square 
$\vert\langle\Psi_{GS}\vert{\hat{M}}_s^{l}\vert \Psi_{1T}\rangle\vert^2$ exhausts 
the sum in Eq. (\ref{stag-magn}) by more than 98.7\%. (This is the value
found by exact diagonalization for the related spin-$1/2$ Heisenberg model on the 
square lattice in Ref. \cite{Peter-88}. As similar result is expected for
approximately $U/4t>u_0\approx 1.3$.)

The special properties with respect to the lattice symmetry group of the
lowest energy eigenstates contributing to the linear Goldstone modes of the
corresponding $S_s=1$ spin-wave spectrum reveal the space-symmetry breaking 
of the $N_a^2\rightarrow\infty$ ground state. In the present case 
of the half-filled Hubbard model on the square lattice the translation symmetry is broken. 
Hence both the $\vec{k}=[0,0]$ and $\vec{k}=[\pi,\pi]$ momenta appear among the
lowest energy eigenstates contributing to the linear Goldstone modes of the
$S_s=1$ spin-wave spectrum.

The $c$ and $s1$ fermion description can be used to derive such a spin-wave spectrum. 
For $x=0$ and $m=0$ it is generated in Ref. \cite{companion} in terms of simple 
two-$s1$-fermion-hole processes. The investigations of that reference confirm that at $x=0$ 
the spin-wave spectrum includes linear Goldstone modes at momenta $\vec{k}=[0,0]$ and $\vec{k}=[\pi,\pi]$.
That the transverse spin-spin correlation function contains gapless poles, as predicted
by the Goldstone Theorem, is consistent with in the limit $N_a^2\rightarrow\infty$ the ground state breaking the
continuous spin $SU(2)$ rotational invariance of the Hamiltonian. The corresponding spin-wave 
spectrum is plotted in Fig. 1 of Ref. \cite{companion} for the half-filled Hubbard model on the square lattice in the thermodynamic
limit $N_a^2\rightarrow\infty$. (As mentioned above, for large $N_a^2$ the lowest $\vec{k}=[\pi,\pi]$ excitation energy 
vanishes as $1/N_a^2$.) The occurrence of linear Goldstone modes at momenta $\vec{k}=[0,0]$ and $\vec{k}=[\pi,\pi]$ 
in the theoretical spin-wave spectrum derived in Ref. \cite{companion} explicitly confirms that the results 
of the $c$ and $s1$ fermion description used in the studies of this paper are fully consistent with for $x=0$, $m=0$, and
temperature $T=0$ long-range antiferromagnetic setting in as $N_a^2\rightarrow\infty$.

Importantly, for $U/4t\approx 1.525$ and $t\approx 295$ meV the spin-wave spectrum of the 
parent compound La$_2$CuO$_4$ (LCO) \cite{LCO-neutr-scatt} is quantitatively described by the 
corresponding theoretical spectrum derived in Ref. \cite{companion} in terms of simple spinon pair 
breaking $s1$ fermion processes. Within the present status of 
the scheme used in the studies of that reference one cannot calculate explicitly matrix elements of the two-electron 
spin-triplet operator between energy eigenstates and corresponding spectral-weight distributions.
The $x=0$ and $m=0$ results of Ref. \cite{companion} on the spin-wave 
spectrum of the parent compound LCO profit from combination of the $c$ and $s1$ fermion
description with the complementary method of Ref. \cite{LCO-Hubbard-NuMi}. They reveal that  
the microscopic mechanisms that generate the coherent spectral-weight spin-wave energy spectrum are 
in terms of spinon pair breaking $s1$ fermion processes very simple. Indeed the two-spinon $s1$ fermion description
renders a complex many-electron problem involving summation of an infinite set of ladder diagrams
\cite{LCO-Hubbard-NuMi} into a non-interacting two-$s1$-fermion-hole spectrum, described by simple 
analytical expressions.

Within the semi-classical description of spin waves, they can be pictured as long wave-length
twists of the order parameter. In turn, within the present quantum description a spinon bond
pair of a spin-singlet two-spinon $s1$ fermion is broken, giving rise to two independent spin-up spinons
or two independent spin-down spinons. All remaining spinons in the problem remain confined with spin-neutral two-spinon $s1$ fermions.
The two deconfined spinons are invariant under the electron - rotated-electron unitary transformation.
Thus their spin-triplet excite-state occupancies correspond to an isolated vanishing-energy and
vanishing-momentum mode below a continuum of two-$s1$-fermion-hole excitations. 
Indeed, the momenta $\pm{\vec{q}}$ corresponding to the spinon relative motion in the 
spinon pair of the broken $s1$ fermion are transferred over
to two holes, respectively, that emerge in the $s1$ band. Under spin-triplet excitations such a
two-spinon $s1$ fermion breaking may introduce an extra momentum contribution
$\delta {\vec{q}}$. It corresponds to the motion of the center of mass of the spinon broken pair.  
In the latter case the two emerging $s1$ band holes have momenta ${\vec{q}}\,'$ and ${\vec{q}}\,''$
given in Eqs. (\ref{qq-dqdq}) and (\ref{q-dq}). 

As found in Ref. \cite{companion}, the processes associated with most momenta ${\vec{q}}\,'$ and ${\vec{q}}\,''$
lead to an incoherent background of spin spectral weight. In turn, the spin coherent spectral-weight
distribution refers to the spin-wave spectrum. The processes that generate such a coherent spin weight are
such that one of the $s1$ band momenta ${\vec{q}}\,'$ and ${\vec{q}}\,''$ belongs to the boundary line
and the other points in a nodal direction. That some of the spin spectral weight is incoherent is
consistent with the sub-lattice magnetization obeying the inequality $m_{AF}<1$, rather than
reading $m_{AF}=1$, as for the classical N\'eel state. In turn, we argue below that
the spin spectral weight is fully incoherent for spin excitations of $m=0$ and $x>0$ ground states. 

In the thermodynamic limit $N_a^2\rightarrow\infty$ a large number of low-lying energy eigenstates 
$\vert \Psi_{i}\rangle$ with momentum $\vec{k}=\vec{\pi}$ and quantum numbers $S_{\eta}=0$, $S_c=N_a^2/2$,
$S_s= 1$, and $S_s^{x_3}=0,\pm 1$ that contribute to the sum of Eq. (\ref{stag-magn}) converge to the 
corresponding $N_a^2\gg1$ finite ground state. To illustrate the mechanism of spontaneous symmetry 
breaking, we consider for simplicity that the lowest-energy spin triplet state $\vert \Psi_{1T}\rangle$ 
belonging to that set of energy eigenstates gives rise to such a symmetry breaking. Indeed,
for intermediate and large $U/4t$ values such a state exhausts the sum in Eq. (\ref{stag-magn}) by more 
than 98.7\%. However, the lower broken symmetry of the final ground state is the same independently of the number of
low-lying states considered in the analysis of the problem. In the presence of 
a small staggered field ${\vec{B}}_s$ in the $x_3$-direction, a new ground state emerges due to the 
additional term $M_s^{x_3} B_{x_3}$ in the Hamiltonian. This state,
\begin{equation}
\vert \Psi_{b-GS}\rangle \approx C_{GS}\vert \Psi_{GS}\rangle+C_{1T}\vert \Psi_{1T}\rangle \, ,
\label{Psi-b}
\end{equation}
has a finite staggered magnetization, 
\begin{equation}
M_s = \langle\Psi_{b-GS}\vert{\hat{M}}_s^z\vert \Psi_{b-GS}\rangle =
2C_{GS}C_{1T}\langle\Psi_{GS}\vert{\hat{M}}_s^z\vert \Psi_{1T}\rangle \, .
\label{M-s}
\end{equation}
In turn, $\langle\Psi_{GS}\vert{\hat{M}}_s^z\vert \Psi_{GS}\rangle=0$
and $\langle\Psi_{1T}\vert{\hat{M}}_s^z\vert \Psi_{1T}\rangle=0$.
The new ground state $\vert \Psi_{b-GS}\rangle$ has not a well defined spin $S_s$, yet
it has the same quantum numbers $S_{\eta}=0$
and $S_c=N_a^2/2$ as the spin-singlet ground state $\vert \Psi_{GS}\rangle$.

In the thermodynamic limit $N_a^2\rightarrow\infty$, the system assumes the largest 
possible magnetization for arbitrarily small staggering field ${\vec{B}}_s$, so that 
$C_{GS}=C_{1T}=1/\sqrt{2}$ and thus $M_s = \langle\Psi_{GS}\vert{\hat{M}}_s^z\vert \Psi_{1T}\rangle$.
The state $\vert \Psi_{b-GS}\rangle$ of Eq. (\ref{Psi-b}) is contained in a reduced subspace spanned by 
$\vert\Psi_{GS}\rangle$ and $\vert\Psi_{1T}\rangle$. In such a reduced subspace the unity partition operator reads 
$\approx \vert\Psi_{GS}\rangle\langle\Psi_{GS}\vert +\vert\Psi_{1T}\rangle\langle\Psi_{1T}\vert$.
We then obtain the following staggered magnetization squared,
\begin{equation}
M_s^2 = \langle\Psi_{b-GS}\vert{\hat{M}}_s^z\vert \Psi_{b-GS}\rangle^2  = 
\langle\Psi_{GS}\vert({\hat{M}}_s^z)^2\vert \Psi_{GS}\rangle \, .
\label{M-s-2}
\end{equation}
It follows that the staggered magnetization squared in the $N_a^2\rightarrow\infty$ ground state 
$\vert \Psi_{b-GS}\rangle$ with broken symmetry is identical to the long-range antiferromagnetic order 
of the correlation function in the spin-singlet $N_a^2\gg 1$ ground state $\vert \Psi_{GS}\rangle$,
consistently with Eq. (\ref{stag-magn-t-limit}). Indeed, the studies of Ref. \cite{general} reveal
that for a number of sites $N_a^2\gg 1$ large but finite the $m=0$ ground states are
spin-singlet states. For the present case of a $m=0$ and $x=0$ ground state this
agrees with an exact theorem introduced and proved in Ref. \cite{Lieb-89}.

The new ground state $\vert \Psi_{b-GS}\rangle$ of Eq. (\ref{Psi-b}) is a superposition
of states with different spin $S_s$ and thus breaks the Hamiltonian
spin $SU(2)$ symmetry contained in its global $SO(3)\times SO(3)\times U(1)=[SU(2)\times SU(2)\times U(1)]/Z_2^2$
symmetry. Specifically, the spin rotational symmetry $SU(2)$ is
spontaneously broken to $U(1)$ by the formation of the staggered magnetization.
In turn, the new ground state has the same quantum numbers $S_{\eta}=0$
and $S_c=N_a^2/2$ as the spin-singlet ground state $\vert \Psi_{GS}\rangle$.
Therefore, the corresponding $\eta$-spin symmetry $SU(2)$ and $c$ fermion
symmetry $U(1)$, respectively, are not broken. A similar result is obtained if
besides $\vert \Psi_{GS}\rangle$ and $\vert \Psi_{1T}\rangle$, the new ground
state $\vert \Psi_{b-GS}\rangle$ contains a larger set of low-lying energy eigenstates 
$\vert \Psi_{i}\rangle$ with momentum $\vec{k}=\vec{\pi}$ and quantum numbers $S_{\eta}=0$, $S_c=N_a^2/2$,
$S_s= 1$, and $S_s^{x_3}=0,\pm 1$ other than $\vert \Psi_{1T}\rangle$.
Hence rather than $SO(3)\times SO(3)\times U(1)$, the
symmetry of new ground state $\vert \Psi_{b-GS}\rangle$
is $[U(2)\times U(1)]/Z_2^2=[SO(3)\times U(1)\times U(1)]/Z_2=[SU(2)\times U(1)\times U(1)]/Z_2^2$. 

Consistently with the studies of the $x=0$ two-spinon $s1$ fermion pairing energy
of Ref. \cite{companion}, in the Hubbard model on the square lattice state configurations such that 
$N_{a_{\eta}}^2=[N_a^2-2S_c]=0$ and thus $N_{a_s}^2=N_a^2$ exist below an energy scale 
$\mu^0\equiv \lim_{x\rightarrow 0}\mu$. Such an energy scale refers to the spin degrees of freedom yet 
it equals one half the charge Mott-Hubbard gap. The latter gap defines the range $\mu \in (-\mu^0,\mu^0)$
of the chemical potential $\mu$ at $x=0$ and $m=0$ \cite{general,companion}. The energy scale
$\mu^0$ may be used as order parameter of the long-range antiferromagnetic order.
It equals the excitation energy below which such an order exists in the limit $N_a^2\rightarrow\infty$
at $m=0$, $x=0$, and zero temperature $T=0$. The limiting behaviors of $\mu^0$ are approximately the following,
\begin{eqnarray}
\mu^0 & \approx & 32\,t\,e^{-\pi\sqrt{4t\over U}} \, ,
\hspace{0.25cm} U/4t\ll 1 \, ,
\nonumber \\
& \approx & {2e^1\,t\over \pi}\sqrt{1+(U/4t -u_0)}  \, ,
\hspace{0.25cm} u_0 \leq U/4t\leq u_1 \, ,
\nonumber \\
& \approx  & [U/2 - 4t] \, ;
\hspace{0.25cm}  U/4t\gg 1 \, .
\label{Delta-0-s}
\end{eqnarray}
It vanishes $\mu^0\rightarrow 0$ for $U/4t\rightarrow 0$ and is finite and
an increasing function of $U/4t$ for $U/4t$ finite. For $U/4t\rightarrow\infty$ 
it behaves as $\mu^0\approx U/2\rightarrow\infty$. This is why in that limit, when the spin degrees of freedom
of the half-filled Hubbard model are described by the isotropic spin-$1/2$ Heisenberg model,
the state configurations for which $N_{a_{s}}^2=N_a^2$ exist at any finite 
energy.

The behavior $\mu_0 (U/4t)\approx \mu_0 (u_0)\sqrt{1+(U/4t -u_0)}$ 
reported in Eq. (\ref{Delta-0-s}) is expected to be a good approximation for the 
intermediate $U/4t$ range $U/4t\in (u_0,u_1)$ of interest for the square-lattice
quantum liquid studies of Ref \cite{companion}. Here $u_0\approx 1.302$ and $u_1\approx 1.600$. The approximate magnitude 
$\mu_0 (u_0)\approx [2e^1/\pi]\,t$ is that consistent with the relation $\mu_0 (u_0)\approx \mu_0 (u_*)/\sqrt{1+(u_* -u_0)}$.
The value $U/4t=u_* = 1.525$ is that at which the studies of that reference lead by a completely different method to 
$\mu_0 (u_*)\approx 566$\,meV for $t=295$\,meV and $U/4t\approx u_* =1.525$. The use of $\mu_0 (u_0)\approx [2e^1/\pi]\,t$ 
in the formula $\mu_0 (U/4t)\approx \mu_0 (u_0)\sqrt{1+(U/4t -u_0)}$ leads for $t=295$\,meV, $U/4t\approx u_* =1.525$,
and $u_0=1.302$ to nearly the same magnitude, $\mu_0 (u_*)\approx 565$\,meV. In turn,
the value $U/4t=u_0=1.302$ is that at which the energy parameter $2\Delta_0$ is found below to reach its maximum 
magnitude. That energy parameter such that $2\Delta_0<\mu_0$
for $U/4t>0$ is the energy below which the short-range incommensurate-spiral spin order 
considered below survives for $0<x\ll 1$, $m=0$, and zero temperature $T=0$. 

We make the reasonable assumption that at $T=0$ 
the energy parameter $\mu^0$ and the sub-lattice magnetization
$m_{AF}$ of Eq. (\ref{m-AF}) scale in the same way as follows,
\begin{equation}
\mu^0 = U\,m_{AF}\,\alpha^0 \, .
\label{mu0-ep0}
\end{equation}
A naive spin-density wave mean-field approach \cite{bag-mech} leads to the
relation $\mu^0 = [U/2]\,m_{AF}$. In turn, in the
case of the relation (\ref{mu0-ep0}) the coefficient $\alpha^0 \equiv \mu^0/[U\,m_{AF}]$ 
has the following limiting behaviors,
\begin{eqnarray}
\alpha^0 & = & 0.500 \, ,
\hspace{0.25cm} U/4t\ll 1 \, ,
\nonumber \\
& \approx & 0.603 \, ,
\hspace{0.25cm} U/4t = u_0 \approx 1.302 \, ,
\nonumber \\
& \approx & 0.536 \, ,
\hspace{0.25cm} U/4t = u_* \approx 1.525 \, ,
\nonumber \\
& \approx & 0.519 \, ,
\hspace{0.25cm} U/4t = u_1 \approx 1.600 \, ,
\nonumber \\
&  =  & {1\over 2m_{AF}^{HM}} \approx 0.814 \, ,
\hspace{0.25cm} U/4t\gg 1 \, .
\label{epsil-0}
\end{eqnarray}
The physical quantities $m_{AF}$ of Eq. (\ref{m-AF}), $U\,m_{AF}$, and
$\mu^0$ of Eq. (\ref{Delta-0-s}) are increasing functions of $U/4t$.
In turn, $\alpha^0 = \mu^0/U\,m_{AF}$ is a monotonous function of $U/4t$
with both minima and maxima. In the limit $U/4t\rightarrow 0$ the effects of all fluctuations vanish and the 
value $\alpha^0 = 1/2$ is that predicted by mean-field theory relation $\mu^0 = [U/2]\,m_{AF}$. 
As a function of $U/4t$, the coefficient $\alpha^0$ first increases until reaching a maximum at a $U/4t$
value below $u_0$. Interestingly, in the intermediate range $U/4t\in (u_0,u_1)$ it is a decreasing function
of $U/4t$. It reaches a minimum value larger than $0.500$ and smaller than $0.519$ at a $U/4t$
magnitude laying between $U/4t=u_1\approx 1.6$ and $U/4t=2$. It then becomes an increasing
function of $U/4t$, reaching the value $\alpha^0 \approx 0.814$ for $U/4t\gg 1$. Since
$\mu^0/U\rightarrow 1/2$ as $U/4t\rightarrow\infty$, note that the corresponding $U/4t\gg 1$ expression
$\alpha^0 =1/[2m_{AF}]$ would give $\alpha^0 =1/2$ if the $m_{AF}$ value was that 
of the classical N\'eel state $m_{AF}=1$, rather than that of the
spin-$1/2$ Heisenberg model $m_{AF}=m_{AF}^{HM}\approx 0.6142$.

Consistently with for the Hubbard model on a square lattice $\mu^0$ being the energy below 
which the long-range antiferromagnetic order exists in the limit $N_a^2\rightarrow\infty$
at $x=0$, $m=0$, and zero temperature $T=0$, 
the limiting behaviors $\mu^0 \approx 32\,t\,e^{-2\pi\sqrt{t/U}}$ and $\mu^0 \approx  U/2$
given in Eq. (\ref{Delta-0-s}) for $U/4t\ll 1$ and $U/4t\gg 1$, respectively,
are those of the zero-temperature spin gap of Eq. (13) of Ref. \cite{Hubbard-T*-x=0}.
Within the operator description used in this paper the $c$ fermions and spinon occupancies
generate the state representations of the groups $U(1)$ and spin $SU(2)$, respectively, in the model
global $SO(3)\times SO(3)\times U(1)=[SU(2)\times SU(2)\times U(1)]/Z_2^2$ symmetry.
In turn, the studies of Ref. \cite{Kopec} isolate strongly fluctuating modes generated by 
the Hamiltonian (\ref{H}) Hubbard term according to the charge $U(1)$ and spin $SU(2)$ symmetries.
Within the gauge transformation of that reference, the strongly correlated problem 
is casted into a system of noninteracting ``$h$ fermions'' submerged in the bath of 
strongly fluctuating $U(1)$ and $SU (2)$ gauge potentials. Those couple
to fermions via hopping term plus the Zeeman-type contribution with a massive field 
$\varrho (\vec{r}_j\tau)$. Within the description of that reference a $U(1)$ and $SU(2)$ gauge transformation 
is used to factorize the charge and spin contribution to the original electron operator 
in terms of the emergent gauge fields. The $U(1)$ charge $h$ fermions and $SU(2)$ spins of 
Ref. \cite{Kopec} refer to the $c$ fermions and spinons, respectively, of our description. In what the relation of the 
energy scale of Eq. (\ref{Delta-0-s}) to the results of that reference is concerned, in the $x=0$ 
antiferromagnetic phase the above massive field $\varrho (\vec{r}_j\tau)$ assumes the staggered form,
\begin{equation}
\varrho (\vec{r}_j\tau) = \mu_0\,e^{i\vec{\pi}\cdot\vec{r}_j} \, .
\label{rho-mu0}
\end{equation}
The description of Ref. \cite{Kopec} is valid for large $U/4t$ values, so that $\mu_0\approx U/2$,
as given in Eq. (\ref{Delta-0-s}). In turn, the $c$ fermions and spinons of this paper
are directly related to the rotated electrons of the description of Ref. \cite{general}, whose double occupancy is a good
quantum number for $U/4t>0$. This is why the energy scale $\mu_0$ of Eq.
(\ref{rho-mu0}) is here well defined for the whole range of $U/4t$ values.
Consistently, $\mu^0 \approx 32\,t\,e^{-\pi\sqrt{4t\over U}}\rightarrow 0$ and thus
$\rho (\vec{r}_j\,\tau)\rightarrow 0$ for $U/4t\rightarrow 0$ upon the disappearance 
of the long-range antiferromagnetic order, as given in Eq. (\ref{Delta-0-s}).

\subsection{Short-range incommensurate-spiral spin order at $x=0$ and $0< T\ll T_0^*$ and for $0<x\ll 1$ and $0\leq T\ll T_0^*$}

Here it is argued that the spin effective lattice being identical to
the original lattice is a necessary condition for the emergence
of a ground-state long-range antiferromagnetic order in the limit
$N_a^2\rightarrow\infty$. Moreover, it follows from the property II of Section III-A that
since the ground-state rotated-electron occupancy configurations
are for finite values of $U/4t$ more ordered than those of
the electrons generating the same state, a lack of long-range antiferromagnetic
order of the rotated-electron spins in the limit $N_a^2\rightarrow\infty$ for
$N_{a_{s}}^2/N_a^2<1$ would imply a similar lack of 
long-range antiferromagnetic order for the spins of the
original electrons as $N_a^2\rightarrow\infty$. The inequality $N_{a_{s}}^2/N_a^2<1$ 
applies to the $x>0$ and $m=0$ ground states. For those the spin effective
lattice is different from the original lattice, in contrast to that of the
absolute $x=0$ and $m=0$ ground state.

\subsubsection{The general effects of hole doping accounting for the hidden global $U(1)$ symmetry}

Quantum-Monte Carlo methods, when applicable, are the only unbiased tools for 
quantitative studies of the effects of a small hole concentration $x$ in the physics of 
the Hubbard model on the square lattice. Unfortunately, some of the interactions
that become active for $x>0$ cannot be studied by Quantum-Monte Carlo
simulations due to the well-known ``sign problem''.
While part of our results are argued on phenomenological grounds, taking
account for the effects of the hidden global $U(1)$ symmetry found recently for the
Hubbard  model on any bipartite lattice in Ref. \cite{bipartite} introduces a new
scenario and framework, which may be useful for future quantitative studies of 
the square-lattice model. In addition, such a new scenario allows the preliminary 
qualitative discussion of the problem presented in this paper.

In accordance to a general theorem proved in Ref. \cite{Taka}, at half filling the terms
of the Hamiltonian (\ref{HHr}) expansion in $t/U$ with odd powers in $t$ vanish due to 
the particle-hole symmetry and the resulting invariance of the spectrum under 
$t\rightarrow -t$. For instance, the $x=0$ and $m=0$ Hamiltonian terms of 
Eq. (\ref{Heff}) result only from the terms of order $t^2$ and $t^4$ of the Hamiltonian
(6) of Ref. \cite{HO-04}. (The $T_0$ and $T_{\pm 1}$ operators of that reference include a factor $t$ absent in the
corresponding operators of Eq. (\ref{T-op}).) In turn, for finite hole concentration 
$x>0$ and vanishing spin density $m=0$ the expansion in powers of $t/U$ of the Hamiltonian (\ref{HHr}) 
involves terms with odd powers in $t$, absent at $x=0$. We argue that the emergence of such new
terms absent at $x=0$ along with related effects associated with changes in the spin effective lattice 
upon ``turning on'' the hole concentration $x$ destroy 
the sub-lattice magnetization $m_{AF}$.

Expression of the Hamiltonian of the Hubbard model on the square lattice (\ref{H})
in terms of rotated-electron creation and annihilation operators leads to the Hamiltonian 
(\ref{HHr}), which has an infinite number of terms. Its expansion up to fifth order in $t/U$
leads for the $x=0$ half-filled Hubbard model in the one- and two-electron subspace
to the Hamiltonian expression given in Eq. (\ref{Heff}), which involves only spinon
operators. In turn, for $x>0$ the Hamiltonian (\ref{HHr}) may be expressed in terms of
$c$ fermion operators, $\eta$-spinon operators, and spinon operators. This is done by the 
use of the expressions provided in Eq. (\ref{c-up-c-down}) of Appendix A for the rotated-electron 
operators in terms of such operators. The uniquely obtained
Hamiltonian also has an infinite number of terms and thus is rather complex.

The problem slightly simplifies for the $x>0$ Hamiltonian (\ref{HHr}) in the one- and two-electron subspace. 
In terms of rotated-electron creation and annihilation operators, its terms generated
up to fourth order in $t/U$ are for $x>0$ and within a unitary transformation the 
equivalent to the $t-J$ model with ring exchange and various correlated hoppings \cite{HO-04}.
Moreover, in the one- and two-electron subspace such an Hamiltonian may be expressed 
in terms of only $c$ fermion operators and spinon operators. The rotated quasi-spin 
operators are given by $q^l_{\vec{r}_j}=s^l_{\vec{r}_j}+p^l_{\vec{r}_j}$ where $l=\pm,z$. Here $s^l_{\vec{r}_j}$
and $p^l_{\vec{r}_j}$ are the spinon local operators and $\eta$-spinon local 
operators, respectively, given in Eq. (\ref{sir-pir}). Since in the one- and two-electron
subspace there are no rotated-electron
doubly occupied sites, the $\eta$-spin effective lattice is empty and the operator
$p^l_{\vec{r}_j}$ plays no active role. However, the obtained Hamiltonian expression 
is for $x>0$ much more involved than that given in Eq. (\ref{Heff}) for $x=0$ and is omitted here.
In addition to involving spinon operators, it contains $c$ fermion operators and
its number of terms is much larger than for $x=0$. Moreover, it involves $c$ fermion - spinon interactions
and both non-frustated isotropic spinon interactions and spinon interactions 
with some degree of frustration. Hence such interactions
cannot be studied by Quantum-Monte Carlo simulations due to the ``sign problem''. 

In such an involved $x>0$ Hamiltonian expression 
the $c$ fermion operators act onto the $c$ effective lattice occupancies and the spinon operators
act onto the spin effective lattice occupancies. (The $c$ fermion operators commute with the spinon
operators.) Also in the $x>0$ Hamiltonian (\ref{HHr}) general expression containing an 
infinite number of terms, the $c$ fermion operators act onto the $c$ effective lattice occupancies,
the $\eta$-spinon operators act onto the $\eta$-spin effective lattice occupancies,
and the spinon operators act onto the spin effective lattice occupancies. 
The state representations of the new hidden global $U(1)$ symmetry are generated
by the $c$ fermion occupancy configurations in the $c$ effective lattice. As mentioned
above, the point is that for each energy eigenstate the relative positions of the sites of 
the $\eta$-spin and spin effective lattices in the original lattice are stored in the
$c$ effective lattice: For $U/4t>0$ the latter lattice is identical to the original
lattice for all $4^{N_a^2}$ energy eigenstates and the $N_c=2S_c$ $c$ fermions occupy the $2S_c$ sites singly occupied
in the original lattice by the rotated electrons of the description introduced in Ref. \cite{general}. We recall
that for $U/4t>0$ the sites of (i) the $\eta$-spin effective lattice and (ii) the spin effective lattice correspond to those
(i) singly occupied and (ii) unoccupied and doubly occupied by rotated electrons.
Hence for each energy eigenstate the relative positions of the sites of 
the (i) $\eta$-spin effective lattice and (ii) spin effective lattice in the original lattice
refer to the relative positions of the sites of the $c$ effective lattice (i) unoccupied by
$c$ fermions and (ii) occupied by $c$ fermions, respectively.

That the relative positions of the sites of the $\eta$-spin and spin effective lattices in the 
original lattice are stored in the $c$ fermion occupancy configurations of the $c$ effective 
lattice is consistent with for $N_a^2\rightarrow\infty$, $N_{a_{\eta}}^2/N_a^2=[1-2S_c/N_a^2]>0$,
and $N_{a_{s}}^2/N_a^2=2S_c/N_a^2>0$ the occupancy configurations of the $\eta$-spinons
and spinons of the general description of Ref. \cite{general} referring to independent 
$\eta$-spin and spin effective lattices, respectively. Moreover, in that limit these lattices can be considered as square
lattices with spacing $a_{\eta}$ and $a_s$, respectively, provided in Eq. (\ref{a-alpha}) of 
Appendix A. For the one- and two-electron subspace the latter spacing
reads $a_{s} = a/\sqrt{1-x}$, as given in Eq. (\ref{NNCC}).
Hence alike for $x=0$ and $m=0$ ground state, for $x>0$ and $m=0$ ground states the 
the $M_s=2S_c$ spin-$1/2$ spinons remain occupying a full spin effective lattice with
$N_{a_{s}}^2=M_s=2S_c$ sites. 

Although in the limit $N_a^2\rightarrow\infty$ the occupancies of the spin, $\eta$-spin, and $c$ 
effective lattices are independent, for the $x>0$ Hubbard model on the square lattice in the
one- and two-electron subspace the $M_s=2S_c$ spin-$1/2$ spinons interact with
each other and with the $c$ fermions. However, as mentioned above 
the expression of the $x>0$ Hamiltonian (\ref{HHr}) in the one- and two-electron subspace
in terms of $c$ fermion operators and spinon operators leads to a very involved
quantum problem. Indeed, the usefulness of the $c$, spin, and $s1$ effective lattices and corresponding
$c$ and $s1$ momentum bands description refers to that Hamiltonian in the one- and two-electron subspace
in normal order relative to the initial $x>0$ and $m=0$ ground state. Fortunately, such a
ground-state normal-ordering simplifies the quantum problem \cite{companion}. It provides implicitly 
and naturally a criterion for the selection of a few dominant Hamiltonian terms expressed 
in terms of $c$ fermion operators and two-spinon $s1$ fermion operators. That problem is studied in
Ref. \cite{companion} in terms of a suitable energy functional valid for intermediate and large
values of $U/4t$. The spin degrees of freedom of such a functional describe both 
the $x=0$ and $x>0$ problems. The model spin spectrum relative to the $x=0$ and $m=0$
ground state is one of the few problems for which there are results from 
controlled approximations  involving summation of an infinite set of ladder diagrams
\cite{LCO-Hubbard-NuMi}. As mentioned above, the spin spectrum provided by the energy functional of
Ref. \cite{companion} quantitatively agrees with both the spin-wave 
spectrum derived in Ref. \cite{LCO-Hubbard-NuMi} and that observed in 
the parent compound LCO \cite{LCO-neutr-scatt}.

Alike for the $x=0$ and $m=0$ ground state, for $x>0$ and
$m=0$ ground states the $M_s=2S_c$ spinons remain confined within $N_{s1}=S_c$
spin-neutral two-spinon $s1$ fermions. The $s1$ fermions are generated from
bosonic spin-neutral two-spinon $s1$ bond particles and thus 
have long-range interactions associated with the effective
vector potential ${\vec{A}}_{s1} ({\vec{r}}_j)$ of Eq. (\ref{A-j-s1-3D}) of Appendix A.
An important property is that for $x>0$ and $m=0$ ground states the $s1$ fermion 
momentum band is full and for one-electron and two-electron excited states 
displays a single hole and none or two holes,
respectively \cite{companion}. The $s1$ - $s1$ fermion  
interactions associated with the effective vector potential of 
Eq. (\ref{A-j-s1-3D}) of Appendix A are stronger than those
that arise between the emerging $s1$ fermions and
pre-existing $c$ fermions. In spite of that, the former do not lead to
$s1$ - $s1$ fermion inelastic scattering. The
obvious reason is that due to phase-space restrictions
associated with the exclusion principle and
energy and momentum conservation requirements
there are no available momentum values
in the $s1$ band for excited-state occupancy configurations. 

Both for the $x=0$ and $m=0$ ground state and $x>0$ and $m=0$ ground states 
the $M_s=2S_c$ spin-$1/2$ spinons occupy a full spin effective lattice with
$N_{a_{s}}^2=M_s=2S_c$ sites. However, an important point is that within the present description there is a qualitative difference 
between the $x=0$ and $x>0$ problems. At $x=0$ the spin effective lattice is identical
to the original lattice whereas for $x>0$ it has a smaller number of sites 
$N_{a_{s}}^2=2S_c<N_a^2$ so that its spacing $a_{s} = a/\sqrt{1-x}$ is larger. Within the description
of the problem used in the studies of this paper, this is one of the
main effects of hole doping. Indeed, the lack of the $\eta$-spin effective lattice, $N_{a_{\eta}}^2=x\,N_a^2=0$, occurring at $x=0$
implies that the spin effective lattice is identical to the original lattice and thus has
the same number of sites $N_{a_{s}}^2=2S_c=N_a^2$ as that lattice. 
We argue that the spin effective lattice being identical to the original lattice
is a necessary condition for the ground-state long-range antiferromagnetic order to emerge as 
$N_a^2\rightarrow\infty$. The concepts of spin effective lattice and $s1$ effective lattice
are only valid in that limit. Only in it are such lattices approximate square lattices with spacing
$a_{s} = a/\sqrt{1-x}$ given in Eq. (\ref{NNCC}) and $a_{s1} \approx \sqrt{2}\,a_s$ 
provided in Eq. (\ref{a-a-s1-sube}), respectively. The incommensurability relative to
the original square lattice spacing $a$ of the effective lattice spacings $a_{s} = a/\sqrt{1-x}$ 
and $a_{s1} \approx \sqrt{2}\,a_s$ is consistent with a ground-state long-range antiferromagnetic order
occurring for $N_a^2\rightarrow\infty$ only at $x=0$. Such an incommensurability allows processes 
that destroy long-range antiferromagnetic order becoming active for $x>0$. Within the expansion in powers 
of $t/U$ of the Hamiltonian (\ref{HHr}), such processes are associated for instance with Hamiltonian 
terms with odd powers in $t$, absent at $x=0$. 

The possible existence of low-lying spin excitations for $x>0$ is not a sufficient condition
for a spontaneously broken symmetry to occur in the limit $N_a^2\rightarrow\infty$. According to 
the results of Ref. \cite{general},
the $m=0$ and $x>0$ ground states are for $N_a^2\gg 1$ large but finite spin-singlet
states. Note though that only if the state obtained by application of the two-electron
spin-flip operator onto such ground states had finite overlap with low-lying spin excitations
would these states acquire a long-range spin order as
$N_a^2\rightarrow\infty$. However, we argue that if low-lying spin excitations
exist for $m=0$ and $x>0$ such an overlap vanishes. It vanishes as well
in the case of non existence of low-lying spin excitations due to a spin gap:
There is no coherent spin spectral weight both for vanishing
and/or finite energy. Therefore, in addition to 
the $m=0$ and $x>0$ ground states remaining spin-singlet
states in that limit, the corresponding spin-triplet spectrum is fully
incoherent: Its sharp spectral features are not $\delta$-function like.
The absence of coherent spin-wave excitations is consistent with 
the lack of long-range spin order in the $m=0$ and $x>0$ ground states
in the limit $N_a^2\rightarrow\infty$.

For a $x>0$ and $m=0$ ground state one has that the number of sites of the
$\eta$-spin effective lattice $N_{a_{\eta}}^2=x\,N_a^2>0$ is non zero 
even if the hole concentration $0<x\ll 1$ is very small.
For such a state the spin effective lattice has a number of sites $N_{a_{s}}^2<N_a^2$
smaller than that of the original lattice. Consistently with the corresponding
effective lattice spacings reading $a_{s} = a/\sqrt{1-x}$ 
and $a_{s1} \approx \sqrt{2}\,a_s$, below further evidence is
provided that for $0<x\ll 1$ the ground state of the Hubbard model
on the square lattice has a short-range incomensurate-spiral spin order.
Moreover, according to the results of Ref. \cite{companion} the ground state
has a short-range spin order for $0<x<x_*$. For $x>x_*$ it is a disordered
state without short-range spin order. Here $x_*>0.23$ for approximately $U/4t>u_0\approx 1.3$.

Finally, we provide further evidence that the form of the $s1$ effective lattice spacing (\ref{a-a-s1-sube})
is for the Hubbard model on the square-lattice consistent with the above mentioned $x=0$ and $x>0$ spin orders,
respectively. That at $x=0$ and $m=0$ the spacing of the square $s1$ effective lattice is given by $a_{s1}=\sqrt{2}\,a$
reveals that then its periodicity has increased relative to that of 
the original lattice, which in that case is identical to the spin effective lattice. 
Indeed, at $x=0$ the $s1$ effective lattice is one of the two sub-lattices of
the original lattice and thus refers to a $\sqrt{2}\times\sqrt{2}$
reconstruction in which the periodicity of the spin-sub-system
real-space structure is increased. Such an effect
is consistent with the occurrence of the long-range antiferromagnetic
order for $N_a^2\rightarrow\infty$ at $x=0$ and $m=0$. 
In turn, that for $x>0$ and $m=0$ the square $s1$ effective lattice 
spacing reads instead $a_{s1}\approx \sqrt{2/(1-x)}\,a$ 
is consistent with the emergence of the short-range incommensurate-spiral spin
order. Indeed now the $s1$ effective lattice is one of the two sub-lattices of
the spin effective lattice which for $x>0$ is different from and
incommensurate to the original lattice.

\subsubsection{Quantum and thermal phase transitions}

It is argued above that the lack of finite overlap of the state generated by application
of the two-electron spin-flip operator onto a $x>0$ and $m=0$ ground state
with low-lying excited states is behind such a ground state remaining a spin-singlet
state for $N_a^2\rightarrow\infty$, alike for $N_a^2\gg 1$ large but finite \cite{general}. 
This is so independently of the existence or non existence (spin gap)
of spin-triplet low-lying states. Hence such a ground state has no long-range 
antiferromagnetic order in the limit $N_a^2\rightarrow\infty$ and its symmetry
is that of the Hamiltonian, $SO(3)\times SO(3)\times U(1)=[SU(2)\times SU(2)\times U(1)]/Z_2^2$. 
A necessary condition for the occurrence of such a long-range order is
according to our above analysis that the spin effective lattice is identical to the original 
lattice. That condition is not met by $x>0$ and $m=0$ ground states.
For them the Hamiltonian hidden global $U(1)$ symmetry generator
eigenvalue $S_c$ obeys the inequality $S_c<N_a^2/2$. It then follows that the number of sites
of the $\eta$-spin effective lattice $N_{a_{\eta}}^2=[N_a^2-2S_c]>0$ 
is finite, so that the number of sites of the spin effective lattice 
$N_{a_{s}}^2=2S_c<N_a^2$ is smaller than the number of sites of the
original lattice. For the one- and two-electron subspace 
considered in this paper such inequalities read $N_{a_{\eta}}^2=x\,N_a^2>0$ and 
$N_{a_{s}}^2=(1-x)\,N_a^2<N_a^2$, respectively. 

There is strong evidence of the occurrence in the half-filled Hubbard model on the square lattice
of strong short-range antiferromagnetic correlations for finite temperatures $T>0$ below a crossover temperature called 
$T_x$ in Ref. \cite{Hubbard-T*-x=0}, which here we denote by $T_0^*$. This is consistent with then the system 
being driven into a phase with short-range spin order. Furthermore, that the occurrence of long-range 
antiferromagnetic order as $N_a^2\rightarrow\infty$ requires that $T=0$, $N_{a_{\eta}}^2=0$, and $N_{a_{s}}^2=N_a^2$ is
consistent with the short-range spin order occurring for $m=0$, $0<x\ll 1$, and $0\leq T< T_0^*$ 
having basic similarities to that occurring for $m=0$, $x=0$, and $0<T< T_0^*$. The latter order was studied previously
in Ref. \cite{Hubbard-T*-x=0} for $0<T\ll T_0^*$. 

As further justified below, for both vanishing and finite temperatures a phase displaying a short-range spiral-incommensurate spin 
order is then expected to occur for (i) $m=0$, $0<x\ll 1$, and $0\leq T\ll T_0^*$
and (ii) $m=0$, $x=0$, and $0<T\ll T_0^*$. At $m=0$ and temperatures below $T_0^*$, the system is driven both for 
(i) $0<x\ll 1$ and $0\leq T<T_0^*$ and (ii) $x=0$ and $0<T<T_0^*$ into a renormalized classical regime where the 
$N_a^2\rightarrow\infty$, $x=0$, and $T=0$ long-range antiferromagnetic order is replaced by such a short-range spin order, which
is a quasi-long-range spin order as that studied in Ref. \cite{Principles} for simpler spin systems. 

An interesting physical issue is whether the quantum phase transition separating
the $x=0$ and $m=0$ ground state from the $0<x\ll 1$ ground state at $T=0$ corresponds
to a ``deconfined'' quantum critical point \cite{Senthil-04}. Indeed, both at $x=0$
and for $0<x\ll 1$ the ground-state $M_s=2S_c$ spinons are confined within $N_{s1}=S_c$
spin-neutral two-spinon $s1$ fermions. When expressed in terms of rotated-electron
creation and annihilation operators, the Hamiltonian of the Hubbard
model on the square lattice has for $U/4t$ finite an infinite number
of terms, as given in Eq. (\ref{HHr}). As mentioned above, the Hamiltonian terms generated
up to fourth order in $t/U$ are for $x>0$ and within a unitary transformation the 
equivalent to the $t-J$ model with ring exchange and various correlated
hoppings \cite{HO-04}. In addition to non-frustated 
isotropic spin interactions as those considered for a spin-$1/2$ Heisenberg
model on the square lattice in Refs. \cite{Anders-07,Anders-10}, the present
spinons have some degree of frustration as well. Hence their
interactions cannot be studied by Quantum-Monte Carlo
simulations due to the ``sign problem''. 
A related interesting open question is whether for $x>0$ and $m=0$ the 
the spin degrees of freedom of the present square-lattice quantum liquid refer to a valence-bond solid or some
type of related valence-bond liquid. In either case, that both at $x=0$ and
for $x>0$ the $M_s=2S_c$ spinons are confined within $N_{s1}=S_c$
spin-neutral two-spinon $s1$ fermions strongly suggests that the
corresponding quantum phase transition refers indeed to a ``deconfined'' 
quantum critical point. If this is so, the critical point is characterized 
by deconfined spin-$1/2$ spinons coupled to some emergent 
$U(1)$ gauge field \cite{Senthil-04,Anders-07,Anders-10}.

We denote by $2\Delta_0$ the energy below which the short-range incommensurate-spiral spin order 
with strong antiferromagnetic correlations survives at zero-temperature, $m=0$, and $0<x\ll 1$. That energy parameter has a 
$U/4t$ dependence qualitatively similar to that of the energy scale $2k_B\,T_0^*$, with the 
equality $2\Delta_0\approx 2k_B\,T_0^*$ approximately holding. Except for $U/4t\rightarrow 0$,
such an energy scale has a different origin than the order parameter 
$\mu^0 = U\,m_{AF}\,\alpha^0$ of Eq. (\ref{mu0-ep0}), which is proportional to
the sub-lattice magnetization $m_{AF}$ of Eq. (\ref{m-AF}). Indeed, for $0<x\ll 1$
the lack of a long-range antiferromagnetic order implies that $m_{AF}=0$.

\subsubsection{The $U/4t$ dependence of the $0<x\ll 1$ energy scales}

The energy scale $2\Delta_0\approx 2k_B\,T_0^*$ considered here plays a major role 
in the square-lattice quantum-liquid studies of Ref. \cite{companion}. 
Here we address the problem of its $U/4t$ dependence 
by combining the results obtained from the use of our general description 
with those of the low-temperature approach to the half-filled Hubbard
model on the square lattice of Ref. \cite{Hubbard-T*-x=0}. 
The investigations of that reference focus on temperatures $0<T\ll T_0^*$. 
The energy parameter $2\Delta_0\approx 2k_B\,T_0^*$ refers to the limit 
$2\Delta_0=\lim_{x\rightarrow 0}2\vert\Delta\vert$ of an $x$ dependent
energy scale $2\vert\Delta\vert$ that plays the role of order parameter
of the phase with short-range spin order. For $0<x\ll 1$ and intermediate and
large $U/4t$ values such an energy scale reads,
\begin{equation}
2\vert\Delta\vert  \approx 2\Delta_0\left(1-{x\over x_*^0}\right)
\, , \hspace{0.25cm} 0<x\ll 1 \, , \hspace{0.25cm} U/4t\geq u_0 \approx 1.302 \, .
\label{Delta}
\end{equation}
The linear dependence on $x$ of $[2\Delta_0 -2\vert\Delta\vert] \approx (x/x_*^0)\,2\Delta_0$ 
for $0<x\ll 1$ is justified in Ref. \cite{companion}. Here $x_*^0\approx 2r_s/\pi$, the ratio 
$r_s =2\Delta_0/8W_{s1}^0$ plays an important role in the square-lattice quantum liquid, and
$W_{s1}^0\equiv \lim_{x\rightarrow 0}\,W_{s1}=W_{s1}\vert_{x=0}$ where $W_{s1}$
is the nodal energy bandwidth $W_{s1}$ of the $s1$ fermion dispersion defined in Ref. \cite{companion}. Its maximum 
magnitude is reached at $U/4t=0$. For $U/4t>0$ it decreases monotonously for increasing values of $U/4t$, 
vanishing for $U/4t\rightarrow\infty$. That for $U/4t\rightarrow\infty$ both $W_{s1}\rightarrow 0$ and 
$\vert\Delta\vert\rightarrow 0$ is associated with the full 
degeneracy of the spin configurations reached in that limit. In it the spectrum of the two-spinon composite
$s1$ fermions becomes dispersionless. The limiting behaviors of the $m=0$ energy parameter
$8W_{s1}^0\equiv \lim_{x\rightarrow 0}\,8W_{s1}=8W_{s1}\vert_{x=0}$ contributing to the ratio
$r_s =2\Delta_0/8W_{s1}^0$ read \cite{companion},
\begin{equation}
8W_{s1}^0 = 32t  \, ,
\hspace{0.25cm}  U/4t = 0 
\, ; \hspace{0.50cm} 
8W_{s1}^0 \approx {\pi\over 2}\,{[8t]^2\over U} \, ,
\hspace{0.25cm} U/8^2t\gg 1 \, .
\label{W-s-0}
\end{equation}
In contrast to the $x>0$ energy parameter $2\vert\Delta\vert$, the energy scale $W_{s1}^0$ is well defined 
both at $x=0$ and for $x>0$, having the same magnitude at $x=0$ and for $x\rightarrow 0$.
Its magnitude $W_{s1}^0(u_*)=[49.6/295]\,t\approx 0.168\,t$ 
obtained at $U/4t =u^*=1.525$ in Ref. \cite{companion} is about $12.25$ times smaller than that
found by use of the limiting expression $4\pi\,t^2/U\approx 2.060\,t$ of Eq. (\ref{W-s-0}) at
$U/4t=1.525$. This reveals that such a limiting expression is valid for a smaller range
of very large $U/4t$ values than in 1D. Indeed, $4\pi\,t^2/U\approx 0.168\,t$ 
at $U/4t\approx 18.70$, whereas $W^0_{s1}\approx 0.168\,t$ at $U/4t\approx 1.525$.
Therefore, the relation $W^0_{s1} = J \approx 4\pi\,t^2/U$
is valid for approximately $U/8^2t\gg 1$, as given in Eq. (\ref{W-s-0}). For the Hubbard model on the square lattice the energy scale 
$J \approx 4\pi\,t^2/U$ controls the physics for a smaller $U/4t$ range than
in 1D, which corresponds to very large $U/4t$ values such that $U/8^2t\gg 1$. 
Hence the intermediate-$U/4t$ range plays a major role in the
physics of that model, as confirmed by the related studies of 
Ref. \cite{companion}.

That for $U/4t>0$ the energy scale $2\Delta_0=\lim_{x\rightarrow 0}2\vert\Delta\vert$
has a different origin than the order parameter $\mu^0 = U\,m_{AF}\,\alpha^0$ of Eq. (\ref{mu0-ep0})
is consistent with for $U/4t>0$ their magnitudes being different. However, symmetry arguments related to the
disappearance of the $N_a^2\rightarrow\infty$ and $x=0$ ground-state
long-range antiferromagnetic order for $U/4t\rightarrow 0$
imply that $\lim_{U/4t\rightarrow 0}2\Delta_0=\mu^0$. Indeed
the ratio $2\Delta_0/\mu^0\rightarrow 1$ involving the two energy scales becomes 
one in that limit. Moreover, the energy scale $2\Delta_0$ interpolates between $2\Delta_0= \mu^0\approx 32t\,e^{-\pi\sqrt{4t/U}}$ for
$U/4t\ll 1$ and $2\Delta_0= 8W^0_{s1}\approx \pi\,[8t]^2/U$ for approximately $U/8^2t\gg 1$. It
goes through a maximum magnitude at a $U/4t$ value found below to be approximately given by $U/4t=u_0\approx 1.302$.

The energy parameter $\mu_0$ is an increasing function of $U/4t$.
As given in Eq. (\ref{Delta-0-s}), it behaves as $\mu_0\approx 32\,t\,e^{-\pi\sqrt{4t\over U}}$
for $U/4t\ll 1$ and as $\mu_0\approx  [U - 8t]$ for $U/4t\gg 1$.
In turn, the energy scale $8W_{s1}^0$ is a decreasing function of $U/4t$. According to 
Eq. (\ref{W-s-0}) it is given by $8W_{s1}^0 = 32t$ at $U/4t=0$ and decreases approximately as 
$8W_{s1}^0 \approx \pi\,[8t]^2/U$ for very large $U/4t$. Since the energy scale $2\Delta_0$ interpolates between
these two behaviors, it vanishes both for $U/4t\rightarrow 0$ and $U/4t\rightarrow\infty$
as $2\Delta_0\approx 32t\,e^{-\pi\sqrt{4t/U}}$ and $2\Delta_0 \approx \pi (8t)^2/U$, respectively. 
In these two limits it becomes $\mu_0$ and the energy scale $8W_{s1}^0 \approx\pi (8t)^2/U$ 
associated with the strong $0<x\ll 1$ antiferromagnetic correlations, respectively.
This is consistent with its maximum magnitude being reached at an intermediate $U/4t$ value
$U/4t=u_0$ at which the equality $\mu_0\approx 8W_{s1}^0$ holds. We then define such a $U/4t$ value as that
at which the lines $8W_{s1}^0 =8W_{s1}^0 (U/4t)$ and $\mu_0 = \mu_0 (U/4t)$
cross and thus the equality $\mu_0 (u_0)= 8W_{s1}^0 (u_0)$ holds.
The ratio $r_s =2\Delta_0/8W_{s1}^0$ is parametrized in the following as $r_s=e^{-\lambda_s}$ 
where $\lambda_s = \vert\ln (2\Delta_0/8W_{s1}^0)\vert$ controls such an interpolation 
behavior. 

The energy scale $2\Delta_0$ can then be expressed as,
\begin{equation}
2\Delta_0 = r_s\,8W_{s1}^0 = 8W_{s1}^0\,e^{-\lambda_s} 
\, ; \hspace{0.35cm} 
\lambda_s = \vert\ln (2\Delta_0/8W_{s1}^0)\vert \, ,
\label{Delta-0-gen}
\end{equation}
where $\lambda_s$ has the limiting behaviors,
\begin{eqnarray}
\lambda_s & = & \pi\sqrt{4t/U} \, , \hspace{0.25cm} U/4t \ll 1 
\, ; \hspace{0.50cm}
\lambda_s \approx 4t\,u_0/U \, , \hspace{0.25cm} 
u_{00} \leq U/4t \leq u_1 \, ; \hspace{0.50cm}
\lambda_s = 0 \, , \hspace{0.25cm} U/4t \rightarrow\infty \, ,
\nonumber \\
u_{00} & \approx & (u_0/\pi )^2 \approx 0.171  \, ; \hspace{0.35cm} 
u_0 \approx 1.302 \, ; \hspace{0.35cm} u_1 \approx 1.600 \, .
\label{lambda-s}
\end{eqnarray}
The ratio $r_s =2\Delta_0/8W_{s1}^0$ is an increasing function
of $U/4t$. It changes continuously from $r_s=0$ for $U/4t\rightarrow 0$
to $r_s=1$ for $U/4t\rightarrow\infty$. For $u_{00} \leq U/4t \leq u_1$ it is approximately given by
$r_s\approx e^{-4t\,u_0/U}$ rather than by $r_s\approx e^{-\pi\sqrt{4t/U}}$ for $U/4t\ll 1$. This 
is consistent with for large $U/4t$ values $(1-r_s)$ being proportional to $(1-r_s)\propto 4t/U$ rather than
to $(1-r_s)\propto \sqrt{4t/U}$.

The temperature $T_x$ of Ref. \cite{Hubbard-T*-x=0} that plays the role of 
our temperature $T_0^*\approx 2\Delta_0/2k_B$ is plotted in Fig. 3 of
that reference. Its $U/4t$ dependence is qualitatively correct. $T_x$ vanishes both
in the limits $U/4t\rightarrow 0$ and $U/4t\rightarrow\infty$. It
goes through a maximum magnitude at an intermediate
value $5/4<U/4t<3/2$. Nevertheless, the interpolation function used to produce
it, provided in Ref. 74 of such a paper, is poor for intermediate
values of $U/4t$. However, that does not affect the validity of the results of
Ref. \cite{Hubbard-T*-x=0}. Indeed, the studies of that reference
refer to the temperature range $0<T\ll T_x$ for which the accurate dependence
of $T_x$ on $U/4t$ is not needed. The goal of its Fig. 3 is merely illustrating
qualitatively the $T_x$ behavior over the entire coupling range
\cite{Nicolas}. 

The studies of Ref. \cite{companion} on the square-lattice quantum liquid
refer mostly to intermediate $U/4t$ values. Given the important role plaid
by the energy scale $2\Delta_0\approx 2k_B\,T_0^*$ in the physics of that
quantum liquid, we need a more quantitatively accurate $U/4t$ dependence
of it for intermediate $U/4t$ values. Such a dependence must be consistent with the qualitative physical 
picture of Ref. \cite{Hubbard-T*-x=0}. According to Eqs. (\ref{Delta-0-gen}) and (\ref{lambda-s}) one has that
$2\Delta_0=8W_{s1}^0\,e^{-\lambda_s}$ for $U/4t>0$. Here the parameter $\lambda_s$ 
is a continuous decreasing function of $U/4t$. It is given by
$\lambda_s =\infty$ for $U/4t\rightarrow 0$ and $\lambda_s =0$ for $U/4t\rightarrow\infty$.
This reveals that the magnitude $\lambda_s=1$ separates two physical regimes. Consistently,
$\lambda_s=1$ refers to the $U/4t$ value $U/4t=u_0$ at which
$2\Delta_0$ reaches its maximum magnitude, ${\rm max}\{2\Delta_0\}= 
[\mu_0 (u_0)]\,e^{-1} =8W_{s1}^0(u_0)\,e^{-1}$.

That as found in Ref. \cite{companion} the parameter $x_*^0$ is for
approximately $u_0\leq U/4t\leq u_1$ 
given by $x_*^0=2r_s/\pi=2e^{-4t\,u_0/U}/\pi$ and at $U/4t\approx u_*= 1.525$
reads $x_*^0\approx 0.27$ implies that $u_0\approx 1.3$. Moreover,
in that reference it is found that $8W_{s1}^0 (u_*)=[396.8/295]\,t\approx 1.345\,t$
at $U/4t\approx u_*= 1.525$. Such a result is obtained
in that reference from comparison of the $x=0$ and $m=0$ spin excitation spectra for the 
high symmetry directions found by use of the $c$ and $s1$ fermion description with those
estimated by the controlled approximation of Ref. \cite{LCO-Hubbard-NuMi}.
On combining the equality $\mu_0 (u_0)=8W_{s1}^0(u_0)$
with the magnitude $\mu_0 (u_0)\approx [2e^1/\pi]\,t$ of Eq. (\ref{Delta-0-s}) 
we find $8W_{s1}^0(u_0)\approx [2 e^1/\pi]\,t\approx 1.731\,t$ and
${\rm max}\{2\Delta_0\}= [\mu_0 (u_0)]\,e^{-1} =8W_{s1}^0(u_0)\,e^{-1}\approx 2t/\pi$.

For the intermediate range $U/4t\in (u_0,u_1)$ 
the $U/4t$ dependence of the energy parameter $8W_{s1}^0$ is of the
form $8W_{s1}^0\approx [2e^1/\pi]\,t\,W(U/4t)$. Here $W(u)$ is an unknown 
function of $u=U/4t$ such that $W (u_0)=1$. On combining Eqs. 
(\ref{Delta-0-gen}) and (\ref{lambda-s}) one finds $2\Delta_0 = 8W_{s1}^0\,e^{-4t\,u_0/U}$. Thus 
fulfillment of the maximum condition $\partial 2\Delta_0 (u)/\partial u =0$ at $u=u/4t=u_0$ requires
the function $W( u)$ be given by $W (u)\approx (2-u/u_0)$ for 
$0\leq [(u-u_0)/(u_1-u_0)]\ll 1$. In turn, the use of the above results 
$8W_{s1}^0(u_*)=[396.8/295]\,t\approx 1.345\,t$ and 
$8W_{s1}^0(u_0)\approx [2e^1/\pi]\,t\approx 1.731\,t$ reveals that the
ratio $8W_{s1}^0(u_*)/8W_{s1}^0(u_0)$ may be expressed as
$8W_{s1}^0(u_*)/8W_{s1}^0(u_0)\approx [1-(u_*-u_0)]$. It then follows that
$W (u)\approx [1-(u-u_0)]$ for $u\approx u_* =1.525$. We then
use a suitable interpolation function for $W (u)$, which has these two
limiting behaviors. The energy parameter $8W_{s1}^0$ is
for $U/4t$ intermediate values $U/4t\in (u_0,u_1)$ then given approximately by, 
\begin{eqnarray}
8W_{s1}^0 & \approx & {2 e^1\,t\over \pi}\,W(U/4t) \, ;
\hspace{0.35cm} W (u) \approx 1 - (u-u_0)\,e^{-{u_*-u\over u_*-u_0}\ln (u_0)} \, ,
\hspace{0.25cm} u_0\leq U/4t\leq u_1 \, ,
\nonumber \\
u_0 & \approx & u_* -1 + e^{-1} \pi {198.4\over 295} \approx 1.302 \, ;
\hspace{0.35cm} u_* \approx 1.525 \, .
\label{W-s1-0-u0}
\end{eqnarray}
The value $u_0\approx 1.302$ is that obtained from the equation
$8W_{s1}^0 (u_*) = [2 e^1\,t/\pi]\,[1-(u_*-u_0)]= [396.8/295]\,t \approx 1.731\,t$.

On combining the above results we find the following approximate behaviors for the energy parameter $2\Delta_0$,
\begin{eqnarray}
2\Delta_0 & \approx & 32t\,e^{-\pi\sqrt{4t/U}} \, ,
\hspace{0.25cm}  U/4t\ll 1 \, ,
\nonumber \\
& = & {\rm max}\,\{2\Delta_0\} \approx 2t/\pi \, ,
\hspace{0.25cm}  U/4t = u_0 \, ,
\nonumber \\
& \approx & e^{(1-4t\,u_0/U)}[2t/\pi]\,W(U/4t) \, ,
\hspace{0.25cm} u_0\leq U/4t \leq u_1 \, ,
\nonumber \\
& = & 8W^0_{s1} \approx [\pi\,(8t)^2/2U] \, ,
\hspace{0.25cm} U/8^2t\gg 1 \, .
\label{Delta-0}
\end{eqnarray}
Here $W(U/4t)$ is the interpolation function given in Eq. (\ref{W-s1-0-u0}). Note that
$\partial 2\Delta_0 (u)/\partial u =0$ at $u=u/4t=u_0$, consistently with
$2\Delta_0$ reaching its maximum magnitude $2t/\pi$ at that $U/4t$ value.
The overall $U/t$ dependence of $T_0^*\approx 2\Delta_0/2k_B$ is
similar to that plotted in Fig. 3 of Ref. \cite{Hubbard-T*-x=0} for $T_x$
with the $U/t$ value at which the maximum magnitude is reached shifted from $U/t\approx 5.60$
to $U/t\approx 5.21$. Moreover, that magnitude is lessened from ${\rm max}\,\{T_x\} \approx 0.625\,t/k_B$
to ${\rm max}\,\{T_0^*\}\approx {\rm max}\,\{2\Delta_0/2k_B\}\approx t/[\pi k_B]\approx 0.318\,t/k_B$.

\subsubsection{Short-range incommensurate spiral spin order for $0<x\ll1$ and $T=0$}

Here we provide strong evidence that for intermediate and large values of $U/4t$ 
and small hole concentrations $0<x\ll1$ the short-range spin order of the $m=0$ 
ground state corresponds indeed to that of a spin-singlet incommensurate 
spiral state. In terms of the rotated-electron spins occupancy
configurations the ground state is then a spin-singlet incommensurate 
spiral state for $U/4t>0$, $m=0$, and $0<x\ll1$. That evidence is
found on combining several results. This includes the necessary condition for occurrence of a long-range spin order  
for $N_a^2\rightarrow\infty$ argued in Section III-C to be that the spin effective lattice is identical to the
original lattice and thus $N_{a_s}^2=2S_c=N_a^2$, the result of Ref. \cite{general} that the $m=0$ and $0<x\ll1$ ground state
is a spin-singlet state for $N_a^2\gg 1$ and thus remains so for $N_a^2\rightarrow\infty$, and the results of 
Ref. \cite{Mura} concerning the spin degrees of freedom of a related quantum problem. 

As discussed in Section I, for intermediate and large values of $U/4t$ 
the Hubbard model on the square lattice given in Eqs. (\ref{H}) and (\ref{HHr}) in terms of
electron and rotated-electron operators, respectively, can be mapped 
onto an effective $t-J$ model on a square lattice with $t$, $t'=t'(U/4t)$, 
and $t''=t''(U/4t)$ transfer integrals. The role of the processes associated with 
$t'=t'(U/4t)$ and $t''=t''(U/4t)$ becomes increasingly important
upon decreasing the $U/4t$ value. Reference \cite{Mura} presents
rigorous results on the spin degrees of freedom of the 
$t-J$ model on a square lattice with $t$, $t'$, and $t''$ transfer 
integrals. The investigations of that paper refer to small values of the hole concentration $0<x\ll1$ and spin 
density $m=0$. Their starting point is a suitable action first introduced in Ref. \cite{Wieg}. 
The use in Ref. \cite{Mura} of a staggered CP$^1$ representation for the spin degrees of freedom
allows to resolve exactly the constraint against double occupancy. Within our
description of the problem, this is is equivalent to performing the electron - rotated-electron
unitary transformation. In order to achieve the rigorous result 
that for small hole concentrations there occurs a incommensurate-spiral 
spin order, the effective action for the spin degrees of freedom
is reached after integrating out the charge fermionic degrees of
freedom and the magnetic fast CP$^1$ modes. Importantly, the
dependence on the hole concentration of the coupling constants of the effective
field theory is in Ref. \cite{Mura} obtained explicitly for small $x$.  

We consider the mapping between the above effective $t-J$ model
and the Hubbard model of Eqs. (\ref{H}) and (\ref{HHr})
in the subspace with vanishing rotated-electron double-occupancy.
Accounting for such a mapping, the studies of Ref. \cite{Mura} 
imply that for intermediate and large values of $U/4t$, spin-density $m=0$, and small
hole concentrations $0<x\ll1$ the ground state of the Hubbard model on the square lattice
is an incommensurate spiral state. This is a rigorous result. However the studies of Ref. \cite{Mura} are not 
conclusive on whether for $N_a^2\rightarrow\infty$ and $0<x\ll1$ the $m=0$ incommensurate spiral ground state has short-range
or long-range spin order. 

The necessary condition for occurrence in the limit $N_a^2\rightarrow\infty$ of a ground-state
long-range spin order argued in Section III-C to be that the spin effective lattice is identical to the
original lattice and thus $N_{a_s}^2=2S_c=N_a^2$ is not fulfilled for $0<x\ll1$. We thus argue that 
in the thermodynamic limit $N_a^2\rightarrow\infty$ the
$0<x\ll1$ and $m=0$ incommensurate spiral ground state has no long-range spin order.
Hence such a ground state has the same symmetry and basic properties both for $N_a^2\gg 1$ very 
large but finite and for $N_a^2\rightarrow\infty$. One then combines the result of Ref. \cite{general} that 
for $N_a^2\gg 1$ and thus for $N_a^2\rightarrow\infty$ the $m=0$ and $0<x\ll1$ ground 
state of the Hubbard model on the square lattice (\ref{H}) is a spin-singlet state with that of 
Ref. \cite{Mura} that it is a incommensurate spiral state. This 
consistently implies that for $N_a^2\rightarrow\infty$ and intermediate and large values of $U/4t$ the $m=0$ and $0<x\ll1$
ground state of the Hubbard model on the square lattice is a spin-singlet state with short-range incommensurate-spiral spin 
order and strong antiferromagnetic correlations. 

Finally, a ground-state short-range spin order does not preclude the occurrence 
of a ground-state long-range dimer-dimer order, as in a valence-bond solid.
However, whether the spin degrees of freedom of the square-lattice quantum liquid 
refer for $x>0$ and $m=0$ to a valence-bond solid or a valence-bond liquid 
remains an open issue.
 
\section{Concluding remarks}

In this paper we considered a suitable one- and two-electron subspace
in which the general operator description for the Hubbard model on a square lattice
with $N_a^2\gg 1$ sites introduced in Ref. \cite{general} simplifies.
When acting onto such a subspace the
model refers to a two-component quantum liquid described
in terms of charge $c$ fermions and spin-neutral two-spinon $s1$ fermions.
The one- and two-electron subspace can be divided into smaller subspaces
that conserve $S_c$ and $S_s$. Those are spanned by
energy eigenstates whose generators have simple form in terms of $c$ and $s1$ fermion operators,
as given in Eqs. (\ref{non-LWS}) and (\ref{LWS-full-el}). 
When expressed in terms of $c$ and $s1$ fermion operators, the
Hubbard model on a square lattice in the one- and two-electron
subspace is the square-lattice quantum liquid further studied in Ref. \cite{companion}.
That in such a subspace the $c$ and $s1$ fermion momentum values
are good quantum numbers plays a key role in the investigations of
that reference. The one- and two-electron subspace considered in this paper
contains nearly the whole spectral weight generated from application
of one- and two-electron operators onto the exact ground state.

There is a large consensus that in the thermodynamic limit $N_a^2\rightarrow\infty$
long-range antiferromagnetic order sets in in the $m=0$ ground state of the
half-filled Hubbard model on the square lattice \cite{half-filling,2D-A2,bag-mech,2D-NM,Hubbard-T*-x=0,Kopec}. 
Consistently, in this paper strong evidence is found that provided that in the thermodynamic limit $N_a^2\rightarrow\infty$
a long-range antiferromagnetic order occurs in the ground state of the related isotropic spin-$1/2$ Heisenberg
model on the square lattice, a similar long range order sets in in that limit in the ground state
of the half-filled Hubbard model on the square lattice for $U/4t>0$. Our results indicate that for $U/4t>0$
a ground-state spontaneous symmetry breaking from $SO(3)\times SO(3)\times U(1)$ for
large but finite number of lattice sites $N_a^2\gg1$ to $[U(2)\times U(1)]/Z_2^2=[SO(3)\times U(1)\times U(1)]/Z_2$ 
in the thermodynamic limit $N_a^2\rightarrow\infty$ occurs at $x=0$ and $m=0$ due to emergence
of such a long-range antiferromagnetic order. Our analysis of the problem profits from the 
description of the ground states and low-lying energy eigenstates and corresponding 
state representations of the group 
$SO(3)\times SO(3)\times U(1)$ in terms of occupancy configurations of three effective
lattices. Our results indicate that the spin effective 
lattice being identical to the original lattice and thus $N_{a_{s}}^2=N_a^2$ is a
necessary condition for the occurrence of a ground-state long-range antiferromagnetic order as $N_a^2\rightarrow\infty$. 
The picture which emerges for $N_a^2\rightarrow\infty$ is that of a ground state with long-range antiferromagnetic order for half filling and a
short-range incommensurate-spiral spin order for $0<x\ll 1$ and thus $N_{a_{s}}^2<N_a^2$.

An interesting issue is whether the quantum phase transition separating
the $x=0$ and $m=0$ ground state from the $0<x\ll 1$ ground state at $T=0$ corresponds
to a ``deconfined'' quantum critical point \cite{Senthil-04,Anders-07,Anders-10}. That both at $x=0$ and
for $x>0$ the $M_s=2S_c$ spinons are confined within $N_{s1}=S_c$
spin-neutral two-spinon $s1$ fermions strongly suggests so. This would
imply that the critical point is characterized 
by deconfined spin-$1/2$ spinons coupled to some emergent 
$U(1)$ gauge field. Furthermore, in this paper the general rotated-electron description behind the $c$ and
$s1$ fermion operator representation was also used to provide evidence that the 
Mermin and Wagner Theorem may apply at half-filling to all values $U/4t>0$
of the Hubbard model on the square lattice.

The energy order parameters $\mu^0 = U\,m_{AF}\,\alpha^0$ of Eq. (\ref{mu0-ep0}) and
$2\vert\Delta\vert$ of Eq. (\ref{Delta}) have a different physical origin. The energy scale
$\mu^0$ can be used as order parameter of the $T=0$, $x=0$, and $m=0$ phase with long-range antiferromagnetic
order. Indeed, it is proportional to the $x=0$ and $m=0$ sub-lattice magnetization $m_{AF}$ of 
Eq. (\ref{m-AF}), which vanishes for $x>0$. In turn, the energy order parameter $2\vert\Delta\vert$ 
is associated with the $x>0$ and $m=0$ short-range spin correlations.
However, both $\mu^0$ and $2\vert\Delta\vert$
are at $x=0$ and for $x>0$, respectively, identified in the studies of Ref. \cite{companion} 
with the maximum pairing energy of the $-1/2$ and $+1/2$ spinons
of a composite spin-neutral two-spinon $s1$ fermion. 
The results of that reference extend the $m=0$ and $T=0$
short-range spin order found in this paper for $0<x\ll 1$ to a well-defined range of hole 
concentrations $0<x<x_*$. According to these results, the $x$ dependence 
$2\vert\Delta\vert \approx 2\Delta_0 (1-x/x_*^0)$ given in Eq. (\ref{Delta}) for $0<x\ll1$ is valid for
$x\in (0,x_*)$. This holds provided that approximately $U/4t\in (u_0,u_{\pi})$. Here $u_{\pi}>u_1$ 
where $u_1\approx 1.6$ is the $U/4t$ value at which $r_s=1/2$. For that $U/4t$ range the critical hole concentration
$x_*$ equals the $U/4t$-dependent parameter $x_* ^0 =2r_s/\pi$. The studies 
of Ref. \cite{companion} identify it with a critical hole concentration $x_*\equiv x_* ^0$ 
above which there is no short-range spin order at $T=0$. For $x>x_*$ and $T=0$ a spin disordered state without
short-range order for which the energy scale $2\vert\Delta\vert$ vanishes emerges.
Consistently, $2\vert\Delta\vert \rightarrow 0$ as $0<(x_*-x)\rightarrow 0$.
The short-range incommensurate-spiral spin order discussed here
for $0<x\ll1$ corresponds then to a limiting case of the general short-range 
spin order that according to the investigations of Ref. \cite{companion}
occurs for $0<x<x_*$ and approximately $U/4t\in (u_0,u_{\pi})$. (That
order occurs as well for $U/4t>u_{\pi}$, yet then the critical hole concentration 
$x_*$ may not be given by $x_* ^0 =2r_s/\pi$.)

As confirmed in that reference, for the square-lattice quantum liquid introduced in this paper 
the $c$ and $s1$ fermions play the role of ``quasiparticles''. 
There are three main differences relative to an isotropic Fermi liquid \cite{Pines}.
First, concerning the charge degrees of freedom, the non-interacting limit of the theory refers to $4t^2/U\rightarrow 0$
rather than to the limit of zero interaction $U\rightarrow 0$. Second,
in the $4t^2/U\rightarrow 0$ limit the $c$ fermions
and $s1$ fermions become the holes of the ``quasicharges'' of Ref.
\cite{Ostlund-06} and spin-singlet two-spin configurations of the
spins of such a reference rather than electrons. Indeed, only the
charge dynamical structure factor becomes that of non-interacting spinless
fermions. In turn, the one-electron and spin spectral distributions remain 
non-trivial. Third, for $U/4t>0$ the $s1$ band is full 
for initial $m=0$ ground states and displays a single hole for their one-electron excited states. 
As found in Ref. \cite{companion} its boundary line is anisotropic,
what is behind anomalous one-electron scattering properties. Those involve
the inelastic scattering of $c$ fermions with momenta near the isotropic
$c$ Fermi line with $s1$ fermions with momenta in the vicinity of the 
anisotropic boundary line. 

Concerning the relation to previous results on the Hubbard model on the square lattice and related models
by other authors, as discussed above in this paper our results are consistent with and complementary 
to those of Refs. \cite{Mura,Ostlund-06,Hubbard-T*-x=0,Kopec,LCO-Hubbard-NuMi}.
Elsewhere evidence is provided that upon addition of a weak three-dimensional
uniaxial anisotropy perturbation to the square-lattice quantum liquid, its short-range 
spin order coexists for $N_a^2\rightarrow\infty$, low temperatures, and a well-defined range of 
hole concentrations with a long-range superconducting order.

\begin{acknowledgments}
I thank Daniel Arovas, Nicolas Dupuis, Alejandro Muramatsu, 
Stellan \"Ostlund, Karlo Penc, Pedro D. Sacramento, and Maria Sampaio
for discussions and the support of the ESF Science Program INSTANS and grant PTDC/FIS/64926/2006.
\end{acknowledgments}
\appendix

\section{Results on the $c$ fermion and $s1$ bond-particle description needed for our studies}

For the LWS subspace defined in Ref. \cite{general} the rotated-electron operators of Eq. (\ref{rotated-operators})
can be expressed in terms of the $c$ fermion operators of real-space coordinate
$\vec{r}_j$ and rotated quasi-spin operators $q^{\pm}_{\vec{r}_j}=s^{\pm}_{\vec{r}_j}+p^{\pm}_{\vec{r}_j}$ 
and $q^{x_3}_{\vec{r}_j}=s^{x_3}_{\vec{r}_j}+p^{x_3}_{\vec{r}_j}$ and thus of spinon operators
$s^{\pm}_{\vec{r}_j}$ and $s^{x_3}_{\vec{r}_j}$ and $\eta$-spinon operators
$p^{\pm}_{\vec{r}_j}$ and $p^{x_3}_{\vec{r}_j}$. This is achieved by inverting the relations given in 
Eqs. (\ref{fc+}) and (\ref{rotated-quasi-spin}) with the result,
\begin{equation}
{\tilde{c}}_{\vec{r}_j,\uparrow}^{\dag} =
f_{\vec{r}_j,c}^{\dag}\,\left({1\over 2} +
s^{x_3}_{\vec{r}_j}+p^{x_3}_{\vec{r}_j}\right) + e^{i\vec{\pi}\cdot\vec{r}_j}\,
f_{\vec{r}_j,c}\,\left({1\over 2} - s^{x_3}_{\vec{r}_j} -p^{x_3}_{\vec{r}_j}\right) \, ;
\hspace{0.25cm}
{\tilde{c}}_{\vec{r}_j,\downarrow}^{\dag} =
(s^{-}_{\vec{r}_j}+p^{-}_{\vec{r}_j})\,(f_{\vec{r}_j,c}^{\dag} -
e^{i\vec{\pi}\cdot\vec{r}_j}\,f_{\vec{r}_j,c}) \, .
\label{c-up-c-down}
\end{equation}
The corresponding expressions of the rotated-electron annihilation operators
are trivially obtained from those provided here. 

As given in Eq. (\ref{fc+}), Fourier transform of the $c$ fermion operators $f_{\vec{r}_j,c}^{\dag}$
generates the corresponding momentum $c$ fermion operators $f_{\vec{q}_j,c}^{\dag}$.
Here the discrete momenta $\vec{q}_j$ are good quantum numbers \cite{general}.
In turn, the operators $f_{\vec{q}_j,\alpha\nu}^{\dag}$ of the 
$\alpha\nu$ fermions also introduced in Ref. \cite{general} act onto subspaces with constant values for the set of
numbers $S_{\alpha}$, $N_{\alpha\nu}$, and $\{N_{\alpha\nu'}\}$ for $\nu'>\nu$ and equivalently 
of the numbers $S_c$, $N_{\alpha\nu}$, and $\{N_{\alpha\nu'}\}$ for all $\nu'\neq \nu$. 
Here $\nu =1,...,C_{\alpha}$ is the number of $\eta$-spinon ($\alpha=\eta$) and spinon
($\alpha=s$) pairs confined within a composite $\alpha\nu$ fermion and the maximum $\nu$ value 
$C_{\alpha}$ is expressed below in Eq. (\ref{N-h-an}) in terms of $\alpha\nu$ fermion numbers $N_{\alpha\nu}$.
The above subspaces are spanned by mutually neutral states, that is states with constant values for the numbers 
of $\alpha\nu$ fermions and $\alpha\nu$ fermion holes. Hence  such states 
can be transformed into each other by $\alpha\nu$ band particle-hole processes. As a result
of the transformation that maps the $\alpha\nu$ bond particles into $\alpha\nu$ fermions and
provided that in the thermodynamic limit $N_a^2\rightarrow\infty$ the ratio $N_{\alpha\nu}/N_a^2$ 
is finite for the $\alpha\nu$ branch under consideration, the
latter composite fermions have long-range interactions associated with an effective
vector potential ${\vec{A}}_{\alpha\nu} ({\vec{r}}_j)$ \cite{general,Giu-Vigna,Wang}.
For the one- and two-electron subspace considered in this paper 
the ratio $N_{\alpha\nu}/N_a^2$ remains finite in the limit $N_a^2\rightarrow\infty$
only for the $s1$ fermion branch. For it the effective vector potential ${\vec{A}}_{s1} ({\vec{r}}_j)$ reads \cite{general},
\begin{eqnarray}
{\vec{A}}_{s1} ({\vec{r}}_j) & = & \Phi_0\sum_{j'\neq j}
n_{\vec{r}_{j'},s1}\,{{\vec{e}}_{x_3}\times ({\vec{r}}_{j'}-{\vec{r}}_{j})
\over ({\vec{r}}_{j'}-{\vec{r}}_{j})^2} \, ; \hspace{0.35cm} 
n_{\vec{r}_j,s1} = f_{\vec{r}_j,s1}^{\dag}\,f_{\vec{r}_j,s1} \, ,
\nonumber \\
{\vec{B}}_{s1} ({\vec{r}}_j) & = & {\vec{\nabla}}_{\vec{r}_j}\times {\vec{A}}_{s1} ({\vec{r}}_j)
=  \Phi_0\sum_{j'\neq j}
n_{\vec{r}_{j'},s1}\,\delta ({\vec{r}}_{j'}-{\vec{r}}_{j})\,{\vec{e}}_{x_3}
\, ; \hspace{0.35cm} \Phi_0 = 1 \, .
\label{A-j-s1-3D}
\end{eqnarray}   
Here ${\vec{B}}_{s1}$ is the corresponding fictitious magnetic field,
${\vec{e}}_{x_3}$ is the unit vector perpendicular to the plane, and we use units such 
that the fictitious magnetic flux quantum is given by $\Phi_0=1$. It follows from
the form of the effective vector potential ${\vec{A}}_{s1} ({\vec{r}}_j)$ that
the present description leads to the intriguing situation where the $s1$ 
fermions interact via long-range forces while all interactions in the original Hamiltonian are onsite. 

The theory associated with the operator description introduced in Ref. \cite{general} for the Hubbard
model on the square lattice refers to a well-defined 
vacuum. For hole concentrations $0\leq x<1$ and
maximum spin density $m=(1-x)$ reached at a critical magnetic field
$H_c$ parallel to the square-lattice plane the $c$ fermion operators are 
invariant under the electron - rotated-electron unitary 
transformation. Then there is a fully polarized  
vacuum $\vert 0_{\eta s}\rangle$, which remains
invariant under such a transformation. It reads,
\begin{equation}
\vert 0_{\eta s}\rangle = \vert 0_{\eta};N_{a_{\eta}}^2\rangle\times\vert 0_{s};N_{a_{s}}^2\rangle
\times\vert GS_c;2S_c\rangle \, .
\label{vacuum}
\end{equation}
Here the $\eta$-spin $SU(2)$ vacuum $\vert 0_{\eta};N_{a_{\eta}}^2\rangle$ 
associated with $N_{a_{\eta}}^2$ independent $+1/2$
$\eta$-spinons, the spin $SU(2)$ vacuum $\vert 0_{s};N_{a_{s}}^2\rangle$ 
with $N_{a_{s}}^2$ independent $+1/2$ spinons, and the $c$ $U(1)$
vacuum $\vert GS_c;2S_c\rangle$ with $N_c=2S_c$ $c$ fermions
remain invariant under the electron - rotated-electron unitary transformation. 
The explicit expression of the state $\vert GS_c;2S_c\rangle$ 
of Eq. (\ref{vacuum}) in terms of the vacuum $\vert GS_c;0\rangle$ appearing in
Eq. (\ref{LWS-full-el}) is $\prod_{{\vec{q}}}f^{\dag}_{{\vec{q}},c}\vert GS_c;0\rangle$.
The vacuum $\vert GS_c;0\rangle$ is part of the electron and
rotated-electron vacuum. The form of the latter vacuum is given by Eq. (\ref{vacuum}) with
$N_{a_{\eta}}^2=N_a^2$, $N_{a_{s}}^2=2S_c=0$, and thus
$\vert GS_c;2S_c\rangle$ replaced by $\vert GS_c;0\rangle$. Only for 
a $m=(1-x)$ fully polarized state are the state $\vert GS_c;2S_c\rangle$
and the corresponding $N_c=2S_c$ fermions 
invariant under the electron - rotated-electron unitary transformation for $U/4t>0$. 

The $\alpha\nu$ momentum band of the operators $f_{\vec{q}_j,\alpha\nu}^{\dag}$
is associated with a well-defined $\alpha\nu$ effective lattice. The
number $N_{a_{\alpha\nu}}^2$ of the $\alpha\nu$ band discrete
momentum values exactly equals the number of sites of such
an effective lattice. It is given by, 
\begin{equation}
N_{a_{\alpha\nu}}^2 = [N_{\alpha\nu} + N^h_{\alpha\nu}] \, ,
\label{N*}
\end{equation}
where the number of unoccupied sites reads,
\begin{equation}
N^h_{\alpha\nu} = 
[2S_{\alpha}+2\sum_{\nu'=\nu+1}^{C_{\alpha}}(\nu'-\nu)N_{\alpha\nu'}] =
[N_{a_{\alpha}}^{2} - 
\sum_{\nu' =1}^{C_{\alpha}}(\nu +\nu' - \vert \nu-\nu'\vert)N_{\alpha\nu'}] 
\, ; \hspace{0.35cm} C_{\alpha} = \sum_{\nu =1}^{C_{\alpha}}\nu\,N_{\alpha\nu}\, .
\label{N-h-an}
\end{equation}
For the particular case of $\nu=1$ the expression (\ref{N-h-an})
of the number of unoccupied sites of the $\alpha 1$ effective lattices simplifies to,
\begin{equation}
N^h_{\alpha 1} = [N_{a_{\alpha}}^{2} - 2B_{\alpha}]
\, ; \hspace{0.35cm} B_{\alpha} = \sum_{\nu =1}^{C_{\alpha}}N_{\alpha\nu}
\, ; \hspace{0.25cm}  \alpha = \eta \, , s \, . 
\label{Nh+Nh}
\end{equation}
This number equals that of $\alpha 1$ fermion holes in the $\alpha 1$ band. 

According to the results of Ref. \cite{general}, all sites of the $s1$ effective lattice of $x>0$ and $m=0$ ground states 
are occupied and hence there are no unoccupied sites. In turn, the dominant contributions to the
one-electron and two-electron excitations involve states with one and none or two unoccupied
sites, respectively. For the square-lattice quantum liquid the expression of the related
conserved number $P^h_{s1}$ introduced in that reference simplifies to,
\begin{equation}
P^h_{s1} \equiv e^{i\pi N^h_{s1}}=e^{i2\pi S_{s}}=e^{i2\pi S_{c}}=e^{i\pi N} = \pm 1 \, .
\label{Ns1h-general}
\end{equation}
Here $N$ denotes the number of electrons. For the Hubbard model on the square lattice
in the one- and two-electron subspace considered in this paper the number
$N^h_{s1}$ of unoccupied $s1$ effective lattice sites and thus of $s1$ fermion holes
in the $s1$ momentum band is a good quantum number.

As discussed in Ref. \cite{general}, the $\eta$-spinon, spinon, and $c$ fermion description 
contains full information about the relative positions of the sites of 
the $\eta$-spin and spin effective lattices in the original lattice. Hence it turns out that  
within the $N_a^2\gg 1$ limit and for finite values of
the hole concentration $x$ (and electron density $n=(1-x)$) the 
$\eta$-spin (and spin) effective lattice can be represented by a
square lattice with spacing $a_{\eta}$ (and $a_s$) given by, 
\begin{equation}
a_{\alpha} = {L\over N_{a_{\alpha}}} = {N_a\over N_{a_{\alpha}}}\, a
\, ; \hspace{0.25cm} \alpha = \eta \, , s \, .
\label{a-alpha}
\end{equation}
Moreover, provided that in the thermodynamic limit $N_a^2\rightarrow\infty$ the ratio 
$N_{\alpha\nu}/N_a^2$ is finite, the related $\alpha\nu$ effective lattices can be represented by 
square lattices with spacing,
\begin{equation}
a_{\alpha\nu} = {L\over N_{a_{\alpha\nu}}} = 
{N_a\over N_{a_{\alpha\nu}}}\, a
= {N_{a_{\alpha}}\over N_{a_{\alpha\nu}}}\, a_{\alpha}
\, ; \hspace{0.25cm} 
N_{a_{\alpha\nu}} \geq 1 \, ,
\label{a-a-nu}
\end{equation}
where $\nu = 1,...,C_{\alpha}$ and $\alpha =\eta ,s$. 
In turn, the corresponding $\alpha\nu$ bands whose number of
discrete momentum values is also given by $N_{a_{\alpha\nu}}^2$
are well defined even when $N_{a_{\alpha\nu}}^2$ is
given by a finite small number, $N_{a_{\alpha\nu}}^2=1,2,3,...$

Finally, an important and useful property found in Ref. \cite{general} is that the $\eta\nu$ fermions with any
number $\nu =1,...,C_{\eta}$ of $\eta$-spinon pairs and $s\nu$ fermions with
$\nu =2,...,C_{s}$ spinon pairs whose energy is given by,
\begin{equation} 
\epsilon_{\eta\nu} = 2\nu\mu \, , \hspace{0.25cm} \nu =1,...,C_{\eta}
\, ; \hspace{0.50cm}
\epsilon_{s\nu} = 2\nu\mu_B\,H 
\, , \hspace{0.25cm} \nu =2,...,C_{s} \, ,
\label{invariant-V}
\end{equation}
remain invariant under the electron - rotated-electron
unitary transformation.
Here $H$ denotes the magnitude of a magnetic field
aligned parallel to the square-lattice plane. 
Such $\eta\nu$ fermions and $s\nu$ fermions are non-interacting objects. Hence
their energy is additive in the individual energies
of the corresponding $2\nu$ $\eta$-spinons
and spinons, respectively. For $U/4t>0$ such quantum objects 
refer to the same occupancy configurations in terms
of both rotated electrons and electrons. 


\end{document}